\documentclass[journal]{vgtc}         

\ifpdf
  \pdfoutput=1\relax                   
  \usepackage{graphicx}                
  \DeclareGraphicsExtensions{.pdf,.png,.jpg,.jpeg} 
\else
  \usepackage{graphicx}                
  \DeclareGraphicsExtensions{.eps}     
\fi%

\graphicspath{{figures/}{pictures/}{images/}{./}} 

\usepackage{microtype}                 
\PassOptionsToPackage{warn}{textcomp}  
\usepackage{textcomp}                  
\usepackage{mathptmx}                  
\usepackage{times}                     
\usepackage{cite}                      
\usepackage{tabu}                      
\usepackage{booktabs}                  
\usepackage{amsmath}                   
\usepackage{color}
\usepackage{mathptmx}
\usepackage{amssymb}
\usepackage{enumitem}
\usepackage{multicol}
\usepackage{anyfontsize}

\onlineid{1163}

\vgtccategory{Research}
\vgtcpapertype{Algorithm / Technique}

\title{Wasserstein Distances, Geodesics and Barycenters of Merge Trees}

\author{
Mathieu Pont, Jules Vidal, Julie Delon and Julien Tierny}
\authorfooter{
\item M. Pont, J. Vidal, and J. Tierny are with Sorbonne Université and CNRS.
 E-mail: \{mathieu.pont, jules.vidal, julien.tierny\}@sorbonne-universite.fr.
\item J. Delon is with University of Paris. E-mail: julie.delon@u-paris.fr
}

\shortauthortitle{Pont \MakeLowercase{\textit{et al.}}: 
Wassertein Distances, Geodesics and Barycenters for Merge Trees}

\abstract{This is the abstract.
Unified computational framework.
In contrast to previous work, our framework requires not initial correspondence 
between the trees to process and is less sensitive by design to outliers.\\
line3 -- stress that we used real-life ensembles coming from published 
benchmarks or scivis contests.\\
line4\\
line5\\
line6\\
line7\\
line8\\
line9\\
line10
}

\abstract{
This paper presents a unified computational framework for the estimation of 
distances, 
geodesics 
and barycenters of merge trees.
We extend recent work on the edit distance
\cite{SridharamurthyM20}
and introduce a new metric, called 
the \emph{Wasserstein} distance between merge trees, which is purposely 
designed to enable efficient computations of geodesics and barycenters.
Specifically, our new distance is strictly equivalent to the $L^2$-Wasserstein 
distance between extremum persistence diagrams, but it is restricted to a 
smaller 
solution space, namely, the space of rooted partial isomorphisms between branch 
decomposition trees. 
This 
enables a simple extension of 
existing optimization frameworks \cite{Turner2014} for geodesics 
and barycenters from persistence 
diagrams to merge trees. 
We introduce a task-based algorithm which 
can be generically applied to distance, geodesic, barycenter or 
cluster computation. 
The
task-based nature 
of our approach 
enables 
further accelerations with 
shared-memory 
parallelism. 
Extensive experiments on public ensembles and SciVis 
contest benchmarks demonstrate the efficiency of our approach -- with 
barycenter computations in the orders of minutes for the largest examples -- as 
well as its qualitative ability to generate representative barycenter merge 
trees, visually summarizing the features of interest found in the ensemble.
We 
show
the utility of 
our contributions with dedicated 
visualization applications: feature tracking, temporal reduction and 
ensemble clustering. 
We provide a lightweight C++ implementation 
that can be used to reproduce our results.
}

\teaser{
  \centering
  \includegraphics[width=\linewidth]{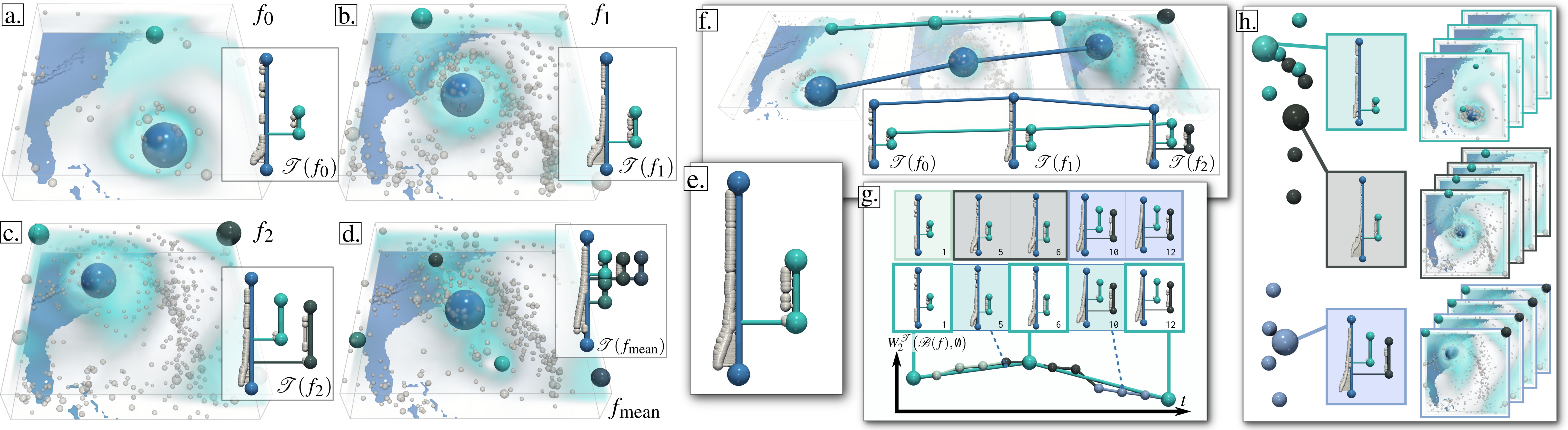}
  \caption{The merge trees of three members (a-c) of the Isabel ensemble (wind 
velocity) concisely and visually encode the number and salience of the 
features of interest found in the data (eyewall and region of high speed 
wind, blue and cyan). They also 
describe how these features are globally connected in the data.
In these trees, branches with a low persistence 
(less than 
$20\%$ 
of the function range) are shown with small white arcs.
The pointwise mean for the three members (d) exhibits 5 salient maxima
(due to distinct eyewall 
locations, blue, cyan and black) and its merge tree is not 
representative of the input 
trees (containing at most 3 large features).
In contrast, the Wasserstein barycenter 
(e) is representative of the input trees, with a number and persistence of 
large branches that better match the input trees (a-c). Our 
framework for distances, geodesics and barycenters enables a 
variety of merge tree based applications, including (f) feature tracking, 
(g) temporal reduction -- \emph{key frames} are automatically identified 
(white insets) and deleted merge trees (blue insets) are 
accurately reconstructed with geodesics 
-- and (h) ensemble clustering and summarization --  the clusters 
and centroids automatically computed by our approach provide a visual summary 
of the main trends of features 
found in the ensemble.
}
	\label{fig:teaser}
}

\vgtcinsertpkg

\begin{document}


\firstsection{Introduction}
\maketitle

\newcommand{\domain}{\mathcal{M}}
\newcommand{\range}{\mathbb{R}}
\newcommand{\sublevelset}[1]{#1^{-1}_{-\infty}}
\newcommand{\superlevelset}[1]{#1^{-1}_{+\infty}}
\newcommand{\Star}{St}
\newcommand{\Link}{Lk}
\newcommand{\simplex}{\sigma}
\newcommand{\face}{\tau}
\newcommand{\lowerlink}{\Link^{-}}
\newcommand{\upperlink}{\Link^{+}}
\newcommand{\Index}{\mathcal{I}}
\newcommand{\offset}{o}
\newcommand{\Natural}{\mathbb{N}}
\newcommand{\criticalSet}{\mathcal{C}}
\newcommand{\diagram}{\mathcal{D}}
\newcommand{\wasserstein}[1]{W^{\diagram}_#1}
\newcommand{\projection}{\Delta}
\newcommand{\hierarchy}{\mathcal{H}}
\newcommand{\decimation}{D}
\newcommand{\xDimD}{L_x^\decimation}
\newcommand{\yDimD}{L_y^\decimation}
\newcommand{\zDimD}{L_z^\decimation}
\newcommand{\xDim}{L_x}
\newcommand{\yDim}{L_y}
\newcommand{\zDim}{L_z}
\newcommand{\Grid}{\mathcal{G}}
\newcommand{\GridD}{\mathcal{G}^\decimation}
\newcommand{\x}{\phantom{x}}
\newcommand{\Mod}{\;\mathrm{mod}\;}
\newcommand{\NN}{\mathbb{N}}
\newcommand{\forwardIntegralLine}{\mathcal{L}^+}
\newcommand{\backwardIntegralLine}{\mathcal{L}^-}
\newcommand{\triangulationOp}{\phi}
\newcommand{\decimationOp}{\Pi}
\newcommand{\isovalue}{w}
\newcommand{\persistence}{\mathcal{P}}
\newcommand{\pointMetric}{d}
\newcommand{\diagramSet}{\mathcal{S}_\mathcal{D}}
\newcommand{\diagramSpace}{\mathbb{D}}
\newcommand{\jointree}{\mathcal{T}^-}
\newcommand{\splittree}{\mathcal{T}^+}
\newcommand{\mergetree}{\mathcal{T}}
\newcommand{\mergetreeSet}{\mathcal{S}_\mathcal{T}}
\newcommand{\branchset}{\mathcal{S}_\mathcal{B}}
\newcommand{\branchspace}{\mathbb{B}}
\newcommand{\mergetreeSpace}{\mathbb{T}}
\newcommand{\editdistance}{D_E}
\newcommand{\wassersteinTree}{W^{\mergetree}_2}
\newcommand{\distanceSequence}{d_S}
\newcommand{\branchtree}{\mathcal{B}}
\newcommand{\branchtreeSet}{\mathcal{S}_\mathcal{B}}
\newcommand{\branchtreeSpace}{\mathbb{B}}
\newcommand{\forest}{\mathcal{F}}
\newcommand{\sequenceSpace}{\mathbb{S}}
\newcommand{\forestMatrix}{\mathbb{F}}
\newcommand{\treeMatrix}{\mathbb{T}}
\newcommand{\normalizedLocation}{\mathcal{N}}
\newcommand{\normalizedWasserstein}{W^{\normalizedLocation}_2}

\newcommand{\julien}[1]{\textcolor{black}{#1}}
\newcommand{\mathieu}[1]{\textcolor{green}{#1}}
\newcommand{\note}[1]{\textcolor{magenta}{#1}}
\newcommand{\cutout}[1]{\textcolor{blue}{#1}}
\renewcommand{\cutout}[1]{}

\newcommand{\revision}[1]{\textcolor{black}{#1}}

\renewcommand{\sectionautorefname}{Sec.}
\renewcommand{\subsectionautorefname}{Sec.}
\renewcommand{\equationautorefname}{Eq.}
\renewcommand{\tableautorefname}{Tab.}

\newcommand{\mycaption}[1]{
\vspace{-1.25ex}
\caption{#1}
\vspace{-2.5ex}
}

Modern datasets, acquired or simulated, are continuously gaining in geometrical 
complexity, thanks to the ever-increasing accuracy of acquisition devices or 
computing power of high performance systems. 
This geometrical complexity makes interactive exploration and analysis 
difficult, which challenges the interpretation of the data by the end users. 
This motivates the definition of expressive data abstractions, capable of 
capturing the main features of the data into concise representations, which 
visually convey the most important information to the users.

\begin{figure}
\centering
\includegraphics[width=0.48\linewidth]{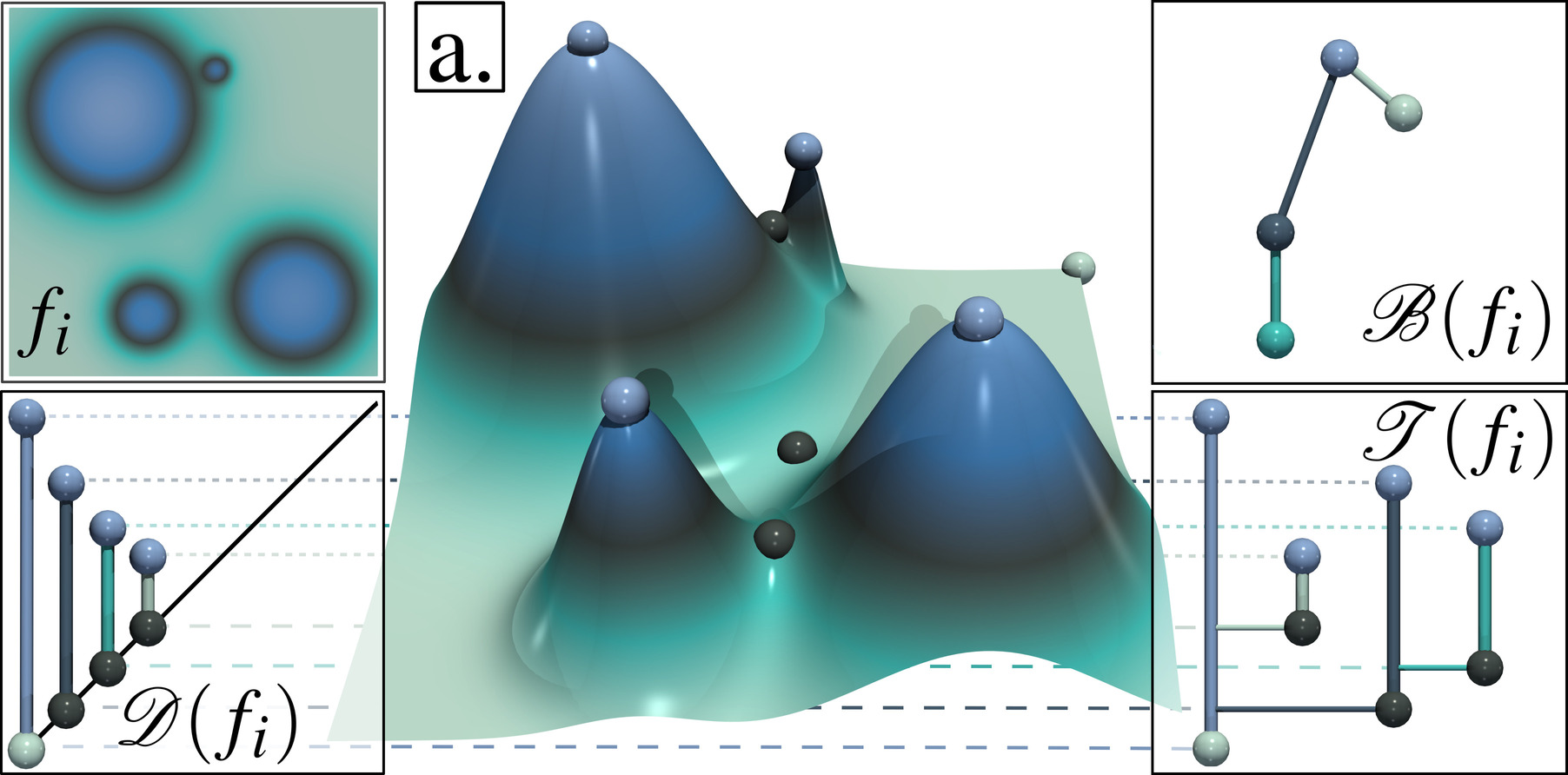}
\hfill
\includegraphics[width=0.48\linewidth]{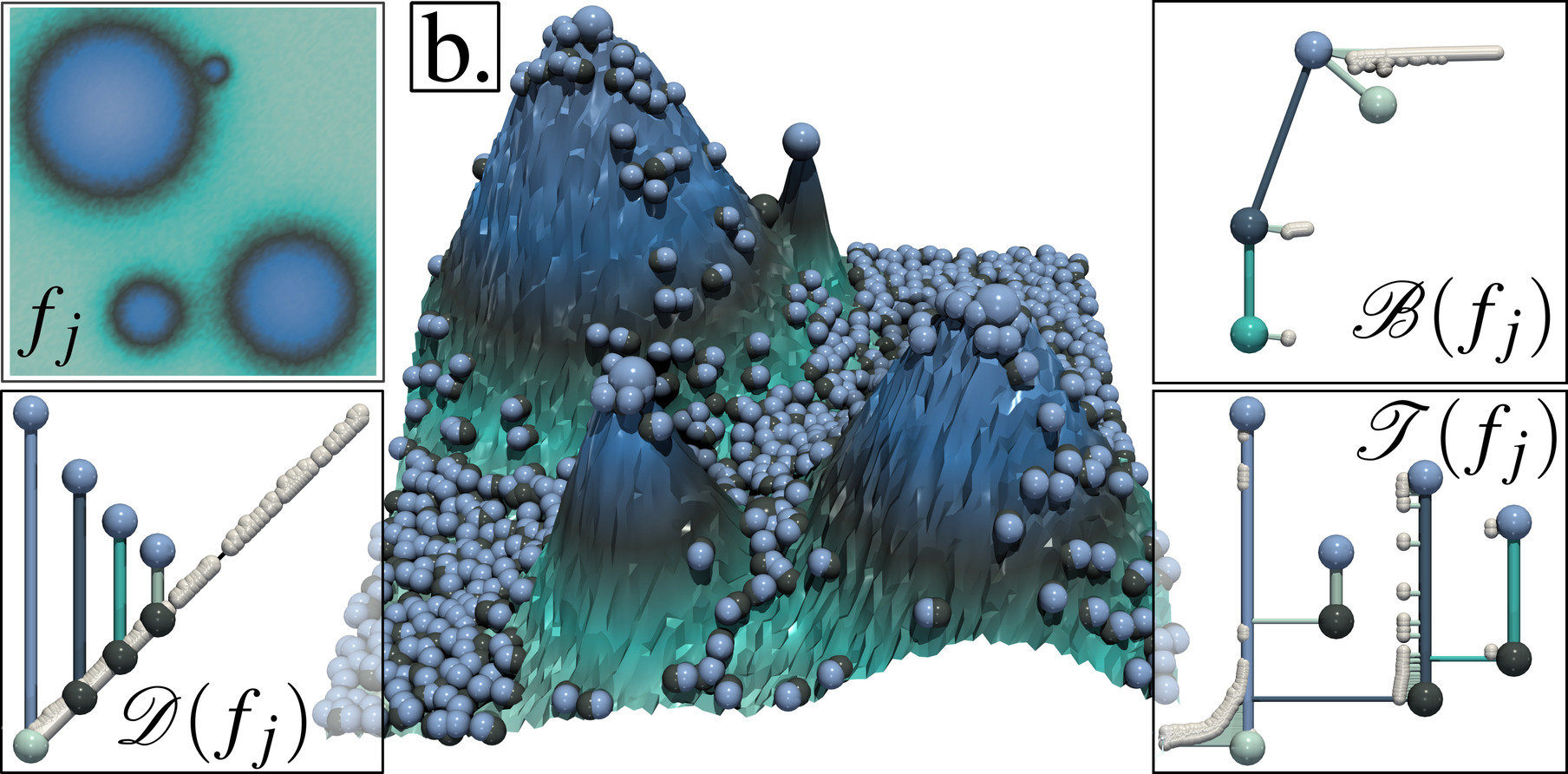}
\mycaption{
Critical points (spheres, white: minima, blue: maxima, other: saddles), 
persistence diagram (bottom left), merge tree (bottom right) and branch 
decomposition tree (top right) of a clean (a) and noisy (b) 2D 
scalar field. 
In both cases, four main hills are clearly 
represented with salient features in the persistence diagram and the 
merge tree. 
Branches with low persistence (less than $10\%$ of the function range) are 
shown with small white arcs.
They 
correspond to 
noisy features in the data (b).}
\label{fig_mergeTree}
\end{figure}

In that context, Topological Data Analysis (TDA) \cite{edelsbrunner09} forms a 
family of generic, robust, and efficient techniques whose utility has been 
demonstrated in a number of visualization tasks \cite{heine16} for revealing 
the implicit structural patterns present in complex datasets. 
Examples of popular application fields include 
turbulent combustion \cite{laney_vis06, bremer_tvcg11, gyulassy_ev14},
 material sciences \cite{gyulassy_vis07, gyulassy_vis15, favelier16},
 nuclear energy \cite{beiNuclear16},
fluid dynamics \cite{kasten_tvcg11}, 
bioimaging \cite{carr04, topoAngler, beiBrain18},
quantum chemistry \cite{chemistry_vis14, harshChemistry, Malgorzata19} or 
astrophysics \cite{sousbie11, shivashankar2016felix}.
\revision{Among the
data abstractions developed in TDA
(see \autoref{sec_relatedWork}), the merge tree \cite{carr00},
which describes the global structure of the connected components of the
sub-level sets of scalar datasets
(\autoref{fig_mergeTree}),
is a
prominent example in the visualization literature \cite{carr04, bremer_tvcg11,
topoAngler}.}


In practice, in addition to the increasing geometrical complexity of 
datasets, users are also confronted to the emergence of \emph{ensemble 
datasets}, where a given phenomenon is not described with only one dataset, but 
with a collection of datasets, called \emph{ensemble members}.
Regarding topological features, a topological data
abstraction such as the merge tree can be computed for each ensemble member 
(possibly in-situ  \cite{insitu, AyachitBGOMFM15}).
\revision{Then, a major challenge for end users is
the interpretation of the resulting ensemble of merge trees. To address this,
 a statistical analysis framework for merge trees is needed, requiring
several key building blocks,
such as: distances (to
compare merge trees), geodesics (to visualize optimum transitions between them),
and barycenters (to visualize one merge tree
\emph{representative} of a set). These building blocks have been well studied
for persistence diagrams \cite{Turner2014,
lacombe2018, vidal_vis19}. However,
persistence diagrams suffer from a
lack of specificity
(\autoref{fig_mergeTreeVSpersistenceDiagram}), which can prevent the
identification of distinct feature trends within the ensemble.}

This paper addresses this problem by introducing a unified computational 
framework for the automatic computation of distances, geodesics, barycenters 
and clusters
of merge trees. 
\cutout{By extending recent work on the edit 
distance \cite{SridharamurthyM20}, we derive a new distance 
metric between 
merge trees which has several advantages over previous work. Our new metric has 
a stronger connection to established metrics between persistence diagrams, 
which facilitates its interpretation. Moreover, it enables a simple approach 
for the computation of geodesics (i.e. length minimizing morphings) between 
merge trees, which itself enables the definition of 
an effective strategy for the automatic computation of barycenters of merge 
trees.}
In particular, we extend recent work on the edit distance
\cite{SridharamurthyM20}
and introduce a new metric, called 
the \emph{Wasserstein} distance between merge trees, which is purposely 
designed to enable efficient computations of geodesics (i.e. length minimizing 
morphings)  and barycenters.
In that regard, our work can be interpreted as an extension of previous work on 
the edit distance \cite{SridharamurthyM20}, to adapt it to the 
optimization strategy previously developed for the computation of barycenters 
of persistence diagrams  \cite{Turner2014}.
We 
present efficient, task-based  algorithms using shared-memory 
parallelism, resulting in the computation of distances, geodesics and 
barycenters in practical times for real-life datasets. 
We illustrate the utility of 
each of our contributions
in dedicated visualization 
tasks. First, we show that our distance computation algorithm can be used for 
a
merge-tree based tracking of features through time. Second, we 
show that our framework for computing geodesics between merge trees can be used 
for the reliable sub-sampling of temporal sequences of merge trees. Third, we 
illustrate the utility of our barycenters for clustering ensemble members based 
on their merge trees, while providing cluster centroids which visually  
summarize the main features of interest present in each cluster.

\subsection{Related work}
\label{sec_relatedWork}
The literature related to our work can be classified into three main 
groups, reviewed in the following: 
\emph{(i)} uncertainty visualization,
\emph{(ii)} ensemble visualization, and
\emph{(iii)} topological methods for ensembles.

\noindent
\textbf{\emph{(i)} Uncertainty visualization:}
Variability in data can be modeled and encoded in several ways. In particular, 
\emph{uncertain} datasets capture variability by modeling each point of the 
domain as a random variable, whose variability is explicitly modeled by an 
estimator of an a priori probability density function (PDF). The analysis 
of uncertain data is a notoriously challenging problem in 
visualization, described in several 
surveys \cite{uncertainty_survey1,unertainty_survey2, uncertainty1, 
uncertainty2, uncertainty3, uncertainty4}. Early techniques focused on 
estimating the entropy of the random variables \cite{uncertainty_entropy}, 
their correlations \cite{uncertainty_correlation} or their gradient variations 
\cite{uncertainty_gradient}. 
\cutout{The positional uncertainty of geometrical objects computed from uncertain data 
also needs to be estimated. For example, }
The 
positional uncertainty of level sets has been studied for several 
interpolation schemes and PDF 
models \cite{uncertainty_isosurface1,uncertainty_isosurface2,
    uncertainty_isosurface3, uncertainty_isosurface4, uncertainty_isocontour1,
uncertainty_nonparam,uncertainty_isocontour2, uncertainty_interp,athwale19}. 
Similarly, the positional uncertainty of critical points has been studied for 
Gaussian \cite{liebmann1,otto1,otto2,petz} or uniform distributions 
\cite{gunther,bhatia,szymczak}. 
A general limitation of existing methods for uncertain data is their 
dependence on the specific PDF model for which they have been designed. This 
reduces their usability for ensemble data, where the PDF estimated from 
the ensemble members
can follow an arbitrary, unknown model. Also, most 
existing techniques for uncertain data do not consider multi-modal PDF models, 
which is however necessary when several, distinct trends are present in the 
ensemble data.

\begin{figure}
\includegraphics[width=\linewidth]{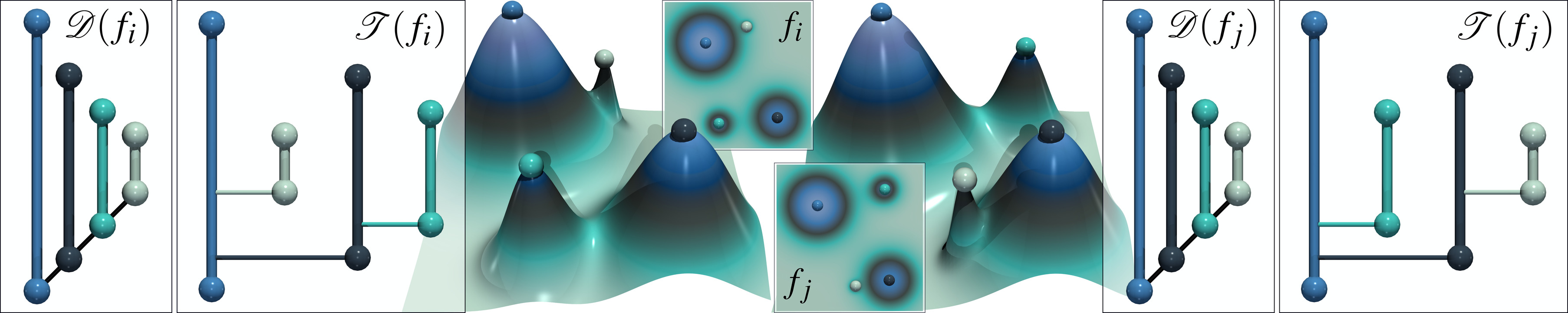}
\mycaption{The persistence diagram, $\diagram(f_i)$,
and the merge tree, $\mergetree(f_i)$,
both 
visually summarize the number, data range and salience of the features of 
interest present in the data. However, the persistence diagram
represents each 
individual feature independently, while the merge tree additionally describes 
how they connect together.
This results in a lack of specificity for 
the persistence diagram which can yield identical 
data 
representations
for 
significantly distinct datasets (from left to right, 
the gaussians with
white and cyan spheres
have been swapped). In contrast, the merge tree captures this nuance and 
produces two distinct data 
representations.}
\label{fig_mergeTreeVSpersistenceDiagram}
\end{figure}

\cutout{
\begin{figure}
\includegraphics[width=\linewidth]{img/states/l2average/l2average.jpg}
\caption{Synthetic ensemble of a pattern with 2 gaussians and additive noise 
(a). The merge tree of the pointwise mean (b) contains 8 persistent features 
although each of the input ensemble members contain only 2 features. The 
Wasserstein barycenter (c) provides a merge tree which is representative of the 
set, with a feature number, range and salience which better describes the input 
ensemble (2 large features).}
\label{fig_limitationPointwiseMean}
\end{figure}}

\noindent
\textbf{\emph{(ii)} Ensemble visualization:}
Another way to model and encode variability in data consists in considering
ensemble datasets. In this setting, the variability is directly encoded by an 
ensemble of empirical observations (i.e. the \emph{members} of the 
ensemble).\cutout{In that 
regard, ensemble visualization methods can be interpreted as a statistical view 
on data variability, whereas uncertainty visualization methods can be considered 
as probabilistic.} Current approaches to ensemble visualization 
typically compute some geometrical objects describing the features of interest 
(level sets, streamlines, etc), for each member of the ensemble. Then, an
aggregation phase estimates a \emph{representative} object for the resulting 
ensemble of geometrical objects. For instance, spaghetti plots 
\cite{spaghettiPlot} are a typical example for studying level-set variability, 
especially for weather data \cite{Potter2009,Sanyal2010}. More specifically, 
box-plots \cite{whitaker2013} describe the variability of contours and curves 
\cite{Mirzargar2014}. For flow ensembles, Hummel et al. \cite{Hummel2013} 
introduce a Lagrangian framework for classification purposes. Clustering 
techniques have been investigated, to identify the main trends, and 
their variability, in ensembles of streamlines \cite{Ferstl2016} and 
isocontours \cite{Ferstl2016b}. However, 
only few approaches 
have applied this overall aggregation strategy to topological objects. 
Favelier et al. 
\cite{favelier2018} and Athawale et al. \cite{athawale_tvcg19} introduced 
approaches for analyzing the variability of critical points and 
gradient separatrices respectively.
Several techniques attempted to 
generate an 
aggregated
contour tree from an ensemble based on overlap-driven 
heuristics \cite{Wu2013ACT,Kraus2010VisualizationOU}. Recently, Lohfink et al.
\cite{LohfinkWLWG20} introduced an approach for the consistent layout of 
multiple contour trees, to support effective visual comparisons between the 
contour trees of the distinct members of an ensemble.
Although the above techniques addressed the visualization of ensembles of 
topological objects, they did not focus explicitly on the computation of 
a \emph{representative} of multiple topological objects, such as barycenters.

\noindent 
\textbf{\emph{(iii)} Topological methods:} Concepts and algorithms from 
computational 
topology \cite{edelsbrunner09} have been investigated, adapted and extended by 
the visualization community for more than twenty years \cite{heine16, 
surveyComparison2021}.
Popular topological representations include the persistence diagram 
\cite{edelsbrunner02, edelsbrunner09}
\revision{(\autoref{sec_background_persistenceDiagrams}),}
which represents the 
population of features of interest in function of their salience, and which 
can be computed via matrix reduction 
\cite{edelsbrunner09, dipha}. The Reeb graph \cite{biasotti08}, which describes 
the connectivity evolution of level sets, has also been widely studied and 
several efficient algorithms have been documented \cite{pascucci07, 
tierny_vis09, parsa12, DoraiswamyN13}, including parallel algorithms 
\cite{gueunet_egpgv19}. 
Efficient algorithms have also been documented for its variants,
the 
merge and contour trees \cite{tarasov98, carr00} 
\revision{(\autoref{sec_background_mergeTrees}),}
and parallel algorithms have 
also been described \cite{MaadasamyDN12, AcharyaN15, 
CarrWSA16, gueunet_tpds19}. The Morse-Smale complex \cite{EdelsbrunnerHZ01, 
EdelsbrunnerHNP03, 
BremerEHP03}, which depicts the 
global behaviour of integral lines, is another popular topological data 
abstraction in visualization \cite{Defl15}. Robust and efficient algorithms 
have been introduced for its computation  \cite{robins_pami11, ShivashankarN12, 
gyulassy_vis18} based on Discrete Morse Theory \cite{forman98}.

\begin{figure}
\includegraphics[width=\linewidth]{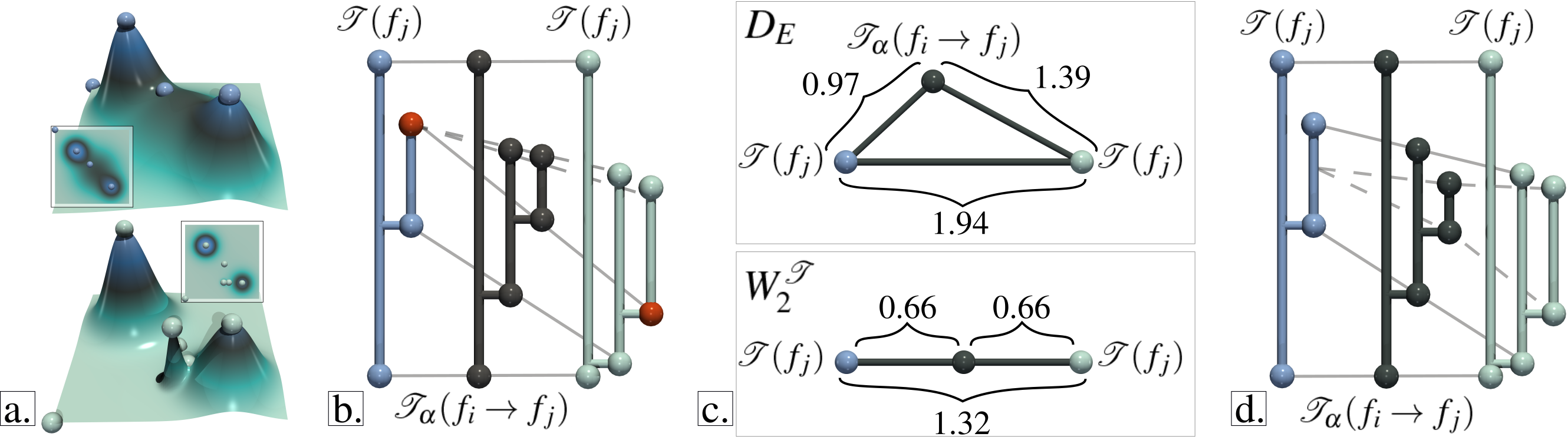}
\mycaption{In this example with
two datasets $f_i$ and $f_j$ (a), the optimal matchings (gray) with regard to 
the edit distance $\editdistance$ \cite{SridharamurthyM20} (b) map a maximum to 
a saddle (red spheres). 
The resulting linear interpolation 
%
$\mergetree_\alpha(f_i \rightarrow f_j)$ 
(b) does not describe a shortest path 
between the input trees (c, top):
$\editdistance\big(\mergetree(f_i), \mergetree_\alpha(f_i \rightarrow f_j)\big) 
+ \editdistance\big(\mergetree_\alpha(f_i \rightarrow 
f_j), \mergetree(f_j)\big) >
\editdistance\big(\mergetree(f_i), \mergetree(f_j)\big)$. In contrast, our new 
metric $\wassersteinTree$ 
enables
linear interpolations (d) which exactly 
coincide with shortest paths (c, bottom). The numbers included in (c) are the 
actual 
values for $\editdistance$ (top) and $\wassersteinTree$ 
(bottom) for this example.}
\label{fig_counterExampleEditDistance}
\end{figure}

Distance 
metrics, which are necessary ingredients for the computation of barycenters, 
have been studied for most of the above objects. Inspired by the literature in 
optimal transport \cite{Kantorovich, monge81}, the Wasserstein distance between 
persistence diagrams \cite{edelsbrunner09} 
\revision{(\autoref{sec_background_persistenceDiagrams})}
and its variant the 
Bottleneck distance \cite{edelsbrunner02} have been extensively studied. They 
are based on a bipartite assignment problem,
for which exact 
\cite{Munkres1957} and approximate \cite{Bertsekas81, Kerber2016} 
implementations are publicly available \cite{ttk17}. Several similarity 
measures have been introduced for Reeb graphs \cite{HilagaSKK01} and their 
variants \cite{SaikiaSW14_branch_decomposition_comparison}. 
However, since these measures are not distance metrics (the 
preservation of the triangle inequality is not specifically enforced), they 
do not seem
conducive to barycenter computation.
Stable distance metrics between Reeb graphs \cite{bauer14} and merge trees 
\cite{morozov14} have been studied from a theoretical point of view but their 
computation, following an exponential time complexity, is not tractable for 
practical datasets in general, except if reliable correspondence labels between 
the nodes of the trees are provided on the input \cite{intrinsicMTdistance}, 
which 
is not practical either for large ensembles. Distances with polynomial 
time computation algorithms have also been investigated.
Similarly to our overall strategy, Beketayev et al. \cite{BeketayevYMWH14} 
focus on a dual representation, the \emph{branch decomposition tree} 
(BDT, \autoref{sec_background_mergeTrees}), but in contrast to our approach, 
they estimate their distances 
by iteratively reducing a target mismatch term, in particular, over a search 
space significantly larger than ours.
Sridharamurthy et al. 
\cite{SridharamurthyM20} specialize efficient algorithms for computing 
constrained edit distances between trees \cite{zhang96} to the 
special 
case of 
merge trees \revision{(see Appendix 1)}, resulting in a distance 
which is computable for 
real-life datasets and 
with acceptable practical stability.
However, it is not conducive to 
simple barycenter computations. 
Indeed, the 
linear interpolation of 
the optimal node 
assignments 
induced by 
this metric 
(\autoref{fig_counterExampleEditDistance}) does not result in
a 
shortest path, and hence generates inaccurate midpoints (i.e. inaccurate 
barycenters 
given two
trees).
This further implies that there is no clear or 
simple strategy for the general
computation of barycenters
according to that metric.
\cutout{In particular, a core 
technical reason 
for this 
is that this metric can 
result in optimal assignments 
involving nodes of different types (for instance matching an extremum node from 
one tree to a 
saddle node in the other). This matching inconsistency not only challenges 
geodesic computation, but it also challenges interpretation, as it corresponds 
in the data to the matching of a peak to a valley, which are typically 
different types of features in the applications.}

Regarding the estimation of a \emph{representative} object from a set of 
topological 
representations, several approaches emerged recently. 
A recent line of work \cite{intrinsicMTdistance, YanWMGW20} 
introduced a framework for computing a \emph{1-center} of a set of merge trees 
(i.e. 
minimizing its maximum distance to the set), according to 
an interleaving distance.
However, as documented by its authors, this 
approach requires pre-existing, reliable correspondence labels between the 
nodes of all the input trees, which is not practical with real-life datasets 
(heuristics need to be considered). 
\revision{Also,}
since they minimize their \emph{maximum} distance to a set, 
1-centers
are typically sensitive to outliers,
which prevents 
their usage for estimating 
trends or supporting clustering tasks (which typically focus on densities 
rather than maximum distances).
\revision{This is further evaluated in \autoref{sec_quality}.}
In contrast, our approach 
focuses on the estimation of \emph{barycenters} (instead of 1-centers) and 
computes a tree which minimizes its \emph{average} 
distance to an ensemble of merge trees (instead of its maximum distance), which 
is less sensitive to outliers, 
which better captures trends and which supports clustering tasks. Moreover, the 
node correspondences between the barycenter and the input trees are 
automatically estimated by our approach via an assignment optimization present 
at the core of our distance estimation. Thus our 
method does not require input correspondences, which makes it readily 
applicable to real-life ensembles. Several 
methods
\cite{Turner2014, 
lacombe2018, vidal_vis19} have been introduced for the automatic 
estimation of barycenters 
\revision{(\autoref{sec_background_persistenceDiagrams})}
of persistence diagrams (or 
vectorized variants 
\cite{Adams2015, Bubenik15}). However, 
the persistence diagram can lack 
specificity in its 
\revision{data characterization} 
(\autoref{fig_mergeTreeVSpersistenceDiagram}). This limitation is 
addressed 
by our work which focuses instead on merge trees.


\subsection{Contributions}
This paper makes the following new contributions:
\begin{enumerate}[itemsep=-0.75ex]
\vspace{-1ex}
 \item{\emph{A practical distance metric between merge trees:}}
 We extend recent work on the edit distance \cite{SridharamurthyM20} and 
 introduce a new distance  between merge trees, which, in contrast to previous 
work,
 is purposely designed to enable efficient computations of geodesics and 
barycenters.
 It can be computed efficiently, it has acceptable practical stability and it 
has a strong connection to established metrics, which eases its interpretation. 
Specifically, 
it
can be interpreted as a variant of the $L^2$-Wasserstein distance for 
persistence diagrams, for which we constrain the underlying 
search space
to account for the 
additional structural information provided by the merge tree.
 \item{\emph{A simple approach for computing geodesics between merge trees:}}
 Given our new metric, we present a simple approach for 
computing geodesics between merge trees. 
It \revision{uses}
a simple linear interpolation of the assignments resulting from our 
new metric, enabling the exact computation of geodesics in linear time.
\revision{This follows from previous work on persistence diagram 
geodesics \cite{Turner2014} and it is made possible 
%
thanks to a new, local 
normalization procedure, guaranteeing the topological consistency of the 
interpolated trees.}
 \item{\emph{An approach for computing barycenters between merge trees:}}
 \revision{Our method for geodesics between merge trees enables a 
straightforward adaptation of}
previous optimization strategies for 
persistence diagram barycenters
\cite{Turner2014}\revision{, resulting,}
to our knowledge, \revision{in} the 
first approach for the computation of barycenters of merge 
trees.
 \item{\emph{Unified computational framework:} 
 We present a unified computational framework for the estimation of distances, 
geodesics, barycenters, and clusters of merge trees. In particular, we 
introduce an efficient, task-based algorithm \revision{adapted from previous 
work on edit distances \cite{zhang96, SridharamurthyM20}, which is} 
generically applicable to any of 
the above tasks. Our algorithm supports shared-memory parallelism, allowing for 
further accelerations in practice.}
 \item{\emph{Applications:} We illustrate the utility of each of our 
contributions with dedicated visualization tasks, 
including feature 
tracking, temporal reduction and ensemble clustering and summarization.}
 \item{\emph{Implementation:} We provide a lightweight C++ implementation of 
our algorithms that can be used for reproduction purposes.}
\end{enumerate}

\section{Preliminaries}
\label{sec_preliminaries}
This section presents the theoretical background of our work. It contains 
definitions adapted from the Topology ToolKit \cite{ttk17}. 
We refer the 
reader to textbooks \cite{edelsbrunner09} for an introduction to 
computational topology.

\subsection{Input data}
\label{sec_inputData}
The input data is an ensemble of $N$ piecewise linear (PL) scalar fields
$f_i : \domain \rightarrow \range$, with $i \in \{1, \dots,  N\}$, defined on a 
PL $d$-manifold $\domain$, with 
$d\leq3$ in our applications. The \emph{sub-level set} of $f_i$, noted 
$\sublevelset{{f_i}}(\isovalue)=\{p \in \domain~|
~f_i(p) < \isovalue\}$, is defined as the pre-image of  $(-\infty, \isovalue)$ 
by 
$f_i$. 
The \emph{super-level set} of $f_i$ is defined symmetrically:
$\superlevelset{{f_i}}(\isovalue)=\{p \in \domain~|
~f_i(p) > \isovalue\}$.
As $\isovalue$ continuously increases, the topology of 
$\sublevelset{{f_i}}(\isovalue)$ changes at specific vertices of $\domain$, 
called the \emph{critical points} of $f_i$ \cite{banchoff70}. 
In practice, $f_i$ is enforced to contain only isolated, non-degenerate 
critical points \cite{edelsbrunner90, edelsbrunner03}.
Critical points are classified by their \emph{index} 
$\Index_i$: $0$ for minima, $1$ for $1$-saddles, $d-1$ for $(d-1)$-saddles and 
$d$ for maxima.

\subsection{Persistence diagrams}
\label{sec_background_persistenceDiagrams}
The persistence diagram is a visual summary of the topological features (i.e. 
connected components, independent cycles, voids) of 
$\sublevelset{{f_i}}(\isovalue)$. Specifically, each topological feature of  
$\sublevelset{{f_i}}(\isovalue)$ can be associated with a unique pair of 
critical points $(c, c')$, corresponding to its \emph{birth} and \emph{death}. 
The Elder rule \cite{edelsbrunner09} states that critical points can be 
arranged according to this observation in a set of pairs, such that each 
critical point appears in only one pair $(c, c')$, with $f_i(c) < f_i(c')$ and 
$\Index_i(c) = \Index_i(c') - 1$. 
For instance, if two connected components of $\sublevelset{{f_i}}(\isovalue)$ 
meet at a critical point $c'$, the \emph{\revision{younger}} component
(created
last, in $c$) \emph{dies}, in favor of the \emph{\revision{older}} one (created
first). Then 
the persistence diagram, noted $\diagram(f_i)$, embeds each pair to a single 
point in 2D at coordinates $\big(f_i(c), f_i(c')\big)$.
The \emph{persistence} of a pair
is given by 
its height $f_i(c') - f_i(c)$. 
Then, the 
persistence diagram provides a visual overview of the features of interest 
of a dataset (\autoref{fig_mergeTree}), where salient features stand out 
from the diagonal while pairs corresponding to noise are located in the
vicinity
of the diagonal. \revision{Note that, in addition to its interest as a visual 
summary, the persistence diagram captures all the information about the 
persistent homology groups of the data 
\cite{edelsbrunner09}.}

\begin{figure}  
  \centering
%
\includegraphics[width=\linewidth]{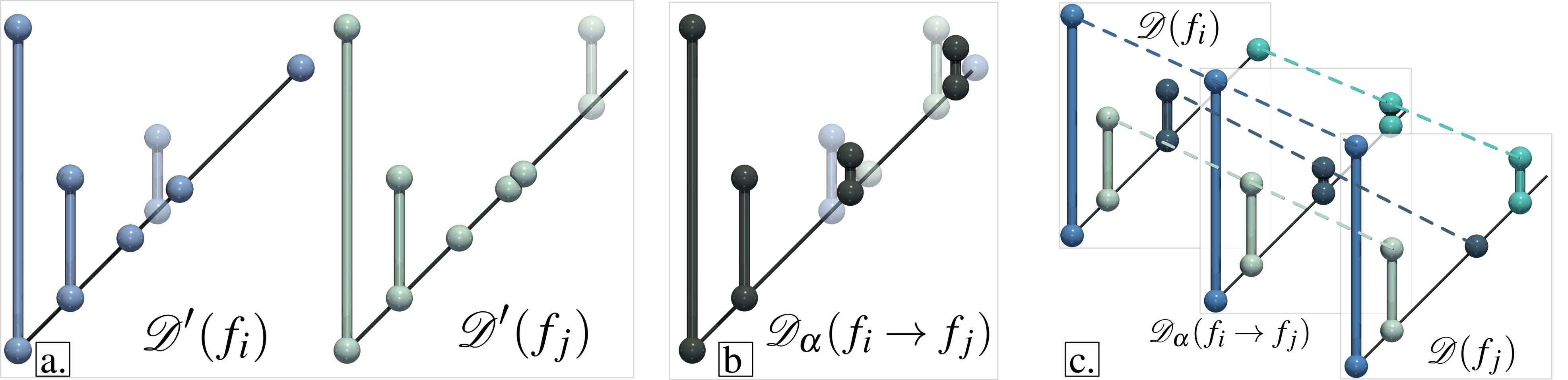}
  \mycaption{Distance and geodesic computation between two persistence diagrams 
$\diagram(f_i)$ and $\diagram(f_j)$, given the metric $\wasserstein{2}$. 
The sets $\overline{P_i}$ and $\overline{P_j}$ are shown in transparent (a).
The matching induced by 
the optimal partial assignment $\phi$ is shown with dashed 
lines (c). For two diagrams, the Wasserstein barycenter (b) is given, 
thanks to the $\pointMetric_2$ distance in the birth/death space, by the 
arithmetic mean of the matched points. Then, the linear interpolation of the 
matchings (c) describes a geodesic \cite{Turner2014}.}
  \label{fig_diagramInterpolation}
\end{figure}

Given two diagrams $\diagram(f_i)$ and $\diagram(f_j)$, a pointwise distance, 
noted $\pointMetric_q$ (with $q > 0$), can be introduced in the 2D birth/death 
space
between two points
\revision{$p_i = (x_i, y_i) \in \diagram(f_i)$} and
\revision{$p_j = (x_j, y_j)
\in 
\diagram(f_j)$}:
\vspace{-1ex}
\begin{eqnarray}
\label{eq_pointMetric}
\revision{\pointMetric_{q}(p_i, p_j) = (|x_j - x_i|^q + |y_j - y_i|^q)^{1/q}
= \| p_i - p_j
\|_q.}
\end{eqnarray}
\vspace{-2ex}

\noindent
By convention,
\revision{$\pointMetric_q(p_i, p_j)$} is set to zero if both
\revision{$p_i$ and
$p_j$}
exactly lie on the diagonal \revision{($x_i = y_i$ and $x_j = y_j$)}.
Let \revision{$P_i$} be a subset of the \revision{off-diagonal} points of
$\diagram(f_i)$
and \revision{$\overline{P_i}$} its
complement \revision{(i.e. the other off-diagonal points of $\diagram(f_i)$ not 
in $P_i$)}. Let \revision{$(\phi, \overline{P_i}, \overline{P_j})$} be a
\emph{partial
assignment} between $\diagram(f_i)$ 
and $\diagram(f_j)$, i.e. a bijective map between \revision{$P_i$} and a subset
\revision{of off-diagonal points} \revision{$P_j$} of
$\diagram(f_j)$, with complement \revision{$\overline{P_j}$}
(\autoref{fig_diagramInterpolation}a). Then, the 
$L^q$-Wasserstein 
distance, noted 
$\wasserstein{q}$, 
can be
introduced as:
\vspace{-1ex}
\begin{eqnarray}
\label{eq_wasserstein}
\wasserstein{q}\big(\diagram(f_i), \diagram(f_j)\big)
= 
\label{eq_wasserstein_mapping}
& \underset{(\phi, \revision{\overline{P_i}, \overline{P_j}}) \in \Phi}\min&
\Big(
\sum_{\revision{p_i \in P_i}}
\pointMetric_q\big(\revision{p_i,
\phi(p_i)}\big)^q\\
\label{eq_wasserstein_destruction}
& + & \sum_{\revision{p_i \in \overline{P_i}}} \pointMetric_q\big(\revision{p_i,
\projection(p_i)}\big)^q\\
\label{eq_wasserstein_creation}
& + & \sum_{\revision{p_j \in \overline{P_j}}} \pointMetric_q\big(
  \revision{\projection(p_j),
  p_j}
  \big)^q
%
\Big)^{1/q}
\end{eqnarray}
\vspace{-2ex}

\noindent
where $\Phi$ is the set of all possible partial assignments 
mapping each 
point 
$\revision{p_i}\in \diagram(f_i)$ to a point $\revision{\phi(p_i) = p_j} \in
\diagram(f_j)$
(line \ref{eq_wasserstein_mapping}), or to its 
diagonal projection, $\revision{\projection(p_i) =
(\frac{x_i+y_i}{2},\frac{x_i+y_i}{2})}$,
denoting the removal of the corresponding feature
from $\diagram(f_i)$ or $\diagram(f_j)$
(lines \ref{eq_wasserstein_destruction} and \ref{eq_wasserstein_creation}). 
Intuitively, the Wasserstein metric optimizes a matching 
between 
the two diagrams, and evaluates their distance 
given
the resulting mismatch.
In practice, 
$\diagram(f_i)$ and $\diagram(f_j)$ are 
\emph{augmented} into $\diagram'(f_i)$ and $\diagram'(f_j)$
\cite{Kerber2016}, 
by injecting the 
diagonal projections of one diagram into the other 
(\autoref{fig_diagramInterpolation}a):
\vspace{-1.5ex}
\begin{eqnarray}
  \nonumber
 \diagram'(f_i) = \diagram(f_i) \cup \{ 
\projection(\revision{p_j}) ~ | ~ \revision{p_j \in
P_j}
\}\\
\nonumber
  \diagram'(f_j) = \diagram(f_j) \cup \{ 
\projection(\revision{p_i}) ~ | ~ \revision{p_i \in
P_i}
\}.
\end{eqnarray}
\vspace{-2ex}

\noindent
This augmentation 
(\autoref{fig_diagramInterpolation}a) preserves the 
distance,
while making the 
assignment problem balanced, 
and thus 
easily 
solvable with traditional 
algorithms \cite{Munkres1957, 
Bertsekas81}
(with \revision{$P_i' = \diagram'(f_i), P_j' =
\diagram'(f_j)$} and \revision{$\overline{P_i'} = \overline{P_j'} =
\emptyset$}).


Given a set $\diagramSet = \{\diagram(f_1), \dots, \diagram(f_N)\}$ of 
persistence diagrams, let $F(\diagram, \alpha)$ be the Fr\'echet energy of the 
set, under the metric 
$\wasserstein{2}$, with the coefficients $\alpha = \{\alpha_1, \alpha_2, \dots, 
\alpha_N\}$, such that
$\alpha_i \in [0, 1]$ and
$\sum_{i} \alpha_i = 1$:
\vspace{-1ex}
\begin{eqnarray}
\label{eq_frechet_energy}
F(\diagram, \alpha) = 
 \sum_{\diagram(f_i) \in \diagramSet}
        \alpha_i
        \wasserstein{2}\big(\diagram, \diagram(f_i)\big)^2.
\end{eqnarray}
\vspace{-2.5ex}

Then the diagram $\diagram^* \in \diagramSpace$ (where $\diagramSpace$ is the 
space of persistence diagrams) which minimizes $F(\diagram, \alpha)$ is called 
the \emph{Wasserstein barycenter} of the set $\diagramSet$ (or its Fr\'echet 
mean
 under the metric $\wasserstein{2}$).
In practice, the coefficients $\alpha_i$ are all set to the same value 
($\alpha_i = 1 / N$, $\forall i \in \{1, \dots, N\}$). 
When $N = 2$ and $\alpha_1 = \alpha_2 = 0.5$ 
(\autoref{fig_diagramInterpolation}b), $\diagram^*$ becomes 
a midpoint 
between 
$\diagram(f_i)$ and 
$\diagram(f_j)$
and the 
set of 
possible values 
for 
$\alpha_1$ and $\alpha_2$ 
(\autoref{fig_diagramInterpolation}c) describes a geodesic in 
$\diagramSpace$ (i.e. 
length minimizing path) 
with regard to the 
$L^2$-Wasserstein
metric \cite{Turner2014}.


\subsection{Merge trees}
\label{sec_background_mergeTrees}
The \emph{join} tree, noted $\jointree(f_i)$, is a visual summary of the 
connected components of $\sublevelset{{f_i}}(\isovalue)$ \cite{carr00}. It is 
a 1-dimensional simplicial complex defined as the quotient space 
$\jointree(f_i) = \domain / \sim$ by the equivalence relation $\sim$
which states that 
$p_1$ and $p_2$ are equivalent 
if 
$f_i(p_1) = f_i(p_2)$ and if $p_1$ and $p_2$ belong to the same connected 
component 
of $\sublevelset{{f_i}}\big(f_i(p_1)\big)$.

The \emph{split} tree (\autoref{fig_mergeTree}), noted $\splittree(f_i)$, is 
defined symmetrically and describes the connected components of the super-level 
set $\superlevelset{{f_i}}(\isovalue)$. Each of these two \emph{directed} trees 
is called a 
\emph{merge} tree, noted generically $\mergetree(f_i)$ in the following. 
Intuitively, these trees track the creation of connected components of 
the sub 
(or super) level sets at their leaves, and merge events at their 
interior nodes. 
These trees are often 
visualized according to a persistence-driven \emph{branch decomposition} 
\cite{pascucci_mr04}, to make the persistence pairs captured by the tree stand 
out. In this context, a \emph{persistent branch} is a monotone path on the tree 
connecting the nodes corresponding to the creation and destruction (according 
to the Elder rule, \autoref{sec_background_persistenceDiagrams})
of a connected component of sub (or super) level set. Then, the branch 
decomposition provides a planar layout of the merge tree, 
where each 
persistent 
branch is 
represented as a vertical segment.
The \emph{branch decomposition tree} (BDT), 
noted 
$\branchtree(f_i)$, is a directed tree 
whose nodes are the persistent branches 
captured 
by the branch decomposition 
and whose arcs denote adjacency relations between them in the 
merge tree.  \revision{In \autoref{fig_mergeTree}, the BDTs (top right 
insets) can be interpreted as the dual of the branch decompositions 
(bottom right insets, with matching colors): each vertical segment in the 
branch decomposition (bottom) corresponds to a node in the BDT (top) and each 
horizontal segment (bottom, denoting an adjacency relation between branches) 
corresponds to an arc in the BDT.}
Intuitively, the 
BDT,
like the persistence diagram, describes the 
population of (extremum-saddle) persistence pairs present in the data. However, 
unlike the persistence diagram, it additionally captures adjacency relations 
between them.

\cutout{As discussed in \autoref{sec_relatedWork}, several metrics have been 
introduced 
for measuring distances between merge trees. We focus in the remainder on the 
edit distance introduced by Sridharamurthy et al. \cite{SridharamurthyM20}, as 
its balance between computability and 
acceptable
stability in practice seems particularly 
promising. The edit distance between two merge trees $\mergetree(f_i)$ and 
$\mergetree(f_j)$, noted $\editdistance\big(\mergetree(f_i), 
\mergetree(f_j)\big)$, is defined as follows.
Let $N_i$ be a subset of the nodes of $\mergetree(f_i)$ and $\overline{N_i} $
its complement. Let $\phi'$ be a \emph{partial} assignment  
between $N_i$ and a subset $N_j$ of the nodes of
$\mergetree(f_j)$ (with complement $\overline{N_j}$).
Then $\editdistance\big(\mergetree(f_i), 
\mergetree(f_j)\big)$ is given by:
\begin{eqnarray}
\label{eq_edit_distance}
 \editdistance\big(\mergetree(f_i), \mergetree(f_j)\big) = 
  & \underset{(\phi', \overline{N_i}, \overline{N_j}) \in \Phi'}\min&  \Big(
    \label{eq_edit_distance_mapping}
    \sum_{n_i \in N_i} \gamma\big(n_i \rightarrow \phi'(n_i)\big)\\
    \label{eq_edit_distance_destroying}
    & + &  \sum_{n_i \in \overline{N_i}} \gamma(n_i \rightarrow \emptyset)\\
    \label{eq_edit_distance_creating}
    & + & \sum_{n_j \in \overline{N_j}} \gamma(\emptyset \rightarrow n_j)
  \Big)
  \label{eq_editDistance}
\end{eqnarray}
%
%
%
where $\Phi'$ is the space of \emph{constrained} partial assignments (i.e. 
$\phi'$ maps disjoint subtrees of $\mergetree(f_i)$ to disjoint subtrees of 
$\mergetree(f_j)$) and where $\gamma$ refers to the cost 
for:
\emph{(i)} mapping a node $n_i \in \mergetree(f_i)$ to a node $\phi'(n_i) = n_j
\in \mergetree(f_j)$ (line \ref{eq_edit_distance_mapping}), 
\emph{(ii)} deleting a node 
$n_i \in \mergetree(f_i)$ (line \ref{eq_edit_distance_destroying}) and
\emph{(iii)} creating 
a node $n_j \in \mergetree(f_j)$ (line \ref{eq_edit_distance_creating}),
$\emptyset$ 
being the empty tree.}
%

%

\cutout{Zhang \cite{zhang96} introduced a polynomial time algorithm for 
computing a 
constrained sequence of edit operations with minimal edit distance 
(\autoref{eq_edit_distance_mapping}), and showed that the resulting distance is 
indeed a 
metric if each cost $\gamma$ for the above three edit operations is itself 
a metric (non-negativity, identity, symmetry, triangle inequality).
Sridharamurthy et al. \cite{SridharamurthyM20} exploited this property to 
introduce their metric,
by defining a distance-based cost model, inspired by the Wasserstein distance 
between persistence diagrams (\autoref{sec_jackground_persistenceDiagrams}),
where $p_i$ and $p_j$ stand for the persistence pairs \emph{containing} the
nodes $n_i \in \mergetree(f_i)$ and $n_j \in \mergetree(f_j)$:
%
\begin{eqnarray}
  \gamma(n_i \rightarrow n_j) & = & \min\big(\pointMetric_\infty(p_i, p_j),
    \gamma(n_i \rightarrow \emptyset) + \gamma(\emptyset \rightarrow n_j)\big)\\
  \gamma(n_i \rightarrow \emptyset) & = & \pointMetric_\infty\big(p_i,
\projection(p_i)\big)\\
  \gamma(\emptyset \rightarrow n_j) & = & \pointMetric_\infty\big(p_j,
\projection(p_j)\big)
\end{eqnarray}
In our work, as described next, we introduce an alternative 
edit distance
which adheres even further to the Wasserstein distance between persistence 
diagrams, 
to ease geodesic computation for merge 
trees.}





\section{Wasserstein Distances Between Merge Trees}
\label{sec_metric}

This section introduces our new distance metric between merge trees, which 
is specifically designed for the subsequent 
%
computation of geodesics 
(\autoref{sec:geodesics}) and barycenters (\autoref{sec:barycenters}). 
For this, we bridge the gap between the edit 
distance between merge trees \cite{SridharamurthyM20} 
and 
existing work addressing 
the computation of geodesics and barycenters for persistence diagrams according 
to the $L^2$-Wasserstein distance \cite{Turner2014}.

\subsection{Overview}
\label{sec_metric_overview}
The end goal of our work is the computation of barycenters of merge trees.
\revision{For this, we extend}
the edit 
distance $\editdistance$ \cite{SridharamurthyM20} \revision{(formalized in
Appendix 1,
additional
material)}, to make it fit
the
optimization strategy used for barycenters of persistence diagrams 
\cite{Turner2014}. 
\revision{Our} key idea consists in transforming 
$\editdistance$
such that it becomes strictly equivalent to the $L^2$-Wasserstein distance of 
persistence diagrams, but given a \emph{restricted} space 
of 
possible 
assignments,
constrained by the structure of the 
input trees $\mergetree(f_i)$ and $\mergetree(f_j)$, hence its name 
Wasserstein distance between merge trees.
%
Then, thanks to this 
compatibility with the $L^2$-Wasserstein distance, the assignments 
resulting from
our metric can be 
directly used
for interpolation-based 
geodesic and barycenter computations (Secs. \ref{sec:geodesics} and 
\ref{sec:barycenters}).
Overall, our strategy involves four major modifications to the 
edit 
distance 
$\editdistance$ \cite{SridharamurthyM20}, detailed in the remainder of this 
section:
\vspace{-1ex}
\begin{enumerate}[leftmargin=.3cm, itemsep=-1ex]
 \item{To consider assignments between persistence pairs instead of merge tree 
nodes, 
we consider an edit distance between 
the 
BDTs
$\branchtree(f_i)$ and $\branchtree(f_j)$
(\autoref{sec_background_mergeTrees}) instead of the input merges trees
$\mergetree(f_i)$ and $\mergetree(f_j)$ (as done with $\editdistance$). This is 
described in 
\autoref{sec_our_distance_definition}.}
\item{
We constrain the assignment search 
space to the space of \emph{rooted partial isomorphisms}. Specifically, 
similarly to $\editdistance$, we enforce the assignment of disjoint 
subtrees of $\branchtree(f_i)$ to disjoint subtrees of $\branchtree(f_j)$. 
Moreover, in contrast to  $\editdistance$, we additionally extend this 
constraint by enforcing the destruction of entire subtrees upon the destruction 
of their root.
%
These two constraints together enforce assignments describing isomorphisms 
between rooted subtrees of $\branchtree(f_i)$ and $\branchtree(f_j)$. Such 
isomorphisms pave the way for interpolation-based geodesics.  This is described 
in Secs.
\ref{sec_our_distance_definition}, \ref{sec_algorithm} and
\ref{sec_geodesicDefinition}.}
\item{We introduce a cost model based on the Euclidean distance 
$\pointMetric_2$ to enable geodesic computation 
by
linear interpolation of the assignments in the 2D 
birth/death space. This is 
described in Secs. \ref{sec_our_distance_definition} and 
\ref{sec_geodesicDefinition}.}
\item{We finally extend our metric with 
a 
local normalization 
term, which enforces
nested birth-death values, along the interpolation of the 
assignments, for nested  branches.
This is described in 
\autoref{sec_normalization}.}
\end{enumerate}

\subsection{Definition and properties}
\label{sec_our_distance_definition}
Given two input merge trees, $\mergetree(f_i)$ and $\mergetree(f_j)$, we first 
consider their  
BDTs
$\branchtree(f_i)$ and 
$\branchtree(f_j)$ (\autoref{sec_background_mergeTrees}). Let
$\revision{B_i}$ be a
subset 
of the nodes 
of $\branchtree(f_i)$ and $\revision{\overline{B_i}}$ its complement. Note
that each node in $\revision{B_i}$  corresponds to a persistence pair of
$\diagram(f_i)$.
Let $\revision{(\phi', \overline{B_i}, \overline{B_j})}$ be a partial
assignment between
\revision{$B_i$} and a subset
\revision{$B_j$} of the
nodes of $\branchtree(f_j)$ (with complement \revision{$\overline{B_j}$}). Then
we
introduce the $L^2$-Wasserstein distance $\wassersteinTree$ between 
the 
BDTs
$\branchtree(f_i)$ and 
$\branchtree(f_j)$ of 
the merge 
trees $\mergetree(f_i)$ and $\mergetree(f_j)$ as:
\vspace{-1.5ex}
\begin{eqnarray}
\label{eq_our_distance}
 \wassersteinTree\big(\branchtree(f_i), \branchtree(f_j)\big) = 
  & \hspace{-.35cm} \underset{(\phi', \revision{\overline{B_i},
\overline{B_j}}) \in
\Phi'}\min &
    \hspace{-.35cm} \Big( 
    \label{eq_our_distance_mapping}
    \sum_{\revision{b_i \in B_i}} \gamma\big(\revision{b_i \rightarrow
\phi'(b_i)}\big)^2\\
    \label{eq_our_distance_destroying}
    & + &  \sum_{\revision{b_i \in \overline{B_i}}} \gamma(\revision{b_i}
\rightarrow \emptyset)^2\\
    \label{eq_our_distance_creating}
    & + & \sum_{\revision{b_j \in \overline{B_j}}} \gamma(\emptyset \rightarrow
\revision{b_j})^2
  \Big)^{1/2}
  \label{eq_editDistance}
\end{eqnarray}
\vspace{-2ex}

\noindent
where $\Phi'$ is the space of constrained partial assignments 
mapping 
disjoints subtrees of $\branchtree(f_i)$ to disjoint subtrees of 
$\branchtree(f_j)$, \revision{\emph{and}} mapping entire subtrees to the empty
tree $\emptyset$
if their root is itself mapped to $\emptyset$.
Then, 
given 
the $k^{th}$ direct child of \revision{$b_i$}, noted  \revision{$b_i^k$},
it follows that \revision{$b_i^k$} either maps through $\phi'$ to a direct
child of
$\revision{\phi'(b_i)} \in \branchtree(f_j)$ (then \revision{$b_i, b_i^k \in
B_i$}) or to
the empty
tree $\emptyset$ (then the subtree rooted in \revision{$b_i^k$}, noted
\revision{$\branchtree(f_i, b_i^k)$}, belongs to \revision{$\overline{B_i}$}).
This further
implies 
%
%
%
%
%
%
that the 
 rooted subtrees 
 $\revision{B_i} \subseteq \branchtree(f_i)$ and $\revision{B_j = \phi'(B_i)}
\subseteq \branchtree(f_j)$ 
are isomorphic and we call $(\phi', \revision{\overline{B_i}, \overline{B_j}})$
a \emph{rooted partial isomorphism}.
Unlike $\editdistance$ (see Appendix 1) but similarly to 
$\wasserstein{2}$ 
(\autoref{eq_wasserstein}), the cost of each  operation (mapping, line 
\ref{eq_our_distance_mapping}, destruction, line 
\ref{eq_our_distance_destroying}, and creation, line 
\ref{eq_our_distance_creating}) is squared, and the square 
root of the sum of the squared costs is considered as the overall distance.

Next, we define the edit  costs as follows (we recall that 
each branch $\revision{b_i} \in \branchtree(f_i)$
exactly coincides with 
a persistence pair 
$\revision{p_i} \in \diagram(f_i)$):

\vspace{-1.5ex}
\begin{eqnarray}
  \label{eq_our_mapping_cost}
  \nonumber
  \gamma\big(\revision{b_i \rightarrow \phi'(b_i)}\big) & = &
\pointMetric_2\big(\revision{b_i,
\phi'(b_i)}\big)\\
 \label{eq_our_mapping_cost}
  \label{eq_our_deletion_cost}
  \gamma(\revision{b_i} \rightarrow \emptyset) & = &
\pointMetric_2\big(\revision{b_i,
\projection(b_i)}\big)\\
\nonumber
  \label{eq_our_creation_cost}
  \gamma(\emptyset \rightarrow \revision{b_j}) & = & \pointMetric_2\big(
    \projection\revision{(b_j),
    b_j}
    \big).
\end{eqnarray}
\vspace{-2ex}

Note that the expression of the $L^2$-Wasserstein distance $\wassersteinTree$ 
between merge trees (\autoref{eq_our_distance}) is therefore identical to the 
expression of the Wasserstein distance between persistence diagrams 
(\autoref{eq_wasserstein}) for $q = 2$, at the notable exception of the 
search space of the partial assignments $\Phi' \subset \Phi$, which is 
constrained to rooted partial isomorphisms.
$\wassersteinTree$ is indeed a distance metric (the proof is included in 
Appendix 2, supplemental material): it
is 
non-negative and symmetric, it preserves the 
identity of indiscernibles as well as the triangle inequality.
Moreover, since $\Phi' \subset \Phi$, it 
follows that $\wassersteinTree\big(\branchtree(f_i), \branchtree(f_j)\big) \geq 
\wasserstein{2}\big(\diagram(f_i), \diagram(f_j)\big)$, which was 
one of the main motivations of
our work (i.e. to 
exploit the merge tree to define a 
more
discriminative metric, 
\autoref{fig_mergeTreeVSpersistenceDiagram}). 
Similarly to Sridharamurthy et al. \cite{SridharamurthyM20}, we mitigate saddle 
swap instabilities in a preprocessing step, by merging adjacent saddles in the 
input trees if their 
difference in scalar value is smaller than a 
threshold $\epsilon_1 \in [0, 1]$ \revision{(relative to the largest 
difference between adjacent saddles)}. Then,  when $\epsilon_1 =
1$, it follows that $\wassersteinTree\big(\branchtree(f_i), 
\branchtree(f_j)\big) 
=
\wasserstein{2}\big(\diagram(f_i), 
\diagram(f_j)\big)$. This simple merging strategy significantly improves 
the practical stability of 
$\wassersteinTree$, as empirically studied in 
\autoref{sec_quality} (\autoref{fig_practicalStability}).

\cutout{Similarly to Sridharamurthy et al. \cite{SridharamurthyM20}, we believe 
that 
the theoretical investigation of the stability of $\wassersteinTree$ goes 
beyond the scope of this paper and we leave it for future work. However, a few 
practical observations can already be formulated. Similarly to $\editdistance$, 
since it is based on merge trees, $\wassersteinTree$ is also subject to 
\emph{saddle swap instabilities} \cite{SridharamurthyM20}: under a slight 
perturbation $\epsilon$ of the 
input data $f_i$ into $f_j$ ($||f_i - f_j||_\infty \leq \epsilon$),
two adjacent saddle nodes in 
$\mergetree(f_i)$ can switch positions in $\mergetree(f_j)$. Such swaps have a 
non-negligible impact on the branch decomposition and therefore, on the derived 
search space $\Phi'$, and thus on $\wassersteinTree$. To mitigate such 
instabilities in practice, we apply the same strategy as Sridharamurthy et al. 
\cite{SridharamurthyM20}: in a preprocessing step, adjacent saddles in each 
input tree whose (relative) difference in scalar value is smaller than a
threshold
$\epsilon_1 \in [0, 1]$ are merged into a single multi-saddle, effectively
canceling the
effect of the order of the merged saddles on the branch decomposition. Note 
that in practice, in our implementation,
we maintain the original 
persistence of each pair involved in such a saddle merge.
Thus, when $\epsilon_1 \rightarrow 
1$, it follows that
$\wassersteinTree\big(\branchtree(f_i), 
\branchtree(f_j)\big) \rightarrow \wasserstein{2}\big(\diagram(f_i), 
\diagram(f_j)\big)$. This simple merging strategy significantly improves 
the stability of 
$\wassersteinTree$ in practice, as empirically demonstrated in 
\autoref{sec_quality} (\autoref{fig_practicalStability}).}

\begin{figure}
\includegraphics[width=\linewidth]{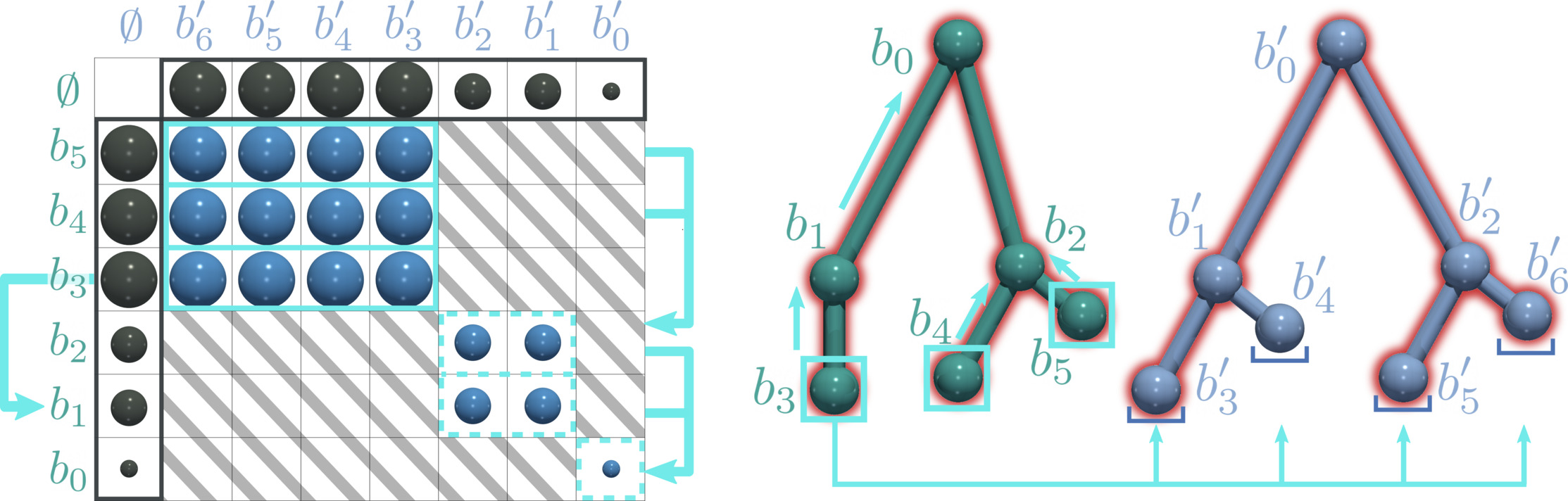}
\mycaption{The exploration of the space $\Phi'$ \julien{(a candidate 
is highlighted in red, right)} relies on the evaluation of 
the sparse matrix $\treeMatrix$ of subtree distances (left). Our task-based 
algorithm optimizes the parallel computation of independent terms. Spheres of 
equal radius in $\treeMatrix$ denote independent terms and arrows between the 
lines of $\treeMatrix$ indicate task dependence (equivalently illustrated with 
arrows in the input BDTs, right).}
%
%
\label{fig_parallel_algorithm}
\end{figure}

\subsection{Computation}
\label{sec_algorithm}
This section describes our algorithm 
for the recursive 
exploration of the search space $\Phi'$ (\autoref{eq_our_distance}).
It is \revision{based on the same recursive traversal as Zhang's algorithm 
\cite{zhang96}},
which we 
simplify as our 
search space is significantly more constrained.
\revision{Specifically, as  detailed in Appendix 3, our distance
evaluation between subtrees (\autoref{eq_algorithm_start}) involves fewer
solutions 
and it is restricted to
subtrees rooted at identical depth only.}
%
%


Given the subtree $\branchtree(f_i, b)$ of $\branchtree(f_i)$ (rooted in $b$)
and $b^k$ the $k^{th}$ direct child of $b$ in 
$\branchtree(f_i, b)$, the distance 
between the subtree $\branchtree(f_i, 
b)$ and the empty tree $\emptyset$ is then obtained  recursively by:
\vspace{-1ex}
\begin{eqnarray}
\label{eq_destroying_things}
 \wassersteinTree\big(\branchtree(f_i, b), \emptyset\big) = 
  \Big(
    \gamma(b \rightarrow \emptyset)^2 
    + \sum_{k} \wassersteinTree\big(\branchtree(f_i, b^k), \emptyset\big)^2
  \Big)^{1/2}.
\end{eqnarray}
\vspace{-2.5ex}

\noindent
The first step of our algorithm 
consists in evaluating
$\wassersteinTree\big(\branchtree(f_i, b), 
\emptyset\big)$ with 
\autoref{eq_destroying_things}
for all branches $b \in 
\branchtree(f_i)$ (and similarly for 
$\branchtree(f_j)$).

Next, let $\forest(f_i, b)$ be the \emph{forest} of $b$ in $ \branchtree(f_i)$:
$\forest(f_i, b)$ is the set of all the subtrees rooted at the $k$ direct 
children
of $b$: $\forest(f_i, b) = \{\branchtree(f_i, b^1), \branchtree(f_i, b^2), 
\dots, \branchtree(f_i, b^k)  \}$. Then the distance between two subtrees 
$\branchtree(f_i, b_i)$ and $\branchtree(f_j, b_j)$
\revision{is set to $+\infty$ when $b_i$ and $b_j$ have distinct depths (gray
crosshatching lines in \autoref{fig_parallel_algorithm}, left). Otherwise
(spheres in \autoref{fig_parallel_algorithm}, left), it}
is obtained recursively by:
\vspace{-1ex}
\begin{eqnarray}
  \label{eq_algorithm_start}
  \wassersteinTree\big(\branchtree(f_i, b_i), \branchtree(f_j, b_j)\big) &=&
    \Big(
      \gamma(b_i \rightarrow b_j)^2\\
      \nonumber
      &+& \wassersteinTree\big(\forest(f_i, b_i), \forest(f_j, b_j)\big)^2
    \Big)^{1/2}.
\end{eqnarray}
\vspace{-2ex}

\cutout{which involves the cost of assigning the branch $b_i$ to $b_j$, plus a
local 
assignment problem involving all their subtrees. In particular, this 
equation enforces that if two branches $b_i$ and $b_j$ are matched through
$\phi'$ (\autoref{eq_our_distance}), then 
the direct subtrees of
$\forest(f_i, b_i)$ can only be matched to direct subtrees
$\forest(f_j, b_j)$, or destroyed (as specified in the description of $\Phi'$,
\autoref{sec_our_distance_definition}).}
Let \revision{$F_i$} be a subset of the forest
$\revision{\forest(f_i, b_i)}$
and \revision{$\overline{F_i}$} its
complement.
The distance between two forests 
is then given recursively by:
\vspace{-1ex}
\begin{eqnarray}
  \nonumber
 \wassersteinTree\big(\forest(\revision{f_i, b_i}), \forest(\revision{f_j,
b_j})\big) =
  & 
  \hspace{-.25cm}
  \underset{(\phi'', \revision{\overline{F_i}, \overline{F_j}}) \in \Phi''}\min&
  \hspace{-.25cm}
  \Big( 
    \sum_{\revision{f_i \in F_i}} \wassersteinTree\big(\revision{f_i,
\phi''(f_i)}\big)^2\\
    \nonumber
    & + &  \sum_{\revision{f_i \in \overline{F_i}}}
      \wassersteinTree\big( \revision{f_i}, \emptyset \big)^2\\
    \nonumber
    & + & \sum_{\revision{f_j \in \overline{F_j}}}
      \wassersteinTree\big(\revision{f_j}, \emptyset\big)^2
  \Big)^{1/2}
  \label{eq_editDistance}
\end{eqnarray}
\vspace{-2ex}

\noindent
where $(\phi'', \revision{\overline{F_i}, \overline{F_j}})$ becomes the solution
of a local, partial assignment 
problem 
between forests, mapping \revision{$F_i$} to a subset
\revision{$F_j \in \forest(f_j,
b_j)$}
(with 
complement \revision{$\overline{F_j}$}) or to the empty tree $\emptyset$, and
which can
be solved with traditional 
assignment algorithms \cite{Munkres1957, 
Bertsekas81} (see 
\autoref{sec_background_persistenceDiagrams}). Then, the overall distance
$\wassersteinTree$ between the two input merge trees is obtained by estimating 
\autoref{eq_algorithm_start} at the roots of $\branchtree(f_i)$ and
$\branchtree(f_j)$, and solving recursively the local assignment problems 
between forests \revision{(the recursion returns are illustrated with arrows 
in \autoref{fig_parallel_algorithm})}. Note that
if 
$\epsilon_1 = 1$ 
(\autoref{sec_our_distance_definition}),
all the branches of 
$\branchtree(f_i)$ and 
$\branchtree(f_j)$ get attached to the roots and the 
recursive local assignment problems between forests (above)
become only one, large, assignment problem between all branches. 
Thus, 
when $\epsilon_1 = 1$, this algorithm indeed becomes 
strictly equivalent to the resolution of 
the assignment
problem 
involved in 
$\wasserstein{2}$ (\autoref{sec_background_persistenceDiagrams}).

\cutout{Finally, note that our algorithm indeed simplifies the approach by 
Zhang 
\cite{zhang96} used by Sridharamurthy et al. \cite{SridharamurthyM20}.
In our work, the destruction of a node (a branch) $b_j \in \branchtree(f_j)$
necessarily implies the destruction of its 
subtrees, i.e. of its
forest $\forest(f_j, b_j)$. Thus, the admissible
solutions in \cite{zhang96, SridharamurthyM20} consisting in deleting 
$b_j$ and mapping a subtree $\branchtree(f_i,
b_i)$ to one of the subtrees of $b_j$ in the forest $\forest(f_j, b_j)$
are no longer admissible 
given our overall solution space $\Phi'$. The removal of such solutions 
drastically simplifies the evaluation of the startup equation (being the
minimum
of three solutions in  \cite{SridharamurthyM20}, Eq. 12) to 
\autoref{eq_algorithm_start} in our work (containing only one expression to
evaluate).}

\subsection{Parallelism}
\label{sec_distanceParallelization}

Similarly to Zhang \cite{zhang96}, Eqs. \ref{eq_destroying_things} and 
\ref{eq_algorithm_start} can be estimated recursively.
\cutout{, by initiating the 
computation at the root of the trees, which will effectively descend down to
the
leaves through the recursive calls,  evaluating the terms of Eqs. 
\ref{eq_destroying_things} and \ref{eq_algorithm_start} at the leaves and then
returning from the recursion, to estimate these terms at the interior nodes, 
from the leaves up to the roots.}
To avoid redundant 
computations, the 
distances between the 
forests $\forest(f_i, \revision{b_i})$ and $\forest(f_j, \revision{b_j})$ are
stored at the entry
$(\revision{b_i, b_j)}$ of a matrix $\forestMatrix$ (of size
$|\branchtree(f_i)| \times
|\branchtree(f_j)|$), while the distances between the subtrees 
$\branchtree(f_i, \revision{b_i})$ and $\branchtree(f_j, \revision{b_j})$ (used
within the
assignment problems between higher forests) are stored in a matrix $\treeMatrix$
(of
the same size\revision{, see \autoref{fig_parallel_algorithm}}).

In our work, we additionally express this computation in terms of 
\emph{tasks}, to leverage task-based shared memory parallelism.
\revision{In particular, we initiate a task for each independent term of
Eqs. \ref{eq_destroying_things} and \ref{eq_algorithm_start}, which is ready for
computation (see \autoref{fig_parallel_algorithm}), as further detailed in
Appendix 4.
Then, the number of parallel tasks
is
initially bounded by the number of leaves in the input BDTs
(which is typically much
larger than the number of cores) and progressively decreases during the
computation.}

\section{Wasserstein Geodesics Between Merge Trees}
\label{sec:geodesics}

This section introduces our approach for the efficient computation of geodesics 
between merge trees, according to the metric $\wassersteinTree$ 
(\autoref{sec_metric}). For this, we leverage the rooted partial isomorphism 
resulting 
from the distance computation, as well as linear interpolations of 
the matchings, as introduced for persistence diagrams \cite{Turner2014}. 

%
%
  
\begin{figure}
  \centering
\includegraphics[height=0.24\linewidth]{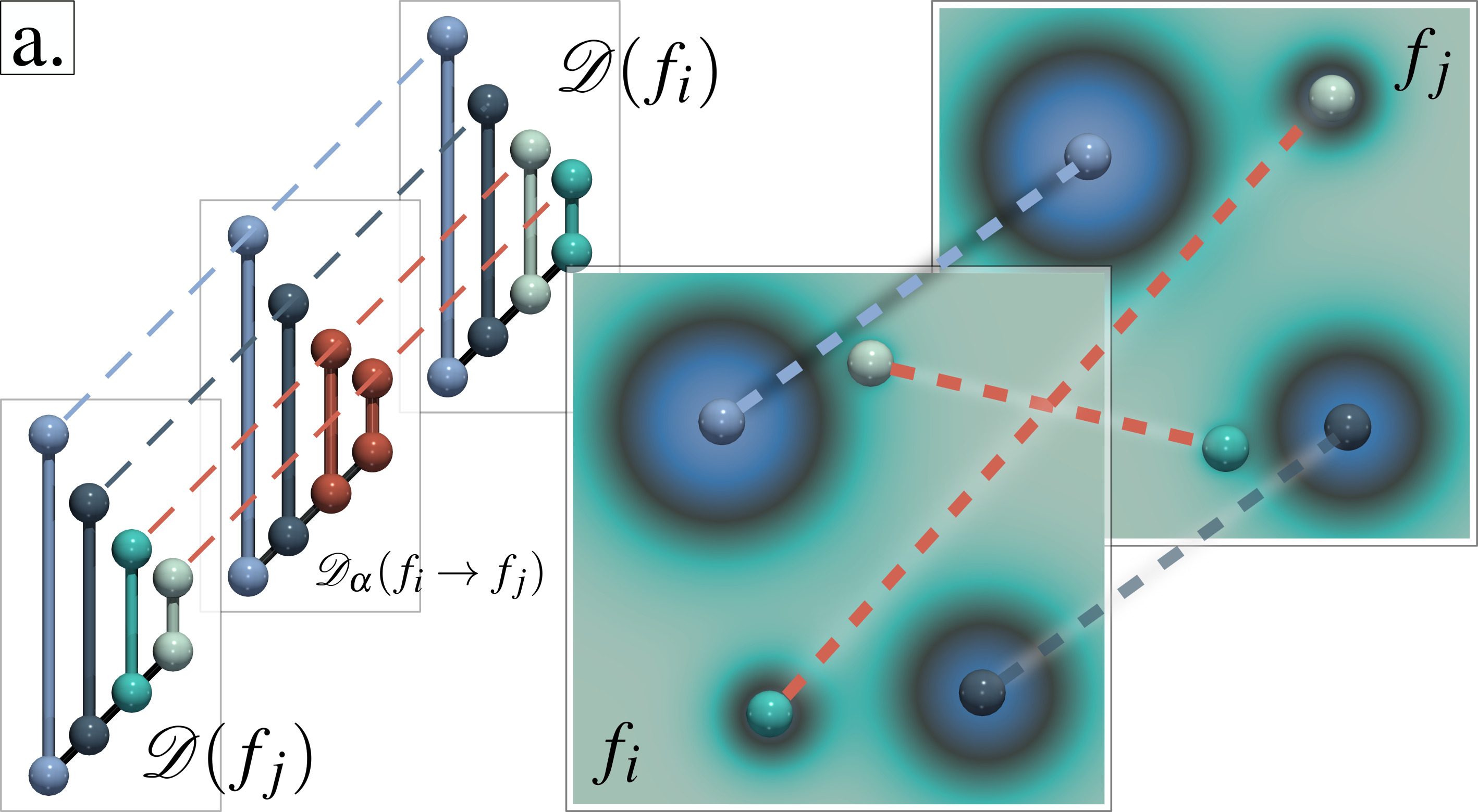}
\hfill
\includegraphics[height=0.24\linewidth]{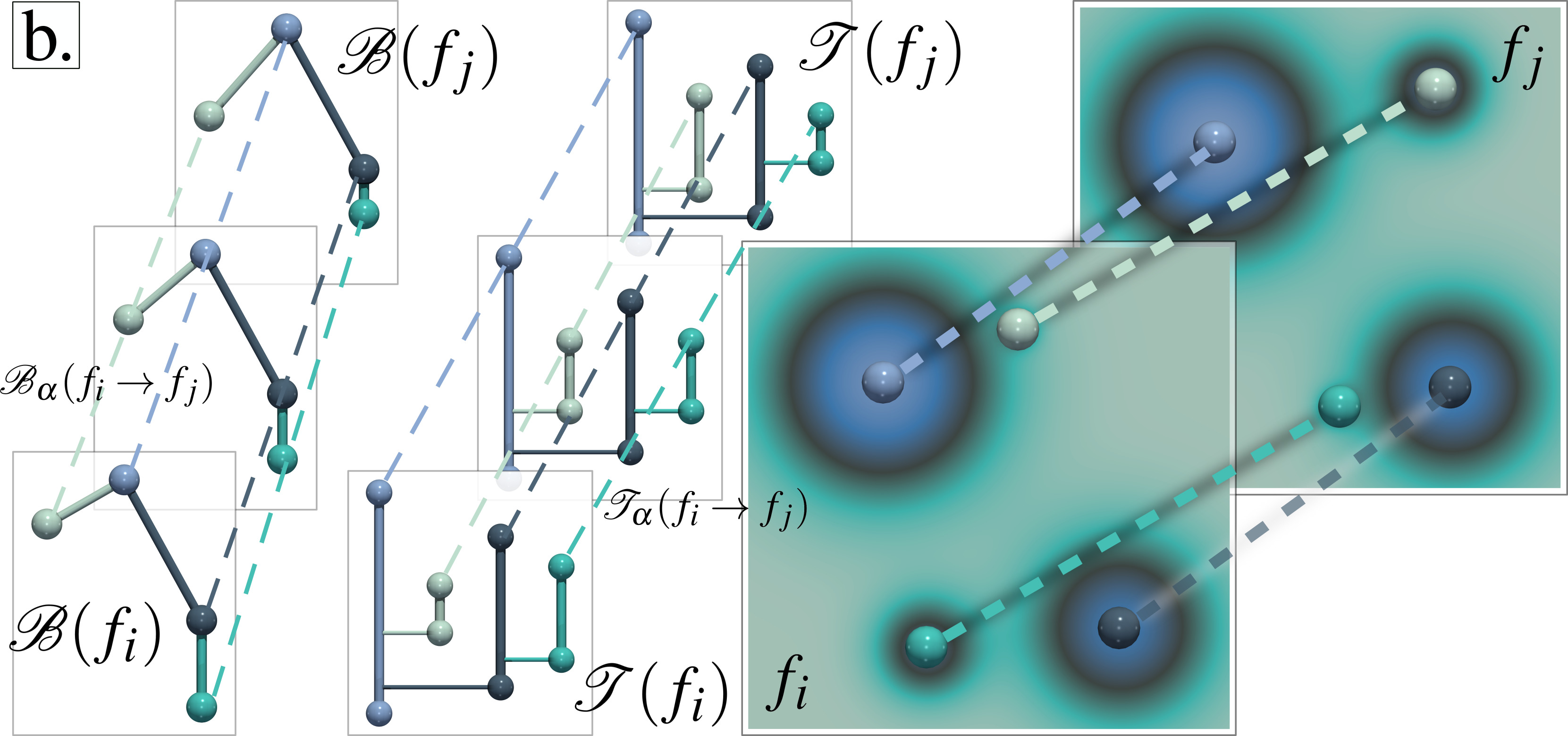}
\mycaption{Our
geodesic computation extends
interpolation-based
geodesics from persistence diagrams (a) to merge trees (b).
The interpolated
BDT
$\branchtree_\alpha(f_i \rightarrow
f_j)$ is obtained by linear interpolation (with local normalization)
of the
partial isomorphism
$\phi'$ in the birth/death space.
In the data, the feature matching (dashed lines) induced by $\phi'$
with $\wassersteinTree$ (b)  better preserves the global structure of the data
than
$\phi$ with $\wasserstein{2}$ (a, red crossing).
}
\label{fig_matchingDiagramVSTree}
\end{figure}

\begin{figure}[b]
\includegraphics[width=\linewidth]{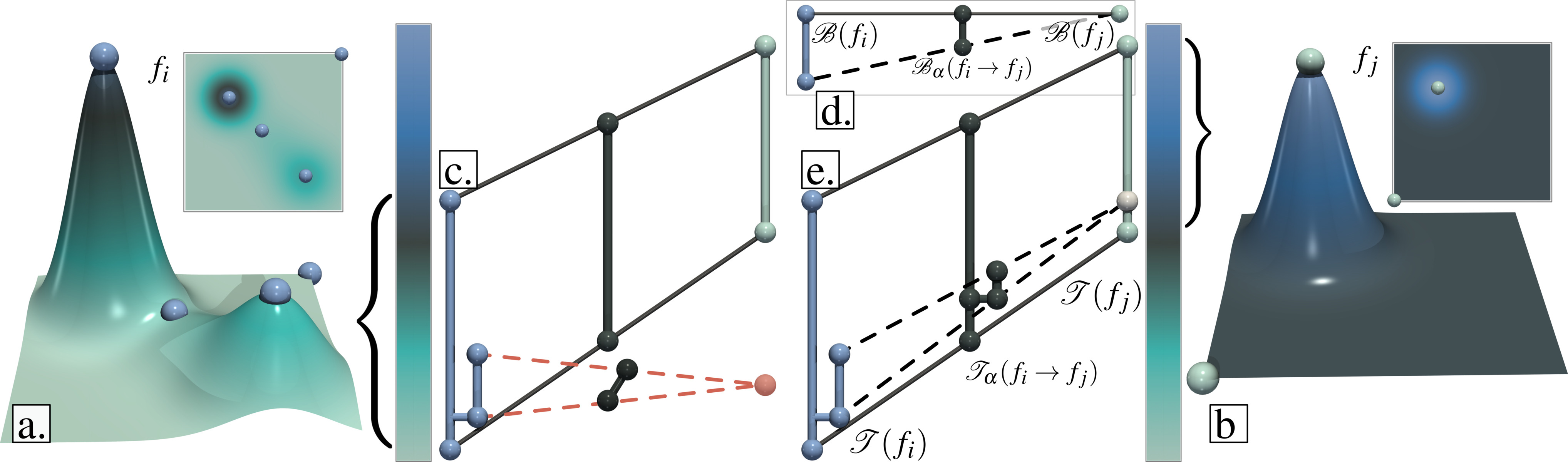}
\mycaption{Given two scalar fields $f_i$ (a) and $f_j$ (b), a simple
interpolation of the birth/death values of the branches
of their BDTs
may
result in inconsistencies
upon
branch destruction \revision{(red)}: the interpolated merge
tree (black) in (c) is disconnected, unlike the interpolated BDT
\textcolor{black}{(d)}.
Our local normalization 
(\autoref{sec_normalization})
addresses this issue by enforcing nested birth/death values for nested
branches.
This results in a valid interpolated merge tree \textcolor{black}{(e)}
whose BDT is indeed equal
to the interpolated BDT (e).}
%
%
\label{fig:slidingterm}
\end{figure}

\subsection{Definition and properties}
\label{sec_geodesicDefinition}
Given two input merge trees 
$\mergetree(f_i)$ and $\mergetree(f_j)$, our 
approach to geodesic computation (\autoref{fig_matchingDiagramVSTree}) 
simply consists 
in linearly interpolating the rooted partial isomorphism $(\phi', 
\revision{\overline{B_i},
\overline{B_j}})$ resulting from the
optimization involved in the computation of 
$\wassersteinTree\big(\branchtree(f_i), \branchtree(f_j)\big)$ 
(\autoref{eq_our_distance}). 
In particular, given the two 
BDTs
$\branchtree(f_i)$ and $\branchtree(f_j)$, the 
interpolated 
BDT,
noted $\branchtree_\alpha(f_i \rightarrow f_j)$ with 
$\alpha 
\in [0, 1]$ such that $\branchtree_0(f_i \rightarrow f_j) = \branchtree(f_i)$ and
$\branchtree_1(f_i \rightarrow f_j) = \branchtree(f_j)$, is 
%
%
obtained by 
considering the union of: 
\vspace{-1.5ex}
\begin{enumerate}[itemsep=-1.5ex]
 \item{the linear
interpolation $B_\alpha \subseteq \branchtree_\alpha(f_i \rightarrow f_j)$, 
between 
the nodes $\revision{B_i} \subseteq \branchtree(f_i)$ and these of $\revision{B_j}
\subseteq \branchtree(f_j)$, given the isomorphism $\phi'$ (the trees 
\revision{$B_i$, $B_j$} and $B_\alpha$ are then isomorphic,
\vspace{-.5ex}
\autoref{fig_matchingDiagramVSTree}):
\begin{eqnarray}
  \label{eq_interpolation_map}
b(\alpha) = (1 - \alpha) b + \alpha \phi'(b) & \forall b \in \revision{B_i}
\end{eqnarray}
}
 \item the linear
 interpolation of the destruction of the 
 subtrees $\revision{\overline{B_i}}$,
noted $\overline{\revision{B_i}^\alpha}
\subseteq \branchtree_\alpha(f_i \rightarrow f_j)$ ($\overline{\revision{B_i}}$ and
$\overline{\revision{B_i}^\alpha}$ are also isomorphic):
\vspace{-.5ex}
\begin{eqnarray}
\label{eq_interpolation_destruction}
 b(\alpha) = (1 - \alpha) b + \alpha \projection(b) & \forall b \in 
\overline{\revision{B_i}}
\end{eqnarray}

\item the 
linear 
interpolation of the creation of the subtrees $\overline{\revision{B_j}}$,
noted 
$\overline{\revision{B_j}^\alpha}
\subseteq \branchtree_\alpha(f_i \rightarrow f_j)$ ($\overline{\revision{B_j}}$ and
$\overline{\revision{B_j}^\alpha}$ are also isomorphic):
\vspace{-.5ex}
\begin{eqnarray}
\label{eq_interpolation_creation}
 b(\alpha) = (1 - \alpha)  \projection(b) + \alpha b & \forall b 
\in 
\overline{\revision{B_j}}.
\end{eqnarray}
\end{enumerate}
\vspace{-1.5ex}

Similarly to the 
distance $\wasserstein{2}$
between persistence diagrams, 
since the edit costs involved in the distance 
$\wassersteinTree$ are Euclidean distances in the birth/death 
space (\autoref{eq_our_mapping_cost}), the interpolated branches $b(\alpha)$ of 
$\branchtree_\alpha(f_i \rightarrow f_j)$ can be efficiently 
computed with the simple linear interpolations described above.
As detailed in Appendix \revision{5},
the resulting interpolated tree 
$\branchtree_\alpha(f_i \rightarrow f_j)$ is indeed on a geodesic 
given $\wassersteinTree$.

\cutout{In particular, given an arbitrary branch decomposition tree 
$\branchtree$, 
similarly to \autoref{eq_frechet_energy},
let $F(\branchtree, \alpha)$ be the Fr\'echet energy:
\begin{eqnarray}
\label{eq_interpolationEnergy}
 F(\branchtree, \alpha) = \alpha \wassersteinTree\big(\branchtree, 
\branchtree(f_i)\big)^2 + (1 - \alpha) \wassersteinTree\big(\branchtree, 
\branchtree(f_j)\big)^2
\end{eqnarray}
As detailed in Appendix 3 (additional material), 
$F\big(\branchtree_\alpha(f_i \rightarrow f_j), \alpha\big)$ is a minimum of 
the above Fr\'echet energy for any $\alpha \in [0, 1]$, which 
confirms that the linear interpolation $\branchtree_\alpha(f_i \rightarrow f_j)$
introduced above
indeed describes a geodesic between $\branchtree(f_i)$ and 
$\branchtree(f_j)$, given $\wassersteinTree$.}


\subsection{From branch decomposition trees to merge trees}
\label{sec_normalization}
The previous section described the computation of geodesics between 
BDTs,
given
$\wassersteinTree$. In this section,
given 
an interpolated 
BDT
$\branchtree_\alpha(f_i \rightarrow 
f_j)$, we describe how to retrieve the corresponding merge tree 
$\mergetree_\alpha(f_i \rightarrow f_j)$ (i.e. a merge tree whose 
BDT
is indeed equal to $\branchtree_\alpha(f_i \rightarrow 
f_j)$).

\begin{figure}[b]
\includegraphics[height=0.18\linewidth]{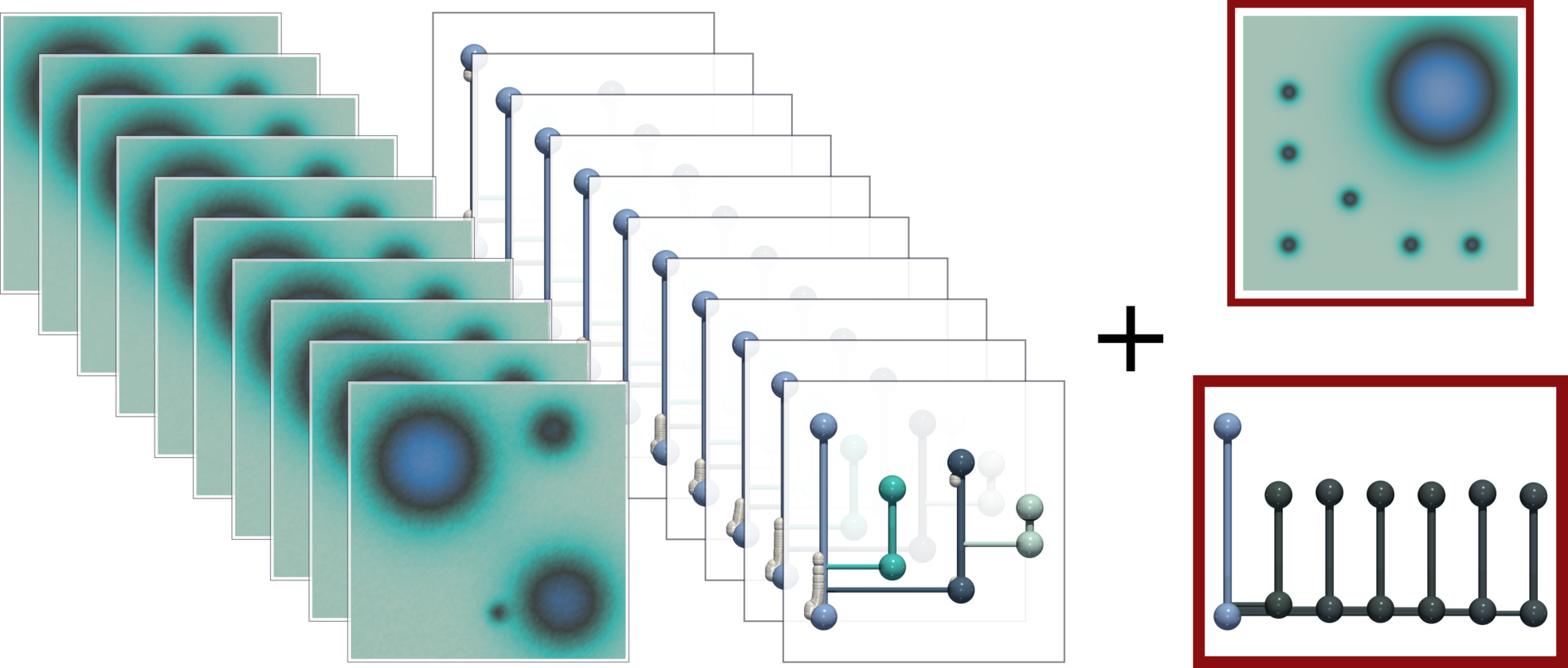}
  \hfill
\includegraphics[height=0.18\linewidth]{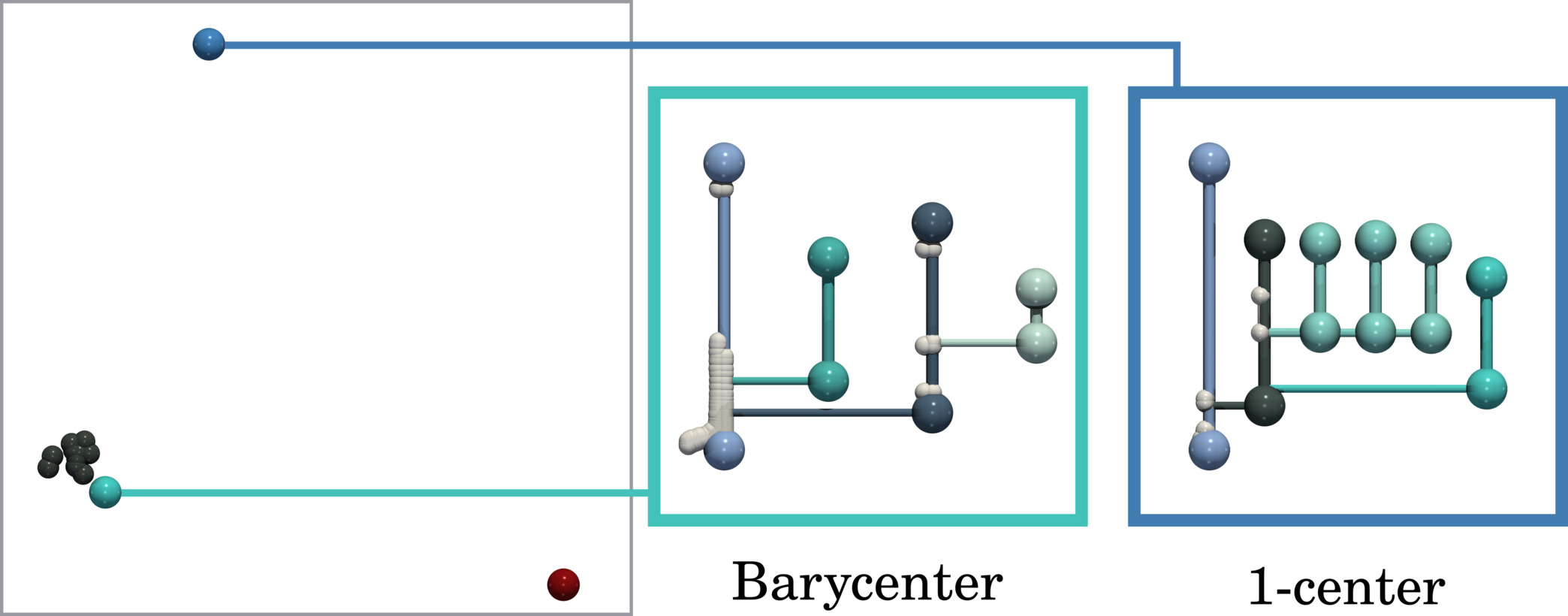}
\mycaption{Visual comparison between our barycenter (\autoref{sec:barycenters}) 
and the 1-center 
of Yan et al. \cite{YanWMGW20, 1centerCode}. 
Left: an ensemble is created with an outlier member $f_j$ (red, 7 persistent 
branches) and 10 noisy versions of a field $f_i$ (4 persistent branches).
Right: planar view of the ensemble computed by multi-dimensional scaling of 
$\wassersteinTree$. The barycenter 
computed with our approach 
(cyan) 
is more similar to the merge trees of 
$f_i$ (same number and persistence of large branches)
and hence better captures the overall trend of the 
ensemble, despite the presence of the outlier  $f_j$ (red sphere).}
\label{fig_barycenter}
\end{figure}

\begin{figure*}
\includegraphics[width=\linewidth]{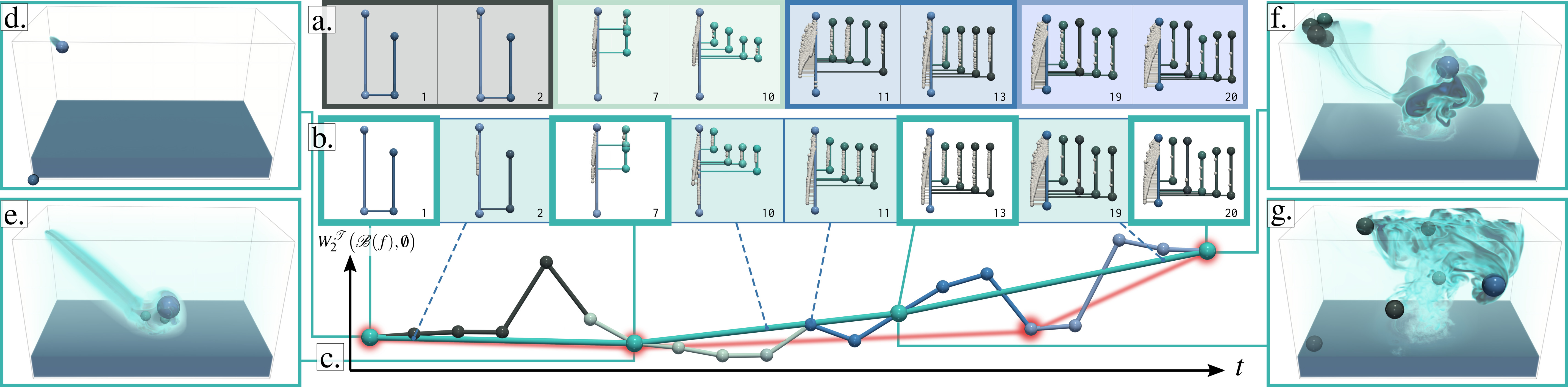}
\vspace{-2ex}
 \mycaption{Geodesic computation for the reduction (b) of temporal sequences 
of merge trees (a).
Our algorithm greedily removes from the sequence the 
trees  that it can accurately estimate 
by geodesic 
computation (trees with blue background (b)\revision{). This reduction is 
also visualized
with the three curves 
in (c), 
plotting
the distance 
$\wassersteinTree$ to the empty tree $\emptyset$ 
over time
(multiple colors: original sequence,
cyan:  
reduction by $\wassersteinTree$, red: reduction
by $\wasserstein{2}$).}
This iterated removal of trees highlights 
\emph{key frames} in the sequence 
(d-g)
corresponding to 
key phases of 
an
 asteroid impact 
simulation \cite{scivis2018}:
initial state (black, time steps 1-5), approach (light green, 6-10), impact 
(blue, 11-15), 
aftermath (light blue, 16-20). In contrast, a similar greedy optimization 
based on the 
distance 
$\wasserstein{2}$
between persistence diagrams (red curve) 
fails at capturing 
the \emph{impact}
phase (blue)
of this sequence.}
\label{fig_temporalReduction}
%
\end{figure*}

\begin{figure}
 \centering 
\includegraphics[width=0.48\linewidth]{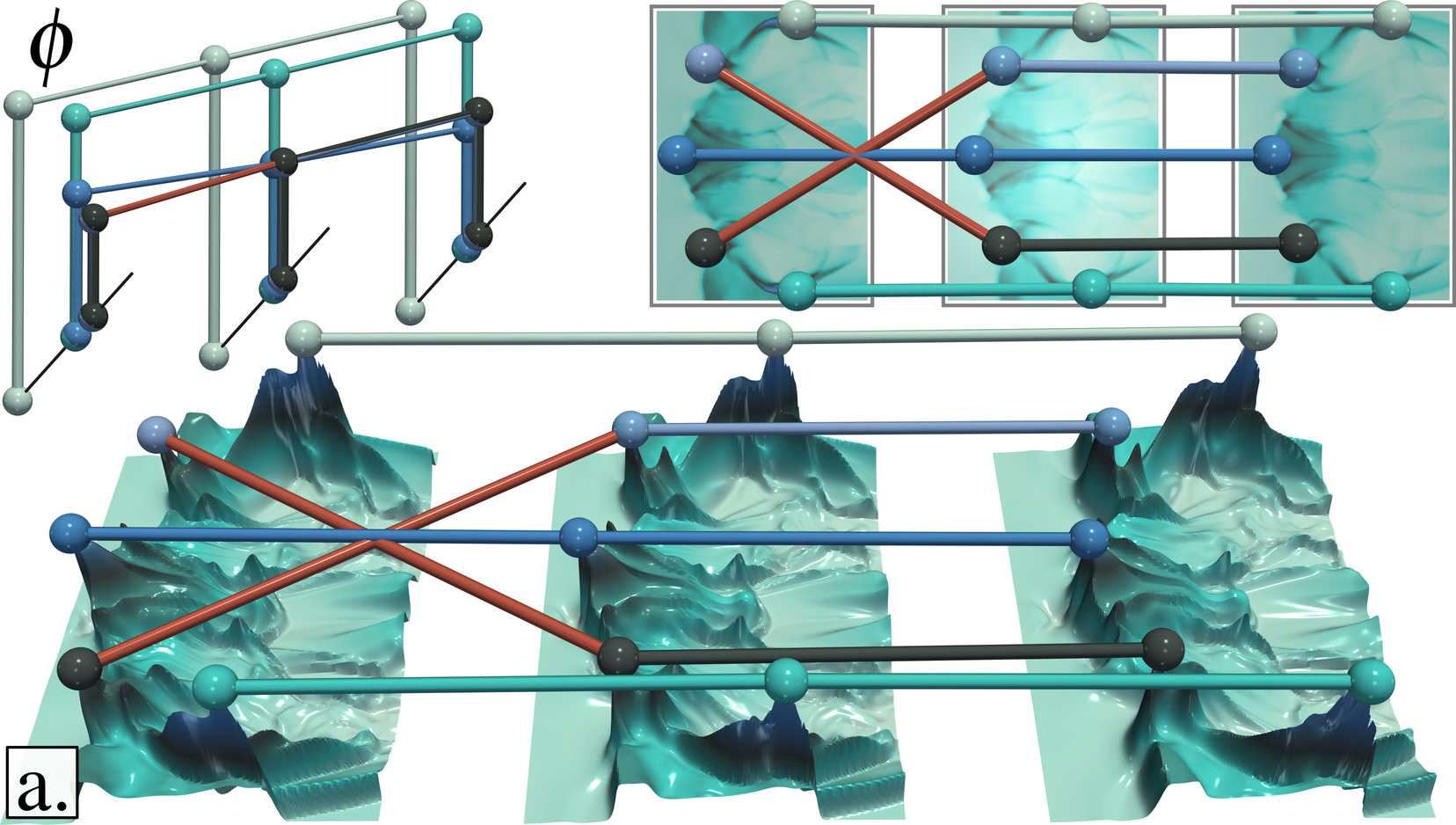}
\hfill
\includegraphics[width=0.48\linewidth]{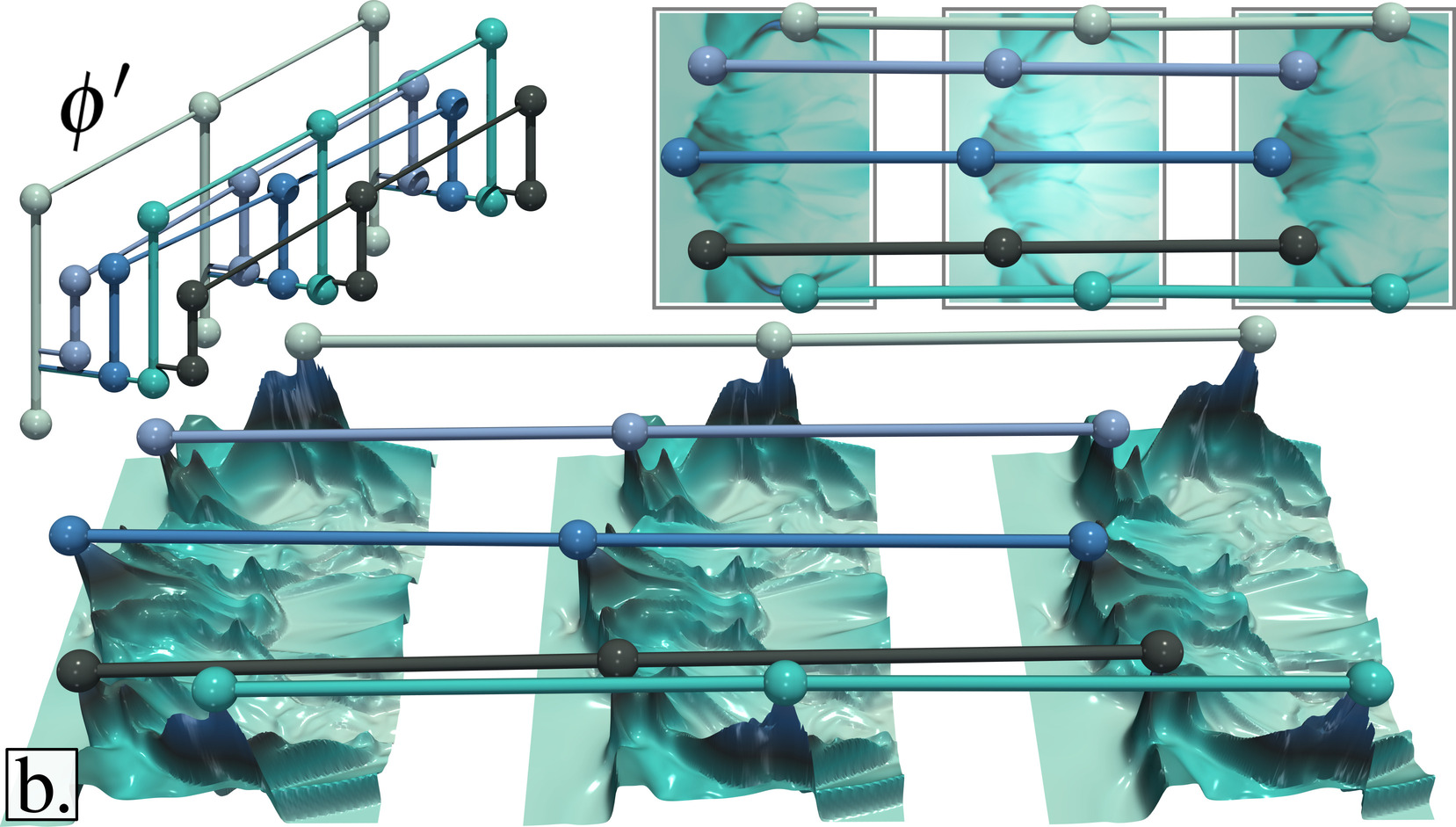}
 \mycaption{Tracking features (the five most persistent maxima, spheres) 
 in time-varying 2D data (ion density during universe formation 
\cite{scivis2008}):
optimal 
 assignment $\phi$ (a) of $\wasserstein{2}$
(\autoref{sec_background_persistenceDiagrams}), optimal  
isomorphism $\phi'$ (b) of $\wassersteinTree$ 
(\autoref{sec_our_distance_definition}). Since 
$\wasserstein{2}$
considers persistence pairs 
individually,
it 
can generate
incorrect
matchings 
resulting in a 
characteristic \emph{crossing} (a, red). 
Our distance improves this aspect (b) thanks to its more constrained search 
space, which better preserves the global structure of the data.}
%
%
\label{fig_timeTracking}
\end{figure}

A requirement for an arbitrary 
BDT
$\branchtree$ to be 
the valid 
BDT
of a merge tree $\mergetree$ is that 
subtrees of $\branchtree$ need to respect a \emph{nesting condition} on their 
birth/death (i.e. $x, y$) values (to respect the Elder rule, 
\autoref{sec_background_persistenceDiagrams}).
In 
particular, given a direct child $b_\alpha^k$ of a branch $b_\alpha \in 
\branchtree_\alpha(f_i \rightarrow f_j)$, we need to guarantee that 
$[x^k_{b_\alpha}, y^k_{b_\alpha}] \subseteq [x_{b_\alpha}, y_{b_\alpha}]$. While
this is guaranteed by construction for the subset $B_\alpha \subseteq 
\branchtree_\alpha(f_i \rightarrow f_j)$ ($B_\alpha$ is isomorphic to $\revision{B_i}$ and
$\revision{B_j}$), this is not necessarily the case for the subsets of
$\branchtree_\alpha(f_i \rightarrow f_j)$ involved in subtree creation or 
destruction ($\overline{\revision{B_i}^\alpha}$ and $\overline{\revision{B_j}^\alpha}$,
\autoref{sec_geodesicDefinition}). In particular,  since the branches involved 
in destructions map independently to the diagonal 
(\autoref{eq_interpolation_destruction}), it is possible that the above nesting 
condition 
is not respected along their interpolation. 
This is 
\revision{shown}
in 
\autoref{fig:slidingterm}\revision{c} (\revision{red interpolation}), where the 
\revision{resulting}
merge tree, $\mergetree_\alpha(f_i \rightarrow 
f_j)$, becomes disconnected and hence invalid (i.e. 
$\branchtree_\alpha(f_i \rightarrow
f_j)$\revision{, \autoref{fig:slidingterm}\revision{d},}
\revision{\emph{is}} 
connected \revision{and} 
not equal to 
the 
BDT
of $\mergetree_\alpha(f_i \rightarrow 
f_j)$ \revision{from \autoref{fig:slidingterm}\revision{c}}).

In the following, 
we introduce a 
pre-processing step for the 
trees $\branchtree(f_i)$ and 
$\branchtree(f_j)$ (together with its inverse post-processing step),
which we call \emph{local normalization}, which addresses this issue and 
guarantees the above nesting condition, even in case of destruction/creation.

Given a direct child $\revision{b_i}^k$ of a branch $\revision{b_i} \in \branchtree(f_i)$, we
consider the following local, birth/death normalization 
$\normalizedLocation(\revision{b_i}^k) = \big(\normalizedLocation_x(\revision{b_i}^k),
\normalizedLocation_y(\revision{b_i}^k)\big)$:
\vspace{-1.5ex}
\begin{eqnarray}
  \nonumber
  \normalizedLocation_x(\revision{b_i}^k) = (x_{\revision{b_i}^k} - x_{\revision{b_i}})/(y_{\revision{b_i}} - x_{\revision{b_i}})\\
  \nonumber
  \normalizedLocation_y(\revision{b_i}^k) = (y_{\revision{b_i}^k} - x_{\revision{b_i}})/(y_{\revision{b_i}} - x_{\revision{b_i}}).
\end{eqnarray}
\vspace{-2.5ex}

Once this 
pre-process 
is 
recursively 
completed,
the Wasserstein distance 
$\wassersteinTree$ between 
the locally normalized 
BDTs,
noted 
$\normalizedLocation\big(\branchtree(f_i)\big)$ 
and $\normalizedLocation\big(\branchtree(f_j)\big)$  is 
computed as described in 
\autoref{sec_algorithm}.
Then, the interpolation of the locally normalized 
BDTs,
noted
$\normalizedLocation\big(\branchtree_\alpha(f_i \rightarrow f_j)\big)$,
is evaluated as described in 
\autoref{sec_geodesicDefinition}. Next, the 
local normalization is recursively reverted to turn 
$\normalizedLocation\big(\branchtree_\alpha(f_i \rightarrow f_j)\big)$
back
into $\branchtree_\alpha(f_i 
\rightarrow f_j)$, by explicitly evaluating 
$\normalizedLocation^{-1}(b_\alpha^\normalizedLocation)$ for each branch 
$b_\alpha^\normalizedLocation \in 
\normalizedLocation\big(\branchtree_\alpha(f_i \rightarrow f_j)\big)$.
Now, even in case of branch destruction, by construction, 
the birth/death interval of 
each interpolated branch 
$\normalizedLocation(b_\alpha)$, noted 
$[\normalizedLocation_x(b_\alpha), 
\normalizedLocation_y(b_\alpha)]$,
is included in $[0, 1]$ 
(since $\projection\big(\normalizedLocation(\revision{b_i})\big)
\subseteq [\normalizedLocation_x(\revision{b_i}), \normalizedLocation_y(\revision{b_i})] \subseteq
[0, 1]$).
Therefore, 
after reverting the 
local normalization, we have the guarantee that $[x^k_{b_\alpha}, 
y^k_{b_\alpha}] \subseteq [x_{b_\alpha}, y_{b_\alpha}]$ for all the branches 
$b_\alpha$ of $\branchtree_\alpha(f_i \rightarrow 
f_j)$.

At this stage, 
$\branchtree_\alpha(f_i \rightarrow 
f_j)$ indeed respects the nesting condition on the birth/death values of all 
its subtrees. Then, 
given the dual relation between merge trees and 
BDTs,
the 
merge tree $\mergetree_\alpha(f_i 
\rightarrow 
f_j)$ can be simply obtained by creating a vertical branch for each node 
$b_\alpha$ of $\branchtree_\alpha(f_i \rightarrow 
f_j)$ and connecting them according the arcs of
$\branchtree_\alpha(f_i \rightarrow 
f_j)$, as illustrated in \autoref{fig:slidingterm} (right). 
The distance $\wassersteinTree$ between 
$\normalizedLocation\big(\branchtree(f_i)\big)$ and 
$\normalizedLocation\big(\branchtree(f_j)\big)$
then still describes a metric 
between $\branchtree(f_i)$ and $\branchtree(f_j)$, such that 
$\branchtree_\alpha(f_i \rightarrow f_j)$ is indeed on a geodesic (see 
Appendix \revision{6}).

\cutout{Note that this process (normalization, followed by an interpolation, 
followed 
by a normalization reversal) is equivalent to a direct interpolation (as 
defined in \autoref{sec_geodesicDefinition}) of our original edit distance 
$\wassersteinTree$, but where the edit costs  
(\autoref{eq_our_mapping_cost}) are re-defined in a locally normalized form:
 \begin{eqnarray}
  \gamma\big(b_i \rightarrow \phi''(b_i)\big)
     = \pointMetric_2^N\big(b_i, \phi''(b_i)\big)
     = \pointMetric_2\Big(\normalizedLocation(b_i),
\phi''\big(\normalizedLocation(b_i)\big)\Big)\\
  \gamma(b_i \rightarrow \emptyset)
     = \pointMetric_2^N\big(b_i, \projection(b_i)\big)
     = \pointMetric_2\Big(\normalizedLocation(b_i),
\projection\big(\normalizedLocation(b_i)\big)\Big)\\
  \gamma(\emptyset \rightarrow b_b) 
     = \pointMetric_2^N\big(b_b, 
\projection(b_b)\big)
     = \pointMetric_2\Big(\normalizedLocation(b_b), 
      \projection\big(\normalizedLocation\big(b_b)\big)\Big)
\end{eqnarray}
As discussed in Appendix 4 (additional material), when considering the above 
locally normalized costs, the distance $\wassersteinTree$ is still 
non-negative, and symmetric and it still preserves the identity of 
indiscernibles and the triangle inequality. Moreover, as detailed in Appendix 5 
(additional material), its linear interpolation (as described in 
\autoref{sec_geodesicDefinition}) still minimizes the Fr\'echet energy 
(\autoref{eq_interpolationEnergy}). Then, overall the branch 
decomposition tree $\branchtree_\alpha(f_i \rightarrow 
f_j)$ interpolated with the above locally normalized costs indeed 
describes a geodesic between $\branchtree(f_i)$ and 
$\branchtree(f_j)$ given $\wassersteinTree$ and $\mergetree_\alpha(f_i 
\rightarrow 
f_j)$ is its valid, corresponding merge tree.}

Note that the local normalization shrinks all the input branches to the 
interval $[0, 1]$, irrespective of their original persistence. To mitigate this 
effect, we introduce a pre-processing step on the 
input BDTs,
which moves, up the trees, subtrees rooted at branches 
with a 
relative persistence smaller than $\epsilon_3$, until 
their persistence relative to their parent becomes smaller than a threshold 
$\epsilon_2$. 
This has the practical effect of reducing the normalized persistence of small 
branches corresponding to small features. Overall, $\epsilon_1$, $\epsilon_2$ 
and $\epsilon_3$ are the only parameters of our approach
and
we use a unique, default set of values ($\epsilon_1 = 0.05$, $\epsilon_2 = 
0.95$ and $\epsilon_3 = 
0.9$) 
in 
our experiments (\autoref{sec_results}). 
In the remainder, we will consider that all the input BDTs are normalized this 
way. 

\section{Wasserstein Barycenters of Merge Trees}
\label{sec:barycenters}

This section introduces our approach for the computation of barycenters of 
merge trees, for the metric $\wassersteinTree$ (\autoref{sec_metric}). 
\cutout{For this, we leverage our approach for geodesic computation 
(\autoref{sec:geodesics}) and adapt existing Fr\'echet energy optimization 
algorithms \cite{Turner2014, vidal_vis19} from 
persistence 
diagrams 
to the case 
of merge trees.}The resulting barycenters will 
serve 
as core tools for clustering ensembles of merge trees 
(\autoref{sec_application}).

\subsection{Definition}
\label{sec_def_frechetMeanBranch}
Let $\branchset = \{\branchtree(f_1), \branchtree(f_2), \dots, 
\branchtree(f_N)\}$ be a set of 
$N$ BDTs.
Similarly to \autoref{eq_frechet_energy}, The Fr\'echet energy,
under the metric $\wassersteinTree$, is given by:
\vspace{-2ex}
\begin{eqnarray}
\nonumber
\label{eq_frechet_energy_trees}
F(\branchtree) = 
 \sum_{\branchtree(f_i) \in \branchset}
        \wassersteinTree\big(\branchtree, \branchtree(f_i)\big)^2.
\end{eqnarray}
\vspace{-3ex}

We call a \emph{Wasserstein barycenter} of $\branchset$,  
a
BDT
$\branchtree^* \in \branchspace$ (where 
$\branchspace$ is the 
space of 
BDTs)
which minimizes $F(\branchtree)$. 
It is 
a centroid of the 
set, i.e. a tree which minimizes the sum of its distances to the set.

\subsection{Computation}
\label{sec_barycenter_optimization}

Our distance 
$\wassersteinTree$ (\autoref{sec_metric}) is identical to 
$\wasserstein{2}$, 
but with 
a 
smaller search space, restricted to rooted partial isomorphisms. 
This 
enabled
an extension of interpolation-based geodesics from persistence 
diagrams to merge trees (\autoref{sec:geodesics}). Given these two components, 
the 
strategy presented by Turner et al. \cite{Turner2014} 
for minimizing the Fr\'echet energy over the space of persistence diagrams can 
be 
directly extended to our framework. For this, we consider an algorithm that 
resembles 
a Lloyd relaxation 
\cite{lloyd82}, and which alternates
an \emph{(i)} \emph{assignment} and an 
\emph{(ii)} \emph{update} procedure. 
First, the candidate $\branchtree$ is initialized at an arbitrary tree of 
$\branchset$. Then the assignment step \emph{(i)} 
computes an optimal
assignment
\revision{$(\phi'_i, \overline{B_{\branchtree}}, \overline{B_i})$}
between $\branchtree$ and each 
tree $\branchtree(f_i) \in \branchset$. Next, 
the update step \emph{(ii)} updates the candidate 
$\branchtree$ to a position in $\branchspace$ which minimizes 
$F(\branchtree)$
under the current set of assignments
\revision{$(\phi'_i,
\overline{B_{\branchtree}}, \overline{B_i})_{i = 1, \dots, N}$}.
This is achieved by 
moving each branch $b \in 
\branchtree$ (in the birth/death space) to the arithmetic mean  of the 
assignments (by generalizing the interpolation defined in Eqs. 
\ref{eq_interpolation_map}, \ref{eq_interpolation_destruction}, and 
\ref{eq_interpolation_creation}, 
to more 
than 
two trees):
\vspace{-2ex}
\begin{eqnarray}
\nonumber
b \leftarrow {1\over{N}}\sum_{i = 1, \dots, N}
\begin{cases}
  \label{eq_barycenter_mean}
  \phi'_i(b) & \text{if } b \in \revision{B_{\branchtree}}\\
  \projection(b) & \text{if } b \in 
\revision{\overline{B_{\branchtree}}}\\
  b & \text{if } b \in 
\revision{\overline{B_i}}.
\end{cases}
\end{eqnarray}
\vspace{-3ex}

This overall assignment/update sequence  is then iterated
(as 
discussed in Appendix 
\revision{7},
each iteration of this sequence decreases 
the Fr\'echet energy constructively).
In our implementation, the algorithm stops and returns the barycenter 
estimation $\branchtree^*$ when the Fr\'echet 
energy decreased by less than $1\%$ between two consecutive iterations. Given
$\branchtree^*$, we obtain its dual merge tree 
$\mergetree^*$ as described in \autoref{sec_normalization}.
\autoref{fig_barycenter} illustrates a barycenter computed with this 
strategy 
for a toy example.



\cutout{Moreover, similarly to the case of $\wassersteinTree$ enables 
interpolation-based geodesics 
our 
framework supports the computation of geodesics with regard to 
$\wassersteinTree$, which therefore makes it readily applicable to the 
optimization strategy presented by Turner et al. \cite{Turner2014} for the 
minimization of the Fr\'echet energy in the case of persistence diagrams, given 
$\wasserstein{2}$.

since our framework 

Thus, our 
approach to barycenter computation readily applies 

the above assignment/update 
strategy, 
but applied on the set $\branchset$ (instead of $\diagramSet$) and considering 
$\wassersteinTree$ (instead of $\wasserstein{2}$). In particular, as detailed 
in Appendix 6 (additional material), for a fixed set of partial isomorphisms 
$\{\phi''_1, \phi''_2, \dots, \phi''_N\}$, the arithmetic mean of the assigned 
points in the birth/death space is indeed a local minimizer of
$F(\branchtree)$.
In practice, we consider that the optimization converged when the energy 
decreased by less than $1\%$ between two consecutive iterations.
Then, given a converged local minimizer $\branchtree^*$, we 
obtain its 
dual merge tree $\mergetree^*$ as described in \autoref{sec_normalization}.}

\cutout{The minimization of \autoref{eq_frechet_energy_trees} involves a large 
assignment problem, involving $N$ (smaller) assignment problems 
(\autoref{sec_algorithm}) between the candidate $\branchtree$ and the trees of 
$\branchset$. 
As discussed by Turner et 
al. \cite{Turner2014} 
for 
persistence diagrams, the minimum of the 
Fr\'echet energy can be found 
by an exhaustive exploration of the 
space of assignments, which is impractical. Instead, 
they introduce a 
gradient descent approach, based on small geodesic jumps,
which we recap here for completeness.

A local minimizer $\diagram^*$ of the Fr\'echet energy 
(\autoref{eq_frechet_energy}) can be efficiently estimated with an
optimization strategy 
\cite{Turner2014} which resembles a Lloyd relaxation 
\cite{lloyd82}, and which alternates an \emph{(i)} \emph{assignment} and an 
\emph{(ii)} \emph{update} procedure. 
First, $\diagram$ is initialized at an 
arbitrary diagram of $\diagramSet$. Then, the assignment step  \emph{(i)}
computes an optimal partial assignment $\phi_i : \diagram \rightarrow 
\diagram(f_i)$ 
between $\diagram$ and each diagram of $\diagramSet$. 
Next, the update step \emph{(ii)}
updates the candidate $\diagram$ to a position in $\diagramSpace$ 
which minimizes the Fr\'echet energy, 
under the current set of assignments $\{\phi_1, \phi_2, \dots, \phi_N\}$. 
This is achieved, similarly to geodesic computation, by replacing each point 
$p_a \in \diagram$ by the arithmetic 
mean (in the birth/death space) of all its assignments $\phi_i(p_a)$. The 
overall 
sequence assignment/update is iterated until convergence of the Fr\'echet 
energy. In particular, once a local minimizer of the Fr\'echet energy is 
obtained for a fixed assignment with the update step (\emph{(ii)}), the 
subsequent assignment step (\emph{(i)}) further improves the assignments if 
possible, hence iteratively decreasing the Fr\'echet energy constructively.

As discussed in \autoref{sec_metric}, our distance between 
BDTs,
$\wassersteinTree$, is identical to the 
$L^2$-Wasserstein distance 
between persistence diagrams, $\wasserstein{2}$, at the notable difference of a 
smaller search space, restricted to partial isomorphisms. Thus, our approach to 
barycenter computation readily applies the above assignment/update strategy, 
but applied on the set $\branchset$ (instead of $\diagramSet$) and considering 
$\wassersteinTree$ (instead of $\wasserstein{2}$). In particular, as detailed 
in Appendix 6 (additional material), for a fixed set of partial isomorphisms 
$\{\phi''_1, \phi''_2, \dots, \phi''_N\}$, the arithmetic mean of the assigned 
points in the birth/death space is indeed a local minimizer of
$F(\branchtree)$. 
In practice, we consider that the optimization converged when the energy 
decreased by less than $1\%$ between two consecutive iterations.
Then, given a converged local minimizer $\branchtree^*$, we 
obtain its 
dual merge tree $\mergetree^*$ as described in \autoref{sec_normalization}.
\autoref{fig_barycenter} illustrates a barycenter computed with this strategy 
for a toy example.}

\cutout{
\subsection{Progressivity}
To accelerate the optimization, Vidal et al. \cite{vidal_vis19} 
introduce a \emph{progressive} computation scheme, which increases assignment 
accuracy between consecutive iterations, and which also progressively populates 
the diagrams with points of decreasing persistence along the iterations. 

We describe in the following a similar strategy for merge trees. Since the 
local assignment problems between forests (\autoref{sec_algorithm}) are 
usually
small (the valence of the nodes in the 
BDTs
is 
rarely greater than 2), these are typically solved very efficiently in our 
approach, and only negligible performance gains can be expected by considering 
a progressively refined assignment accuracy, as done in the case of persistence 
diagrams \cite{vidal_vis19}. Thus, we focus only on persistence-driven 
progressivity in the following. 
Let $\branchtree_\rho(f_i)$ be 
the subset of $\branchtree(f_i)$ containing only branches with a 
relative persistence greater than $\rho \in [0, 1]$. Then, we initialize the 
optimization algorithm described in \autoref{sec_barycenter_optimization} by 
considering the set of trees $\{\branchtree_\rho(f_1), \branchtree_\rho(f_2), 
\dots, 
\branchtree_\rho(f_N)\}$ with \julien{$\rho = 0.2$} and we progressively 
decrease $\rho$ \julien{by $0.02$} after each iteration of the optimization. In 
practice, this results in a fast approximation of a reasonable local minimum of
%
the Fr\'echet energy, for which only very few iterations will be necessary when 
$\rho$ reaches zero.
%
%
%
%
\julien{
- initialement seuil à 50\% (on garde que les branches plus persistantes que 50\%)
- à chaque itération: ajout 2.5\% des paires les plus persistantes par rapport aux débuts (N)
   - i*N/40
- une fois qu'on a tout le monde, si on décroit de moins de 1\%, on s'arrête
}
}

\subsection{Parallelism}
\label{sec_barycenter_parallelism}
The $N$ assignment problems (between the candidate $\branchtree$ and the trees 
of the set $\branchset$, \autoref{sec_barycenter_optimization}) are independent 
and can
be computed in parallel. However, 
this naive strategy
is subject to load imbalance, as 
the input trees 
can have 
different sizes. Hence, each iteration would 
be bounded by the 
sequential
execution 
of the largest of the $N$ assignment problems.

We address this issue 
by leveraging the task-based 
parallelization of our distance computation algorithm 
(\autoref{sec_distanceParallelization}). In particular, we use a single task 
pool for all 
of 
the $N$ assignment problems. Then, the task 
environment picks up at runtime the tasks to compute irrespective of their tree 
of origin, and place them on different threads.
This fine scheduling granularity has the beneficial effect of 
triggering the execution of the tasks 
of a new assignment problem while a first problem is reaching 
completion (and thus exploiting less threads, 
\autoref{sec_distanceParallelization}). This improves thread load 
imbalance and thus increases the overall parallel efficiency.



%

\section{Applications}
\label{sec_application}
The section illustrates the utility of 
our contributions (distances, 
geodesics, and barycenters) in concrete visualization tasks 
(\autoref{fig:teaser}).

\subsection{Branch matching for feature tracking}
Our distance 
(\autoref{sec_algorithm}) relies on the optimization 
of a partial isomorphism between the input
BDTs.
\revision{Then,} the resulting
matchings can
be used to track
features in time-varying data\revision{, as studied for
persistence diagrams \cite{soler_ldav18}}.
\autoref{fig_timeTracking} 
illustrates this
on 
a temporal sequence
\revision{(}SciVis contest 2008 \cite{scivis2008}\revision{)}.
Since
$\wasserstein{2}$ 
considers persistence pairs individually, 
it can generate inconsistent
matchings with a typical 
incorrect \emph{crossing} in the feature tracking (already visible on 
synthetic data, \autoref{fig_matchingDiagramVSTree}). Our distance
$\wassersteinTree$
improves 
this aspect by better preserving the global structure of the data, thanks to 
our more constrained, merge-tree driven, assignment search space.
\revision{Overall, our matchings provide visual hints to the users, to
help them
relate features from distinct time steps.}







\subsection{Geodesics for temporal reduction}
The topological analysis of time-varying data typically requires the 
computation of a topological representation, for instance a merge tree, 
for each time step. Although merge trees are usually orders of 
magnitude smaller 
than the original data, the resulting sequence of 
merge trees can still represent considerable amounts of data. To address this 
issue, we exploit our geodesic computation (\autoref{sec:geodesics}) for the 
reduction of temporal sequences of merge trees. In particular, we greedily 
remove from the sequence, one by one, the merge trees which can be accurately 
reconstructed by simple geodesic computation, until the sequence only contains 
a target number of merge trees (see the detailed algorithm in 
Appendix \revision{8}).
\revision{This enables the reliable visualization of time-varying sequences of
merge trees at greatly reduced storage costs.}
\revision{The} remaining merge trees (white
background, \revision{\autoref{fig_temporalReduction}}) correspond to \emph{key
frames} of the sequence, i.e. time steps
of particular significance in 
terms of the features of interest.
In contrast, a similar
strategy based on persistence diagram interpolation (red curve) fails at 
identifying a key frame in one of the key phases of the sequence (impact, in 
blue).
Also, note that the reduced merge trees 
(reconstructed with geodesics, blue background) are visually highly similar to 
the trees from the input sequence.

%

\cutout{Let $\mathcal{S} = \{\branchtree(f_0), \branchtree(f_1), \dots, 
\branchtree(f_N)\}$ be the input temporal sequence of 
BDTs
(we assume a regular temporal sampling). Let 
$\mathcal{K} \subseteq 
\mathcal{S}$ be a set of key frames. Let $\mathcal{S}' = \{\branchtree'(f_0), 
\branchtree'(f_1), \dots, 
\branchtree'(f_N)\}$ be a \emph{reduced} temporal sequence, where:
\begin{eqnarray}
 \branchtree'(f_i) = \alpha_i \branchtree(f_j) + (1 - \alpha_i) \branchtree(f_k)
\end{eqnarray}
where $\branchtree(f_j)$ and $\branchtree(f_k)$ are two consecutive trees in 
$\mathcal{K}$, such that $j \leq i \leq k$ and $\alpha_i = (k-i)/(k-j)$. 
$\branchtree'(f_i)$ is then on the geodesic between $\branchtree(f_j)$ and 
$\branchtree(f_k)$. We introduce the following distance metric between 
the temporal sequences $\mathcal{S}$ and $\mathcal{S}'$ (see Appendix 7):
\begin{eqnarray}
 \distanceSequence(\mathcal{S}, \mathcal{S}') = \Big(\sum_{i = 0}^{N} 
\wassersteinTree\big(\branchtree(f_i), \branchtree'(f_i)\big)^2\Big)^{1/2}
\end{eqnarray}

Our algorithm for temporal reduction consists in initializing $\mathcal{K}$ 
with the entire input sequence ($\mathcal{K} \leftarrow \mathcal{S}$) and 
then removing greedily, at each iteration, the tree $\branchtree^*$ from 
$\mathcal{K}$
($\mathcal{K} \leftarrow \mathcal{K} - \{\branchtree^*\}$) which minimizes 
$\distanceSequence(\mathcal{S}, \mathcal{S}')$, until $\mathcal{K}$ reaches a 
target size.}


%
%
%

\subsection{Barycenters for topological clustering}
\label{sec_clustering}
To understand the main trends within an ensemble, in terms of features of 
interest, it may be desirable to cluster the ensemble by grouping 
members with a similar \emph{topological profile}. 
For this, we adapt the $k$-means algorithm \cite{elkan03, celebi13} to the 
problem of clustering merge trees. In particular, this can be easily achieved 
by using our merge tree barycenter computation algorithm 
(\autoref{sec:barycenters}) as the centroid estimation routine 
of 
$k$-means, and by using $\wassersteinTree$ (\autoref{sec_metric}) to 
measure the distance between merge trees.
\cutout{In this section, we present 
a variant of $k$-means which 
directly leverages our algorithm for barycenters 
of merge trees (\autoref{sec:barycenters}). 
Initially, $k$ centroids are initialized on $k$ trees of the ensemble 
$\branchset$ (\autoref{sec_def_frechetMeanBranch}) with the \emph{$k$-means++} 
heuristic \cite{celebi13}. Next, each \emph{clustering} iteration consists in 
alternating two sub-routines: \emph{(i)} \emph{assignment} and \emph{(ii)} 
\emph{update}. The assignment step \emph{(i)} assigns each tree 
$\branchtree(f_i)$ to its closest centroid \cite{elkan03} (using 
$\wassersteinTree$, \autoref{sec_metric}). The update step \emph{(ii)} 
re-positions each centroid at the barycenter of its assigned trees, using our 
barycenter algorithm (\autoref{sec_barycenter_optimization}). Finally, the 
clustering iterations are continued until the assignments between the trees and 
the $k$ centroids do not evolve anymore, resulting in the final clustering. The 
above \emph{assignment} and \emph{update} steps use each a single task pool to 
further improve parallel efficiency, as already discussed in 
\autoref{sec_barycenter_parallelism} in the case of barycenters.}
Note that in 
practice, our entire computational framework is implemented in this single 
clustering algorithm (with a unique task pool), as the above clustering 
generalizes the 
barycenter problem ($k = 1$)
as well as the geodesic and distance problems ($N = 2$).

\begin{figure}
\centering
\includegraphics[height=0.23\linewidth]{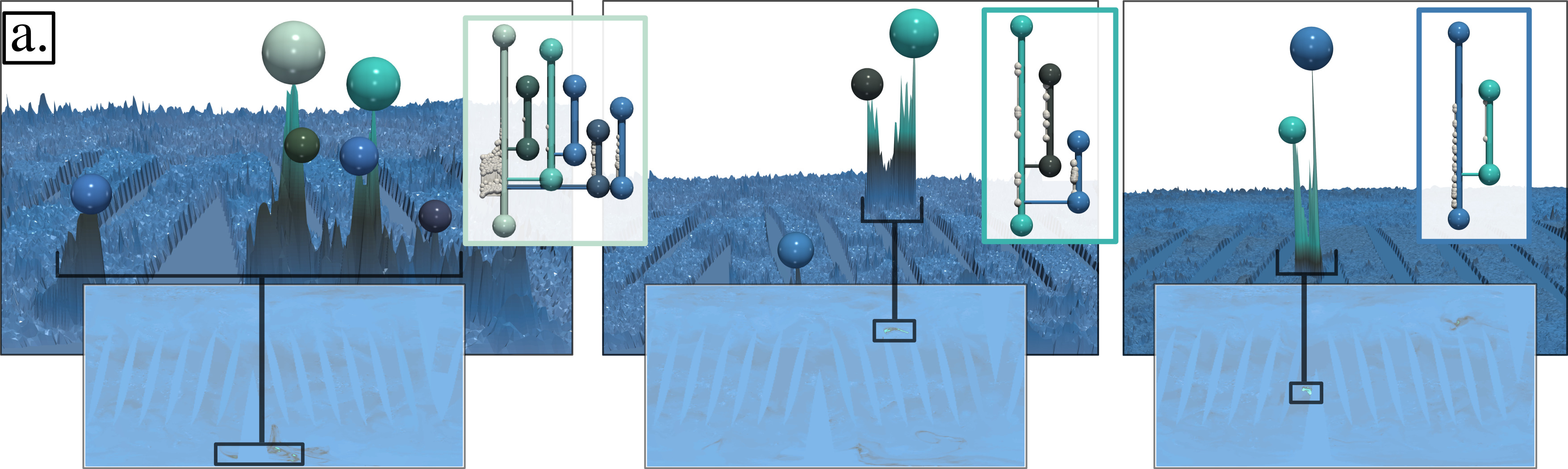}
\hfill
\includegraphics[height=0.23\linewidth]{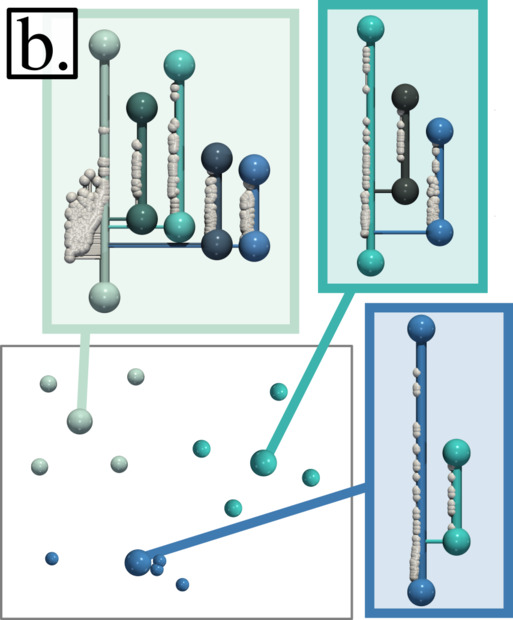}
\mycaption{Three members \textcolor{black}{(a)} of an acquired ensemble, corresponding to  
distinct 
volcanic eruptions \cite{scivis2014}. Our clustering 
approach correctly assigns the members to each cluster (\textcolor{black}{b, }distinct colors in the 
planar view, generated in a post-process by multi-dimensional scaling 
of 
$\wassersteinTree$). 
Our centroids (larger spheres in the planar view) 
provide a visual 
summary of the features of interest \revision{(matching 
colors)} for each cluster.}
%
\label{fig_clusteringResult1}
\end{figure}

\begin{figure}
\includegraphics[width=\linewidth]{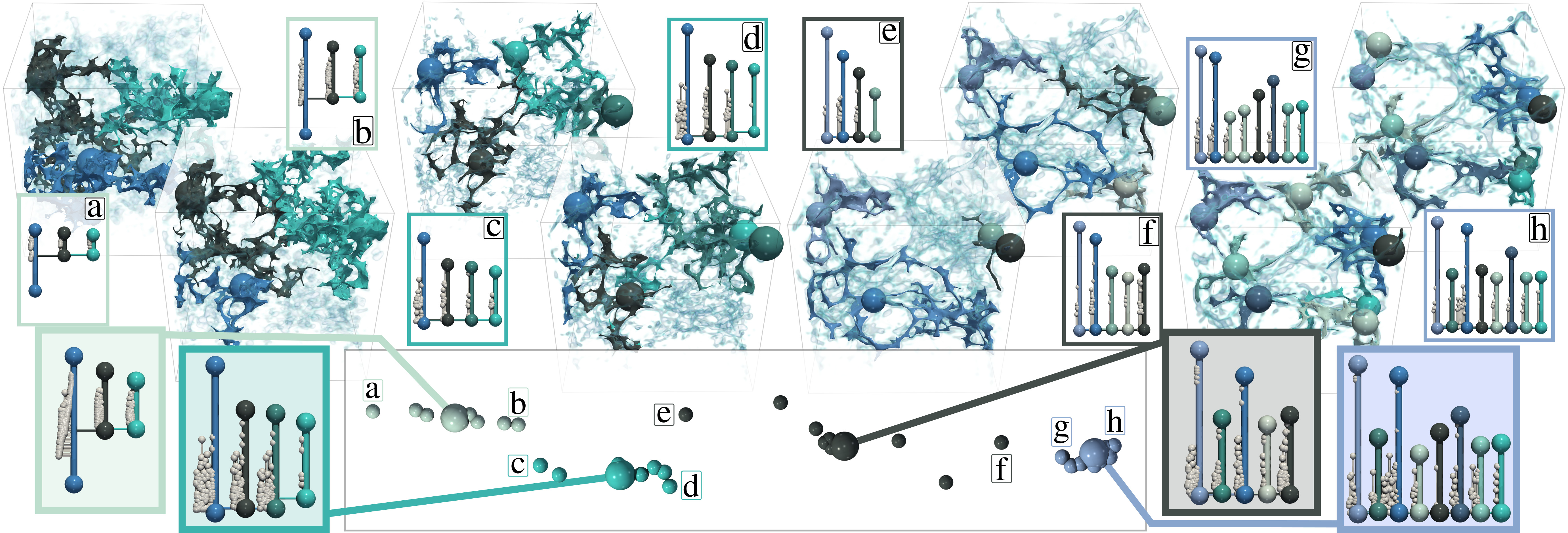}
\vspace{-2ex}
\mycaption{\revision{Eight} members (top) of 
a cosmology ensemble
\cite{scivis2015}, 
and
their merge trees 
(persistent maxima are displayed with matching colors in the data). 
Our clustering 
approach correctly assigns the members to each cluster (distinct colors in 
the bottom
planar view, generated in post-process by multi-dimensional scaling of 
$\wassersteinTree$). Our centroids (large spheres, bottom) provide a visual 
summary which is well representative of the trees in the cluster (same number 
and persistence of large branches\revision{, automatically color-coded based on 
their matching to their centroid}). 
}  
\label{fig_clusteringResult2}
\end{figure}

Figs \ref{fig_clusteringResult1} and \ref{fig_clusteringResult2}
present 
\revision{clustering examples}
obtained with this 
strategy
on an acquired 
\cite{scivis2014} and 
cosmology 
ensemble \cite{scivis2015}. In both 
\revision{cases},
our approach correctly assigns the 
members to each cluster. Moreover, the centroids computed by our algorithm 
provide a visual summary of the features of interest found in each cluster, 
\revision{enabling}
global overviews (Figs. \ref{fig_clusteringResult1}, right, and 
\ref{fig_clusteringResult2}, bottom) 
summarizing the topological profile of each of the main trends found in the 
ensemble.
\revision{In 
both figures, 
the tree branches of 
the
ensemble members are automatically colored 
with the color of 
their matched centroid
branch. This matching visualization 
enables users to 
visually relate the centroid to concrete features in the 
data (\autoref{fig_clusteringResult1})
and 
to compare matching features 
across multiple 
members (i.e. which have been matched 
to the same  centroid branch,
\autoref{fig_clusteringResult2}). Then, the centroid, in addition to 
being a visual summary,
also acts as a reference point for the visual comparison of 
ensemble members.}

%
%

%
%
%

\section{Results}
\label{sec_results}

This section presents experimental results obtained on a computer with 
two Xeon CPUs (3.2 GHz, 2x10 cores, 96GB of RAM). The input 
merge trees were computed with FTM \cite{gueunet_tpds19} and pre-processed to 
discard noisy features 
(persistence simplification threshold:
$0.25\%$ of the data range).
%
We 
implemented our approach in C++ (with the OpenMP task runtime),
as modules for TTK \cite{ttk17, ttk19}.

Our experiments were performed on a variety of simulated and acquired 2D and 3D 
ensembles used in previous work \cite{favelier2018} (vorticity and sea surface 
height) or extracted 
from 
past SciVis contests:
2004 
 (wind velocity magnitude \cite{scivis2004}),
2006 
 (wavefront velocity magnitude \cite{scivis2006}),
2008 
 (ion concentration \cite{scivis2008}),
2014
 (sulfur dioxide concentration \cite{scivis2014}), 
2015
 (dark matter density \cite{scivis2015}),
2016
 (salt concentration \cite{scivis2016}),
2017 
(pressure \cite{scivis2017} ),
2018
 (matter density \cite{scivis2018}). 
 \revision{A detailed specification of these ensembles is provided in
Appendix 9.}






\subsection{Time performance}

The time complexity of our algorithm for exploring the search space 
of $\wassersteinTree$ (\autoref{sec_algorithm}) is similar to that of 
the edit 
distance \cite{zhang96, SridharamurthyM20}. It takes 
$\mathcal{O}(|\branchtree|^2)$ steps in practice, with $|\branchtree|$ the number of nodes 
in the input BDTs 
(in our implementation, 
each local forest assignment problem is solved 
with the efficient \emph{Auction} approximation \cite{Bertsekas81} with default 
parameters). Once $\wassersteinTree\big(\branchtree(f_i), 
\branchtree(f_j)\big)$ is computed, the computation of a point on the geodesic 
(\autoref{sec:geodesics}) between $\branchtree(f_i)$ and $\branchtree(f_j)$ is 
obtained in $\mathcal{O}(|\branchtree|)$ steps. Regarding our barycenter 
computation algorithm (\autoref{sec:barycenters}), each of its iterations takes 
$\mathcal{O}(N |\branchtree|^2)$ steps. 
\autoref{tab_timeSeq} evaluates the practical time performance of our 
computational framework for the barycenter computation (which includes itself 
distance and geodesic computations).
In sequential mode, we observe that the running time is indeed a function 
of the number of ensemble members 
($N$) and the average size of 
the trees ($\branchtree$).
\julien{It is slightly slower for $\wassersteinTree$ than for 
$\wasserstein{2}$, but runtimes remain comparable overall.}
In parallel, speedups 
are the most important for the 
largest examples. However, the iterative nature of our barycenter optimization 
algorithm seems to limit parallel efficiency globally (the end of each 
iteration still constitutes a strong synchronization). For the smaller 
examples, the cost of the task runtime 
seems to become non-negligible in comparison to the actual computation,
resulting in moderate speedups. Still, our 
parallelization significantly reduces runtimes overall, with less than 3 minutes 
of computation on average and at most 15 minutes for the largest examples.




\cutout{It depends on the number of nodes, 
noted $|\branchtree|$ of the input BDTs, as well as their valence (which is 
typically
orders of magnitude
smaller than $|\branchtree|$ in practice and thus considered as 
a small constant).
Then, when $\epsilon_1 = 0$, 
the 
construction of the matrix $\treeMatrix$ requires $\mathcal{O}(|\branchtree|^2)$ 
steps. When $\epsilon_1$ increases, the local forest assignment problems 
grow in size (as discussed in \autoref{sec_algorithm}) and their execution time 
becomes no longer negligible. In particular, the resolution of a local 
assignment between two forests of size $n$ takes $\mathcal{O}(n^3)$ steps in 
the 
worst-case with Munkres algorithm \cite{Munkres1957} ($\mathcal{O}(n^2)$ in 
practice). When $\epsilon_1 = 
1$, we have $\wassersteinTree = \wasserstein{2}$ and $n = |\branchtree|$. Then, 
in the worst case, our algorithm takes $\mathcal{O}(|\branchtree|)^3$ steps. In 
practice, we use the efficient \emph{Auction} approximation \cite{Bertsekas81} 
to accelerate the resolution of these assignments, which results in an overall 
worst-time complexity of $\mathcal{O}(|\branchtree|)^2$.}


%
%
%
%

\subsection{Framework quality}
\label{sec_quality}

\begin{figure}
  \centering
  \includegraphics[width=.975\linewidth]{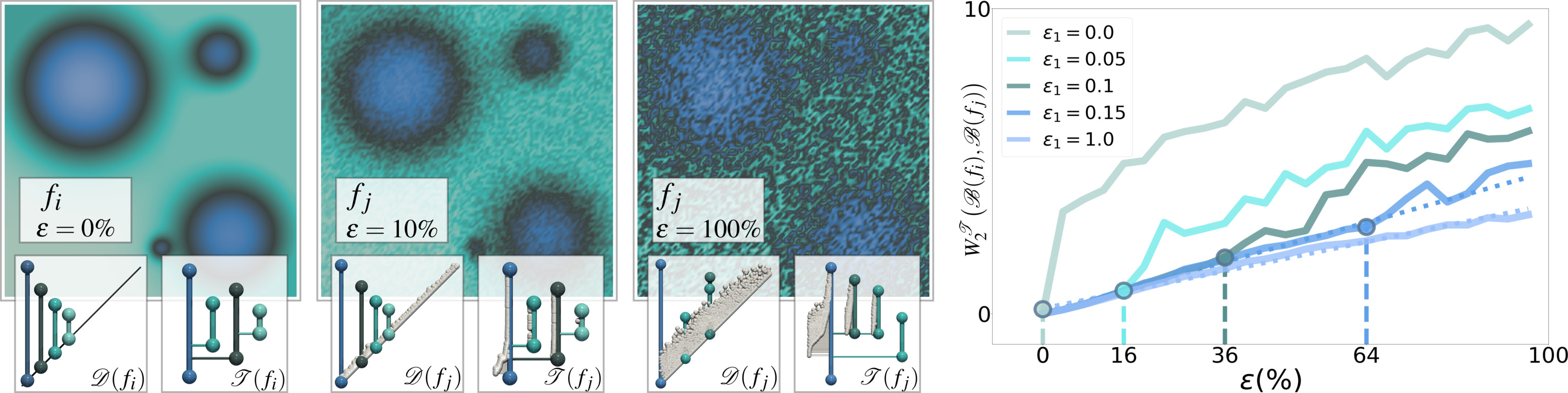}
 \mycaption{Empirical stability evaluation. Given an input scalar field $f_i$, a
noisy version $f_j$ is created by inserting a random noise of increasing 
amplitude $\epsilon$ (left). The evolution of 
$\wassersteinTree\big(\branchtree(f_i), \branchtree(f_j)\big)$ 
with $\epsilon$ (right), for varying values of 
$\epsilon_1$ 
(\autoref{sec_our_distance_definition}), indicates clear transition points 
(colored dots) before which 
$\wassersteinTree$
evolves nearly 
linearly. Before these transition points (i.e. before these noise levels, 
vertical lines), $\wassersteinTree$ is stable.}
\label{fig_practicalStability}
\end{figure}

$\wassersteinTree$ is 
indeed a distance metric (Appendix 2). 
It is more discriminative than 
$\wasserstein{2}$ (i.e. $\wassersteinTree \geq \wasserstein{2}$, 
\autoref{sec_metric}, \autoref{fig_mergeTreeVSpersistenceDiagram}). 
\autoref{fig_practicalStability} evaluates empirically its stability. For this, 
given a scalar field $f_i$, a noisy version $f_j$ is created such that $\|f_i 
-f_j\|_\infty \leq \epsilon$, for increasing values  of $\epsilon$. 
Then, we observe the evolution of 
$\wassersteinTree\big(\branchtree(f_i), \branchtree(f_j)\big)$, as a function of 
$\epsilon$ (\autoref{fig_practicalStability}, right), to estimate how 
$\wassersteinTree$ varies under input perturbations. 
For $\epsilon_1 = 1$, we 
have $\wassersteinTree = \wasserstein{2}$ (\autoref{sec_metric}) and the curve 
evolves nearly linearly ($\wasserstein{2}$ is stable \cite{Turner2014}). For 
other $\epsilon_1$ values, the curves indicate clear \emph{transition} points 
(colored dots) before which $\wassersteinTree$ evolves nearly linearly too. 
This indicates that for reasonable noise levels (smaller than the $\epsilon$ 
value of each transition point, vertical lines), 
$\wassersteinTree$ 
is also 
stable and that only mild increases of $\epsilon_1$ 
result in 
fast shifts of 
these transition points (to an accepted noise level of $64\%$
at
$\epsilon_1=0.15$). This illustrates 
overall 
that the stability of 
$\wassersteinTree$ can indeed be controlled with 
$\epsilon_1$ and 
that 
small values already 
lead to
stable results for reasonable noise levels.
\revision{A detailed empirical analysis of the other two parameters of our
approach ($\epsilon_2$, $\epsilon_3$, \autoref{sec_normalization}) is provided
in Appendix 10 (supplemental material).}

Next, we study the practical relevance of $\wassersteinTree$ by evaluating
our clustering performance.
For 
this, each ensemble of \autoref{tab_timeSeq} is associated with a ground truth 
classification 
(distinct phases of a time-varying phenomenon, distinct input 
parameters, etc), by following the companion specifications 
\cite{scivis2004, scivis2006, scivis2008, scivis2014, scivis2015, scivis2016, 
scivis2017, scivis2018}. Clustering performance is  evaluated with 
accepted scores, namely the normalized mutual information
and 
adjusted rand index
($NMI$, $ARI$). When using our barycenters (\autoref{sec:barycenters}), our 
clustering approach (\autoref{sec_clustering}) 
achieves a perfect classification for all ensembles ($NMI = ARI = 1$). 
These scores decrease to $NMI =0.78$ and $ARI=0.69$ on average when using, 
within $k$-means, a barycenter of persistence diagrams \cite{Turner2014} 
($\epsilon_1 = 1$),
and to $NMI = 
 0.73$ and $ARI=0.56$ when using the 1-center of 
Yan et al. \cite{YanWMGW20} 
(obtained 
with the authors' implementation \cite{1centerCode}, using leaf labels 
generated by our \julien{distance} computation, 
\autoref{sec_metric}). This simply confirms  experimentally that 
$1$-centers in general are not suited for clustering tasks.
A standard clustering approach (multi-dimensional scaling to $kD$ followed by 
$k$-means)
using the distance
$\editdistance$ \cite{SridharamurthyM20} achieves lower average scores than our 
approach, with $NMI =0.89$ and 
$ARI=0.85$ on average.
Overall,
this 
confirms that $\editdistance$ induces more discriminative 
classifiers than $\wasserstein{2}$, and that our metric $\wassersteinTree$ 
further improves that.

\begin{table}
    \caption{Running times (in seconds, 10 run average) of our 
approach for the barycenter 
computation, with respect to $\wasserstein{2}$ ($\epsilon_1 = 1$, 
    \autoref{sec_algorithm}, 
    sequential)
    and to our new metric 
$\wassersteinTree$ (sequential, then with 20 cores).}
    \centering
    \scalebox{0.6125}{
    \makebox[\linewidth]{%
    \begin{tabular}{|l||r|r||r
    ||r|r|r|}
        \hline
        \textbf{Dataset} & $N$ & 
        $|\branchtree|$
        & $\wasserstein{2}$ (1 c.)
& $\wassersteinTree$ (1 c.)
& $\wassersteinTree$ (20 c.)
& Speedup
\\
        \hline
        Asteroid Impact \cite{scivis2018} (3D)  & 7 & 1,295 
        & 514.71 &
450.91 &
93.11 & 4.84
\\ 
        Cloud processes  \cite{scivis2017} (2D) & 12 & 1,209 
        & 54.90 & 
124.99 &
35.14 &
3.55
\\ 
        Viscous fingering   \cite{scivis2016} (3D) & 15 & 118 
        &  5.68 & 
        5.12& 
3.89 & 
1.31
\\ 
        Dark matter
        \cite{scivis2015} (3D) & \revision{40} & 2,592
          & 3,172.37 &
3,083.24 
          & 471.45 
          & 6.53 
\\ 
        Volcanic eruptions  \cite{scivis2014} (2D) & 12 & 811 & 
         171.13 & 
140.02 &
         48.52 & 2.88
\\ 
        Ionization front
        \cite{scivis2008} (2D) & 16 & 135 
        & 10.40 &
12.10 &
        8.20 
        &  1.47
\\ 
        Ionization front
        \cite{scivis2008} (3D) & 16 & 763 
         &  682.76 &
1,277.72 &
         219.61  & 5.81
\\ 
        Earthquake  \cite{scivis2006}  (3D) & 12 & 1,203 
        & 191.54 & 
        509.59 &
        117.31 & 4.34
\\ 
        Isabel  \cite{scivis2004} (3D) & 12 & 1,338 
        & 330.88 & 
        284.19 &
        62.70 & 
        4.53
\\ 
        Starting Vortex  \cite{favelier2018} (2D) & 12 & 124 & 
         7.72
        & 
        5.58 &
        6.11 & 0.91
\\ 
        Sea Surface Height  \cite{favelier2018} (2D) & 48 & 1,787 & 
        4,509.78
& 10,557.07 &
881.49 & 11.97
\\ 
        Vortex Street  \cite{favelier2018} (2D) & 45 & 23 
        &  1.71
        & 1.90
        & 1.44
        &  1.31
\\ 
        \hline
    \end{tabular}
    }
    }
    \label{tab_timeSeq}
\end{table}

\begin{figure}
  \centering
  \includegraphics[width=\linewidth]{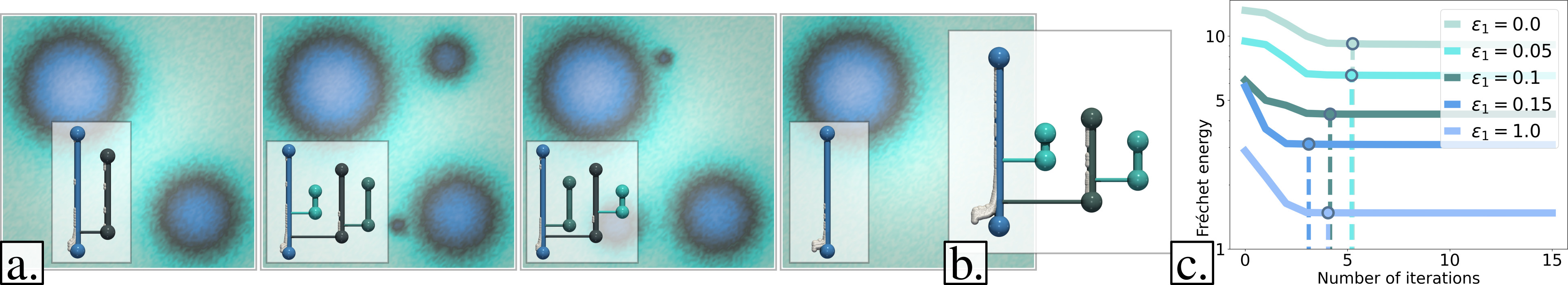}
  \vspace{-2ex} 
 \mycaption{
 Evolution of the Fr\'echet energy for estimating the barycenter (b)
 for an ensemble 
 of 100 noisy variants of four 
fields (a). 
The 
energy
(c) is shown 
for several  $\epsilon_1$ values   
(\autoref{sec_our_distance_definition}). 
In practice, we stop the algorithm 
when 
the energy decreases by less than $1\%$ (vertical lines).}
\label{fig_convergence}
\end{figure}

\julien{\autoref{fig_convergence} shows 
the evolution of the Fr\'echet energy for
our 
barycenter algorithm (\autoref{sec:barycenters}) for various $\epsilon_1$ 
values. In practice, the algorithm stops when the Fr\'echet energy decreases by 
less than $1\%$ between consecutive iterations, which occurs early in the 
process.} 

\autoref{fig_barycenter} provides a visual comparison between our 
barycenter and the $1$-center 
of
Yan et al. \cite{YanWMGW20} 
(obtained 
with the authors' implementation \cite{1centerCode}, using leaf labels 
generated by our \julien{distance} computation, 
\autoref{sec_metric}).
This figure 
confirms the general sensitivity in practice  of $1$-centers to outliers, and 
the 
ability of barycenters to better capture the main trends in the ensemble.
From a qualitative perspective, 
our framework enables the 
computation of 
faithful interpolations of merge trees: the reconstructed trees, blue 
background (\autoref{fig_temporalReduction}), are visually very similar to the 
input trees. Moreover, 
our framework produces barycenters 
(Figs. \ref{fig:teaser}, \ref{fig_clusteringResult1}, and 
\ref{fig_clusteringResult2})
which 
capture well
the main features of the input ensemble:
for each cluster, the 
resulting centroid is visually similar to the input trees of the cluster (same 
number and persistence of large branches). 
Then, our 
clustering framework, 
coupled with our centroids,
provides a faithful visual summary of the features of interest, for 
each of the main trends (i.e. for each cluster) found in the 
ensemble.

\subsection{Limitations}
The search space associated with our metric $\wassersteinTree$ is constrained 
to rooted partial isomorphisms. Then, if a matching exists between two 
BDTs (i.e. if they are not 
both 
destructed when optimizing $\wassersteinTree$), 
it has to match their roots together. In other words, $\wassersteinTree$ nearly 
always 
matches the most persistent branch of the two trees together, which might be 
too restrictive (in particular for feature tracking applications). Note 
however, that $\wasserstein{2}$ 
behaves equivalently: the most 
persistent branch of $\branchtree(f_i)$ corresponds to the component of 
$\sublevelset{{f_i}}(\isovalue)$ created in the global minimum of $f_i$, which 
in principle has infinite persistence 
and which is typically treated separately when evaluating $\wasserstein{2}$. 
Similarly to Sridharamurthy et al. \cite{SridharamurthyM20}, saddle swap 
instabilities are handled in our approach by a pre-processing step
which
merges adjacent saddles (controlled by $\epsilon_1$). An alternative would 
consist in exploring the space of all possible branch decompositions (not 
necessarily persistence-driven), as studied by Beketayev et al.  
\cite{BeketayevYMWH14}. However, the search space would then become 
significantly larger. Moreover, the nesting of birth/death values within 
the BDTs would no longer be guaranteed, which is however a key property which 
we exploit in our framework (\autoref{sec:geodesics}).
When computing barycenters of persistence diagrams,
Vidal et al. \cite{vidal_vis19} showed that 
the optimization
could be drastically 
accelerated by introducing persistence pairs progressively along the 
iterations, while implicitly maintaining previous assignments 
at each 
initialization. We leave the study of such a progressive strategy to future 
work, although the fact that $\wassersteinTree$ handles many small assignment 
problems 
(unlike $\wasserstein{2}$)
indicates that such a strategy may result in only modest gains for merge 
trees. \autoref{fig_practicalStability} provides an empirical evaluation of the 
stability of $\wassersteinTree$. Similarly to  Sridharamurthy et al. 
\cite{SridharamurthyM20}, we believe that the theoretical investigation of the 
stability of $\wassersteinTree$ goes beyond the scope of this paper and we leave 
it for future work.

\section{Conclusion}

In this paper, we presented a computational framework for the estimation of 
distances, geodesics and barycenters of merge trees, with applications to 
feature tracking, temporal reduction and ensemble clustering and summarization. 
Our approach filled the gap between the edit distance \cite{SridharamurthyM20} 
and existing optimization frameworks for persistence diagrams \cite{Turner2014}. 
Our work enables faithful interpolations of merge trees 
(\autoref{fig_temporalReduction}) 
and
the generation of merge trees 
representative of a set (Figs. \ref{fig:teaser}, 
\ref{fig_clusteringResult1}, and 
\ref{fig_clusteringResult2}). Moreover, our task-based algorithm enables 
automatic barycenter computations 
within minutes for real-life ensembles.

A natural direction for future work is the extension of our framework to other 
topological data representations, such as Reeb graphs or Morse-Smale complexes.
However, the question of defining relevant and computable metrics for 
these objects is still an active research debate. Moreover, as illustrated 
by this paper, extending existing metrics to make them conducive to 
efficient geodesic computation further requires additional efforts.
We believe our work is an 
important practical step towards the definition of a larger statistical 
framework on the space of 
merge trees.
%
%
In the future, based on our framework, we 
will 
study
the definition of more sophisticated statistical indexes (for 
instance by investigating a
notion of covariance matrix 
%
for merge trees), to 
support even more advanced analyses of large-scale ensemble data.







\acknowledgments{
{\small This work is partially supported by the
European Commission grant
ERC-2019-COG 
\emph{``TORI''} (ref. 863464, 
\url{https://erc-tori.github.io/}).}}

\clearpage

\bibliographystyle{abbrv-doi}
 
 
\bibliography{template}

\begin{thebibliography}{100}

\bibitem{uncertainty1}
\uppercase{ISO/IEC} \uppercase{G}uide 98-3:2008 uncertainty of measurement -
  part 3: Guide to the expression of uncertainty in measurement
  (\uppercase{GUM}).
\newblock 2008.

\bibitem{AcharyaN15}
A.~Acharya and V.~Natarajan.
\newblock A parallel and memory efficient algorithm for constructing the
  contour tree.
\newblock In {\em IEEE PV}, 2015.

\bibitem{Adams2015}
H.~{Adams}, S.~{Chepushtanova}, T.~{Emerson}, E.~{Hanson}, M.~{Kirby},
  F.~{Motta}, R.~{Neville}, C.~{Peterson}, P.~{Shipman}, and L.~{Ziegelmeier}.
\newblock {Persistence Images: A Stable Vector Representation of Persistent
  Homology}.
\newblock {\em Journal of Machine Learning Research}, 2017.

\bibitem{beiBrain18}
K.~Anderson, J.~Anderson, S.~Palande, and B.~Wang.
\newblock Topological data analysis of functional {MRI} connectivity in time
  and space domains.
\newblock In {\em MICCAI Workshop on Connectomics in NeuroImaging}, 2018.

\bibitem{uncertainty_isosurface1}
T.~Athawale and A.~Entezari.
\newblock Uncertainty quantification in linear interpolation for isosurface
  extraction.
\newblock {\em IEEE TVCG}, 2013.

\bibitem{athwale19}
T.~{Athawale} and C.~R. {Johnson}.
\newblock Probabilistic asymptotic decider for topological ambiguity resolution
  in level-set extraction for uncertain 2d data.
\newblock {\em IEEE TVCG}, 2019.

\bibitem{uncertainty_isosurface2}
T.~Athawale, E.~Sakhaee, and A.~Entezari.
\newblock Isosurface visualization of data with nonparametric models for
  uncertainty.
\newblock {\em IEEE TVCG}, 2016.

\bibitem{athawale_tvcg19}
T.~M. Athawale, D.~Maljovec, C.~R. Johnson, V.~Pascucci, and B.~Wang.
\newblock Uncertainty visualization of 2d morse complex ensembles using
  statistical summary maps.
\newblock {\em CoRR}, abs/1912.06341, 2019.

\bibitem{AyachitBGOMFM15}
U.~Ayachit, A.~C. Bauer, B.~Geveci, P.~O'Leary, K.~Moreland, N.~Fabian, and
  J.~Mauldin.
\newblock Paraview catalyst: Enabling in situ data analysis and visualization.
\newblock In {\em ISAV}, 2015.

\bibitem{banchoff70}
T.~F. Banchoff.
\newblock Critical points and curvature for embedded polyhedral surfaces.
\newblock {\em The American Mathematical Monthly}, 1970.

\bibitem{insitu}
A.~C. Bauer, H.~Abbasi, J.~Ahrens, H.~Childs, B.~Geveci, S.~Klasky,
  K.~Moreland, P.~O'Leary, V.~Vishwanath, B.~Whitlock, and E.~W. Bethel.
\newblock In-situ methods, infrastructures, and applications on high
  performance computing platforms.
\newblock {\em CGF}, 2016.

\bibitem{bauer14}
U.~Bauer, X.~Ge, and Y.~Wang.
\newblock Measuring distance between {R}eeb graphs.
\newblock In {\em SoCG}, 2014.

\bibitem{dipha}
U.~Bauer, M.~Kerber, and J.~Reininghaus.
\newblock Distributed computation of persistent homology.
\newblock In {\em Algorithm Engineering and Experiments}, 2014.

\bibitem{BeketayevYMWH14}
K.~Beketayev, D.~Yeliussizov, D.~Morozov, G.~H. Weber, and B.~Hamann.
\newblock Measuring the distance between merge trees.
\newblock In {\em TopoInVis}. 2014.

\bibitem{Bertsekas81}
D.~P. Bertsekas.
\newblock A new algorithm for the assignment problem.
\newblock {\em Mathematical Programming}, 1981.

\bibitem{harshChemistry}
H.~Bhatia, A.~G. Gyulassy, V.~Lordi, J.~E. Pask, V.~Pascucci, and P.-T. Bremer.
\newblock Topoms: Comprehensive topological exploration for molecular and
  condensed-matter systems.
\newblock {\em J. of Computational Chemistry}, 2018.

\bibitem{bhatia}
H.~Bhatia, S.~Jadhav, P.~Bremer, G.~Chen, J.~A. Levine, L.~G. Nonato, and
  V.~Pascucci.
\newblock Flow visualization with quantified spatial and temporal errors using
  edge maps.
\newblock {\em IEEE TVCG}, 2012.

\bibitem{biasotti08}
S.~Biasotti, D.~Giorgio, M.~Spagnuolo, and B.~Falcidieno.
\newblock Reeb graphs for shape analysis and applications.
\newblock {\em TCS}, 2008.

\bibitem{ttk19}
T.~Bin~Masood, J.~Budin, M.~Falk, G.~Favelier, C.~Garth, C.~Gueunet,
  P.~Guillou, L.~Hofmann, P.~Hristov, A.~Kamakshidasan, C.~Kappe, P.~Klacansky,
  P.~Laurin, J.~Levine, J.~Lukasczyk, D.~Sakurai, M.~Soler, P.~Steneteg,
  J.~Tierny, W.~Usher, J.~Vidal, and M.~Wozniak.
\newblock {An Overview of the Topology ToolKit}.
\newblock In {\em TopoInVis}, 2019.

\bibitem{topoAngler}
A.~Bock, H.~Doraiswamy, A.~Summers, and C.~T. Silva.
\newblock {TopoAngler: Interactive Topology-Based Extraction of Fishes}.
\newblock {\em IEEE TVCG}, 2018.

\bibitem{unertainty_survey2}
G.~Bonneau, H.~Hege, C.~Johnson, M.~Oliveira, K.~Potter, P.~Rheingans, and
  T.~Schultz".
\newblock Overview and state-of-the-art of uncertainty visualization.
\newblock {\em Mathematics and Visualization}, 37:3--27, 2014.

\bibitem{BremerEHP03}
P.~Bremer, H.~Edelsbrunner, B.~Hamann, and V.~Pascucci.
\newblock {A Multi-Resolution Data Structure for 2-Dimensional Morse
  Functions}.
\newblock In {\em Proc. of IEEE VIS}, 2003.

\bibitem{bremer_tvcg11}
P.~Bremer, G.~Weber, J.~Tierny, V.~Pascucci, M.~Day, and J.~Bell.
\newblock Interactive exploration and analysis of large scale simulations using
  topology-based data segmentation.
\newblock {\em IEEE TVCG}, 2011.

\bibitem{Bubenik15}
P.~Bubenik.
\newblock Statistical topological data analysis using persistence landscapes.
\newblock {\em J. Mach. Learn. Res.}, 2015.

\bibitem{carr00}
H.~Carr, J.~Snoeyink, and U.~Axen.
\newblock Computing contour trees in all dimensions.
\newblock In {\em Symp. on Dis. Alg.}, 2000.

\bibitem{carr04}
H.~A. Carr, J.~Snoeyink, and M.~van~de Panne.
\newblock {Simplifying Flexible Isosurfaces Using Local Geometric Measures}.
\newblock In {\em IEEE VIS}, 2004.

\bibitem{CarrWSA16}
H.~A. Carr, G.~H. Weber, C.~M. Sewell, and J.~P. Ahrens.
\newblock {Parallel peak pruning for scalable {SMP} contour tree computation}.
\newblock In {\em LDAV}, 2016.

\bibitem{celebi13}
M.~E. Celebi, H.~A. Kingravi, and P.~A. Vela.
\newblock A comparative study of efficient initialization methods for the
  k-means clustering algorithm.
\newblock {\em Expert Syst. Appl.}, 2013.

\bibitem{Defl15}
L.~De~Floriani, U.~Fugacci, F.~Iuricich, and P.~Magillo.
\newblock Morse complexes for shape segmentation and homological analysis:
  discrete models and algorithms.
\newblock {\em CGF}, 2015.

\bibitem{spaghettiPlot}
P.~Diggle, P.~Heagerty, K.-Y. Liang, and S.~Zeger.
\newblock {\em The Analysis of Longitudinal Data}.
\newblock Oxford University Press, 2002.

\bibitem{DoraiswamyN13}
H.~Doraiswamy and V.~Natarajan.
\newblock Computing reeb graphs as a union of contour trees.
\newblock {\em IEEE TVCG}, 2013.

\bibitem{edelsbrunner09}
H.~Edelsbrunner and J.~Harer.
\newblock {\em Computational Topology: An Introduction}.
\newblock American Mathematical Society, 2009.

\bibitem{EdelsbrunnerHNP03}
H.~Edelsbrunner, J.~Harer, V.~Natarajan, and V.~Pascucci.
\newblock Morse-smale complexes for piecewise linear 3-manifolds.
\newblock In {\em SoCG}, 2003.

\bibitem{EdelsbrunnerHZ01}
H.~Edelsbrunner, J.~Harer, and A.~Zomorodian.
\newblock Hierarchical morse complexes for piecewise linear 2-manifolds.
\newblock In {\em SoCG}, 2001.

\bibitem{edelsbrunner03}
H.~Edelsbrunner, J.~Harer, and A.~Zomorodian.
\newblock Hierarchical {Morse-Smale} complexes for piecewise linear
  2-manfiolds.
\newblock {\em DCG}, 2003.

\bibitem{edelsbrunner02}
H.~Edelsbrunner, D.~Letscher, and A.~Zomorodian.
\newblock Topological persistence and simplification.
\newblock {\em DCG}, 2002.

\bibitem{edelsbrunner90}
H.~Edelsbrunner and E.~P. Mucke.
\newblock Simulation of simplicity: a technique to cope with degenerate cases
  in geometric algorithms.
\newblock {\em ACM ToG}, 1990.

\bibitem{elkan03}
C.~Elkan.
\newblock Using the triangle inequality to accelerate k-means.
\newblock In {\em ICML}, 2003.

\bibitem{favelier2018}
G.~Favelier, N.~Faraj, B.~Summa, and J.~Tierny.
\newblock {Persistence Atlas for Critical Point Variability in Ensembles}.
\newblock {\em IEEE TVCG (IEEE VIS)}, 2018.

\bibitem{favelier16}
G.~Favelier, C.~Gueunet, and J.~Tierny.
\newblock Visualizing ensembles of viscous fingers.
\newblock In {\em IEEE SciVis Contest}, 2016.

\bibitem{scivis2014}
D.~Feng, B.~Hentschel, S.~Griessbach, L.~Hoffmann, and M.~von Hobe.
\newblock {The IEEE SciVis Contest}.
\newblock \url{http://sciviscontest.ieeevis.org/2014/}, 2014.

\bibitem{Ferstl2016}
F.~Ferstl, K.~Bürger, and R.~Westermann.
\newblock Streamline variability plots for characterizing the uncertainty in
  vector field ensembles.
\newblock {\em IEEE TVCG}, 2016.

\bibitem{Ferstl2016b}
F.~Ferstl, M.~Kanzler, M.~Rautenhaus, and R.~Westermann.
\newblock Visual analysis of spatial variability and global correlations in
  ensembles of iso-contours.
\newblock {\em CGF}, 2016.

\bibitem{forman98}
R.~Forman.
\newblock {A User's Guide to Discrete Morse Theory}.
\newblock {\em AM}, 1998.

\bibitem{scivis2016}
C.~Garth, B.~Geveci, B.~Hentschel, J.~Kuhnert, I.~Michel, T.-M. Rhyne, and
  S.~Schröder.
\newblock {The IEEE SciVis Contest}.
\newblock \url{http://sciviscontest.ieeevis.org/2016/}, 2016.

\bibitem{intrinsicMTdistance}
E.~Gasparovic, E.~Munch, S.~Oudot, K.~Turner, B.~Wang, and Y.~Wang.
\newblock Intrinsic interleaving distance for merge trees.
\newblock {\em CoRR}, 1908.00063, 2019.

\bibitem{chemistry_vis14}
D.~Guenther, R.~Alvarez-Boto, J.~Contreras-Garcia, J.-P. Piquemal, and
  J.~Tierny.
\newblock Characterizing molecular interactions in chemical systems.
\newblock {\em IEEE TVCG (IEEE VIS)}, 2014.

\bibitem{gueunet_tpds19}
C.~Gueunet, P.~Fortin, J.~Jomier, and J.~Tierny.
\newblock {Task-Based Augmented Contour Trees with Fibonacci Heaps}.
\newblock {\em {IEEE} TPDS}, 2019.

\bibitem{gueunet_egpgv19}
C.~Gueunet, P.~Fortin, J.~Jomier, and J.~Tierny.
\newblock {Task-based Augmented Reeb Graphs with Dynamic ST-Trees}.
\newblock In {\em EGPGV}, 2019.

\bibitem{gunther}
D.~G{\"u}nther, J.~Salmon, and J.~Tierny.
\newblock Mandatory critical points of 2{D} uncertain scalar fields.
\newblock {\em CGF}, 2014.

\bibitem{gyulassy_ev14}
A.~Gyulassy, P.~Bremer, R.~Grout, H.~Kolla, J.~Chen, and V.~Pascucci.
\newblock Stability of dissipation elements: A case study in combustion.
\newblock {\em CGF}, 2014.

\bibitem{gyulassy_vis18}
A.~Gyulassy, P.~Bremer, and V.~Pascucci.
\newblock {Shared-Memory Parallel Computation of Morse-Smale Complexes with
  Improved Accuracy}.
\newblock {\em IEEE TVCG (IEEE VIS)}, 2018.

\bibitem{gyulassy_vis07}
A.~Gyulassy, M.~A. Duchaineau, V.~Natarajan, V.~Pascucci, E.~Bringa,
  A.~Higginbotham, and B.~Hamann.
\newblock Topologically clean distance fields.
\newblock {\em IEEE TVCG (IEEE VIS)}, 2007.

\bibitem{gyulassy_vis15}
A.~Gyulassy, A.~Knoll, K.~Lau, B.~Wang, P.~Bremer, M.~Papka, L.~A. Curtiss, and
  V.~Pascucci.
\newblock Interstitial and interlayer ion diffusion geometry extraction in
  graphitic nanosphere battery materials.
\newblock {\em IEEE TVCG}, 2015.

\bibitem{heine16}
C.~Heine, H.~Leitte, M.~Hlawitschka, F.~Iuricich, L.~De~Floriani,
  G.~Scheuermann, H.~Hagen, and C.~Garth.
\newblock A survey of topology-based methods in visualization.
\newblock {\em CGF}, 2016.

\bibitem{scivis2015}
B.~Hentschel, B.~Geveci, M.~Turk, and S.~Skillman.
\newblock {The IEEE SciVis Contest}.
\newblock \url{http://sciviscontest.ieeevis.org/2015/}, 2015.

\bibitem{HilagaSKK01}
M.~Hilaga, Y.~Shinagawa, T.~Komura, and T.~L. Kunii.
\newblock Topology matching for fully automatic similarity estimation of 3d
  shapes.
\newblock In {\em ACM SIGGRAPH}, 2001.

\bibitem{Hummel2013}
M.~Hummel, H.~Obermaier, C.~Garth, and K.~I. Joy.
\newblock Comparative visual analysis of lagrangian transport in {CFD}
  ensembles.
\newblock {\em IEEE TVCG}, 2013.

\bibitem{uncertainty2}
C.~R. Johnson and A.~R. Sanderson.
\newblock A next step: Visualizing errors and uncertainty.
\newblock {\em IEEE Computer Graphics and Applications}, 2003.

\bibitem{Kantorovich}
L.~Kantorovich.
\newblock On the translocation of masses.
\newblock {\em AS URSS}, 1942.

\bibitem{kasten_tvcg11}
J.~Kasten, J.~Reininghaus, I.~Hotz, and H.~Hege.
\newblock Two-dimensional time-dependent vortex regions based on the
  acceleration magnitude.
\newblock {\em IEEE TVCG}, 2011.

\bibitem{Kerber2016}
M.~Kerber, D.~Morozov, and A.~Nigmetov.
\newblock Geometry helps to compare persistence diagrams.
\newblock {\em ACM Journal of Experimental Algorithmics}, 2016.

\bibitem{Kraus2010VisualizationOU}
M.~Kraus.
\newblock Visualization of uncertain contour trees.
\newblock In {\em IVTA}, 2010.

\bibitem{lacombe2018}
T.~{Lacombe}, M.~{Cuturi}, and S.~{Oudot}.
\newblock {Large Scale computation of Means and Clusters for Persistence
  Diagrams using Optimal Transport}.
\newblock In {\em NIPS}, 2018.

\bibitem{laney_vis06}
D.~E. Laney, P.~Bremer, A.~Mascarenhas, P.~Miller, and V.~Pascucci.
\newblock Understanding the structure of the turbulent mixing layer in
  hydrodynamic instabilities.
\newblock {\em IEEE TVCG (IEEE VIS)}, 2006.

\bibitem{liebmann1}
T.~Liebmann and G.~Scheuermann.
\newblock Critical points of gaussian-distributed scalar fields on simplicial
  grids.
\newblock {\em CGF}, 2016.

\bibitem{lloyd82}
S.~Lloyd.
\newblock Least squares quantization in {PCM}.
\newblock {\em IEEE Transactions on Information Theory}, 1982.

\bibitem{LohfinkWLWG20}
A.~P. Lohfink, F.~Wetzels, J.~Lukasczyk, G.~H. Weber, and C.~Garth.
\newblock Fuzzy contour trees: Alignment and joint layout of multiple contour
  trees.
\newblock {\em CGF}, 2020.

\bibitem{MaadasamyDN12}
S.~Maadasamy, H.~Doraiswamy, and V.~Natarajan.
\newblock {A hybrid parallel algorithm for computing and tracking level set
  topology}.
\newblock In {\em Proc. of HiPC}, 2012.

\bibitem{uncertainty3}
A.~Maceachren, A.Robinson, S.~Hopper, S.~Gardner, R.~Murray, M.~Gahegan, and
  E.~Hetzler.
\newblock Visualizing geospatial information uncertainty: What we know and what
  we need to know.
\newblock {\em CGIS}, 2005.

\bibitem{beiNuclear16}
D.~Maljovec, B.~Wang, P.~Rosen, A.~Alfonsi, G.~Pastore, C.~Rabiti, and
  V.~Pascucci.
\newblock Topology-inspired partition-based sensitivity analysis and
  visualization of nuclear simulations.
\newblock In {\em IEEE PV}, 2016.

\bibitem{Mirzargar2014}
M.~Mirzargar, R.~Whitaker, and R.~Kirby.
\newblock Curve boxplot: Generalization of boxplot for ensembles of curves.
\newblock {\em IEEE TVCG}, 2014.

\bibitem{monge81}
G.~Monge.
\newblock M\'emoire sur la th\'eorie des d\'eblais et des remblais.
\newblock {\em Acad\'emie Royale des Sciences de Paris}, 1781.

\bibitem{morozov14}
D.~Morozov, K.~Beketayev, and G.~H. Weber.
\newblock Interleaving distance between merge trees.
\newblock In {\em TopoInVis}. 2014.

\bibitem{Munkres1957}
J.~Munkres.
\newblock Algorithms for the assignment and transportation problems.
\newblock {\em Journal of the Society for Industrial and Applied Mathematics},
  1957.

\bibitem{Malgorzata19}
M.~Olejniczak, A.~S.~P. Gomes, and J.~Tierny.
\newblock {A Topological Data Analysis Perspective on Non-Covalent Interactions
  in Relativistic Calculations}.
\newblock {\em International Journal of Quantum Chemistry}, 2019.

\bibitem{scivis2006}
K.~Olsen, S.~Day, B.~Minster, R.~Moore, Y.~Cui, A.~Chourasia, M.~Thiebaux,
  H.~Francoeur, P.~Maechling, S.~Cutchin, and K.~Nunes.
\newblock {The IEEE SciVis Contest}.
\newblock \url{http://sciviscontest.ieeevis.org/2006/}, 2006.

\bibitem{otto1}
M.~Otto, T.~Germer, H.-C. Hege, and H.~Theisel.
\newblock Uncertain 2{D} vector field topology.
\newblock 2010.

\bibitem{otto2}
M.~Otto, T.~Germer, and H.~Theisel.
\newblock Uncertain topology of 3{D} vector fields.
\newblock {\em IEEE PV}, 2011.

\bibitem{uncertainty4}
A.~T. Pang, C.~M. Wittenbrink, and S.~K. Lodha.
\newblock Approaches to uncertainty visualization.
\newblock {\em The Visual Computer}, 1997.

\bibitem{parsa12}
S.~Parsa.
\newblock {A deterministic \emph{o(m log m)} time algorithm for the reeb
  graph}.
\newblock In {\em SoCG}, 2012.

\bibitem{pascucci_mr04}
V.~Pascucci, K.~Cole-McLaughlin, and G.~Scorzelli.
\newblock {Multi-resolution computation and presentation of contour trees}.
\newblock In {\em IASTED}, 2004.

\bibitem{pascucci07}
V.~Pascucci, G.~Scorzelli, P.~T. Bremer, and A.~Mascarenhas.
\newblock Robust on-line computation of {R}eeb graphs: simplicity and speed.
\newblock {\em ACM ToG}, 2007.

\bibitem{scivis2018}
J.~Patchett and G.~R. Gisler.
\newblock {The IEEE SciVis Contest}.
\newblock \url{http://sciviscontest.ieeevis.org/2018/}, 2018.

\bibitem{petz}
C.~Petz, K.~P{\"o}thkow, and H.-C. Hege.
\newblock Probabilistic local features in uncertain vector fields with spatial
  correlation.
\newblock {\em CGF}, 2012.

\bibitem{uncertainty_gradient}
T.~Pfaffelmoser, M.~Mihai, and R.~Westermann.
\newblock Visualizing the variability of gradients in uncertain 2{D} scalar
  fields.
\newblock {\em IEEE TVCG}, 2013.

\bibitem{uncertainty_isosurface3}
T.~Pfaffelmoser, M.~Reitinger, and R.~Westermann.
\newblock Visualizing the positional and geometrical variability of isosurfaces
  in uncertain scalar fields.
\newblock {\em CGF}, 2011.

\bibitem{uncertainty_correlation}
T.~Pfaffelmoser and R.~Westermann.
\newblock Visualization of global correlation structures in uncertain 2{D}
  scalar fields.
\newblock {\em CGF}, 2012.

\bibitem{gerris}
S.~Popinet.
\newblock {Gerris Flow Solver}.
\newblock \url{http://gfs.sourceforge.net/wiki/index.php/Main_Page}, 2006.

\bibitem{uncertainty_isocontour1}
K.~P{\"o}thkow and H.-C. Hege.
\newblock Positional uncertainty of isocontours: Condition analysis and
  probabilistic measures.
\newblock {\em IEEE TVCG}, 2011.

\bibitem{uncertainty_nonparam}
K.~P{\"o}thkow and H.-C. Hege.
\newblock Nonparametric models for uncertainty visualization.
\newblock {\em CGF}, 2013.

\bibitem{uncertainty_isocontour2}
K.~P{\"o}thkow, C.~Petz, and H.-C. Hege.
\newblock Approximate level-crossing probabilities for interactive
  visualization of uncertain isocontours.
\newblock {\em Int. J. Uncert. Quantif.}, 2013.

\bibitem{uncertainty_isosurface4}
K.~P\"{o}thkow, B.~Weber, and H.-C. Hege.
\newblock Probabilistic marching cubes.
\newblock In {\em CGF}, 2011.

\bibitem{uncertainty_entropy}
K.~Potter, S.~Gerber, and E.~W. Anderson.
\newblock Visualization of uncertainty without a mean.
\newblock {\em IEEE Computer Graphics and Applications}, 2013.

\bibitem{Potter2009}
K.~Potter, A.~Wilson, P.~Bremer, D.~Williams, C.~Doutriaux, V.~Pascucci, and
  C.~R. Johnson.
\newblock Ensemble-vis: A framework for the statistical visualization of
  ensemble data.
\newblock In {\em 2009 IEEE ICDM}, 2009.

\bibitem{uncertainty_survey1}
J.~C. Potter~K, Rosen~P.
\newblock From quantification to visualization: A taxonomy of uncertainty
  visualization approaches.
\newblock {\em IFIP AICT}, 2012.

\bibitem{robins_pami11}
V.~Robins, P.~J. Wood, and A.~P. Sheppard.
\newblock {Theory and Algorithms for Constructing Discrete Morse Complexes from
  Grayscale Digital Images}.
\newblock {\em {IEEE} Trans. Pattern Anal. Mach. Intell.}, 2011.

\bibitem{SaikiaSW14_branch_decomposition_comparison}
H.~Saikia, H.~Seidel, and T.~Weinkauf.
\newblock Extended branch decomposition graphs: Structural comparison of scalar
  data.
\newblock {\em CGF}, 2014.

\bibitem{Sanyal2010}
J.~Sanyal, S.~Zhang, J.~Dyer, A.~Mercer, P.~Amburn, and R.~Moorhead.
\newblock Noodles: A tool for visualization of numerical weather model ensemble
  uncertainty.
\newblock {\em IEEE TVCG}, 2010.

\bibitem{uncertainty_interp}
S.~Schlegel, N.~Korn, and G.~Scheuermann.
\newblock On the interpolation of data with normally distributed uncertainty
  for visualization.
\newblock {\em IEEE TVCG (IEEE VIS)}, 2012.

\bibitem{gaussianResampling}
D.~Shepard.
\newblock {A two-dimensioanl interpolation function for irregularly-spaced
  data}.
\newblock In {\em {ACM National Conference}}, 1968.

\bibitem{ShivashankarN12}
N.~Shivashankar and V.~Natarajan.
\newblock {Parallel Computation of 3D Morse-Smale Complexes}.
\newblock {\em CGF}, 2012.

\bibitem{shivashankar2016felix}
N.~Shivashankar, P.~Pranav, V.~Natarajan, R.~van~de Weygaert, E.~P. Bos, and
  S.~Rieder.
\newblock Felix: A topology based framework for visual exploration of cosmic
  filaments.
\newblock {\em IEEE TVCG}, 2016.

\bibitem{soler_ldav18}
M.~Soler, M.~Plainchault, B.~Conche, and J.~Tierny.
\newblock Lifted {W}asserstein matcher for fast and robust topology tracking.
\newblock In {\em IEEE LDAV}, 2018.

\bibitem{sousbie11}
T.~Sousbie.
\newblock The persistent cosmic web and its filamentary structure: Theory and
  implementations.
\newblock {\em Royal Astronomical Society}, 2011.

\bibitem{SridharamurthyM20}
R.~Sridharamurthy, T.~B. Masood, A.~Kamakshidasan, and V.~Natarajan.
\newblock Edit distance between merge trees.
\newblock {\em IEEE TVCG}, 2020.

\bibitem{szymczak}
A.~Szymczak.
\newblock Hierarchy of stable {M}orse decompositions.
\newblock {\em IEEE TVCG}, 2013.

\bibitem{tarasov98}
S.~Tarasov and M.~Vyali.
\newblock Construction of contour trees in 3d in o(n log n) steps.
\newblock In {\em SoCG}, 1998.

\bibitem{scivis2008}
R.~Taylor, A.~Chourasia, D.~Whalen, and M.~L. Norman.
\newblock {The IEEE SciVis Contest}.
\newblock \url{http://sciviscontest.ieeevis.org/2008/}, 2008.

\bibitem{ttk17}
J.~Tierny, G.~Favelier, J.~A. Levine, C.~Gueunet, and M.~Michaux.
\newblock The {T}opology {T}ool{K}it.
\newblock {\em IEEE TVCG (IEEE VIS)}, 2017.
\newblock \url{https://topology-tool-kit.github.io/}.

\bibitem{tierny_vis09}
J.~Tierny, A.~Gyulassy, E.~Simon, and V.~Pascucci.
\newblock Loop surgery for volumetric meshes: Reeb graphs reduced to contour
  trees.
\newblock {\em IEEE TVCG (IEEE VIS)}, 2009.

\bibitem{Turner2014}
K.~{Turner}, Y.~{Mileyko}, S.~{Mukherjee}, and J.~{Harer}.
\newblock {Fr\'echet Means for Distributions of Persistence Diagrams}.
\newblock {\em DCG}, 2014.

\bibitem{vidal_vis19}
J.~Vidal, J.~Budin, and J.~Tierny.
\newblock {Progressive Wasserstein Barycenters of Persistence Diagrams}.
\newblock {\em IEEE TVCG (IEEE VIS)}, 2019.

\bibitem{scivis2004}
W.~Wang, C.~Bruyere, B.~Kuo, and T.~Scheitlin.
\newblock {The IEEE SciVis Contest}.
\newblock \url{http://sciviscontest.ieeevis.org/2004/}, 2004.

\bibitem{whitaker2013}
R.~T. Whitaker, M.~Mirzargar, and R.~M. Kirby.
\newblock Contour boxplots: A method for characterizing uncertainty in feature
  sets from simulation ensembles.
\newblock {\em IEEE TVCG}, 2013.

\bibitem{scivis2017}
T.~Wischgoll, A.~Chourasia, K.~Gorges, M.~Bruck, and N.~Rober.
\newblock {The IEEE SciVis Contest}.
\newblock \url{http://sciviscontest.ieeevis.org/2017/}, 2017.

\bibitem{Wu2013ACT}
K.~Wu and S.~Zhang.
\newblock A contour tree based visualization for exploring data with
  uncertainty.
\newblock {\em IJUQ}, 2013.

\bibitem{surveyComparison2021}
L.~Yan, T.~B. Masood, R.~Sridharamurthy, F.~Rasheed, V.~Natarajan, I.~Hotz, and
  B.~Wang.
\newblock Scalar field comparison with topological descriptors: Properties and
  applications for scientific visualization.
\newblock {\em CGF}, 2021.

\bibitem{1centerCode}
L.~Yan, Y.~Wang, E.~Munch, E.~Gasparovic, and B.~Wang.
\newblock {Source Code for a Structural Average of Labeled Merge Trees for
  Uncertainty Visualization}.
\newblock \url{https://github.com/tdavislab/amt}, 2019.

\bibitem{YanWMGW20}
L.~Yan, Y.~Wang, E.~Munch, E.~Gasparovic, and B.~Wang.
\newblock A structural average of labeled merge trees for uncertainty
  visualization.
\newblock {\em IEEE TVCG (IEEE VIS)}, 2019.

\bibitem{zhang96}
K.~Zhang.
\newblock {A Constrained Edit Distance Between Unordered Labeled Trees}.
\newblock {\em Algorithmica}, 1996.

\end{thebibliography}

\fontsize{8.4}{10.08}\selectfont
\setcounter{section}{0}
\section*{Appendix}
\includegraphics{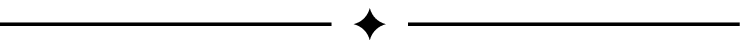}

\section{The edit distance between merge trees \cite{SridharamurthyM20}}

This section formalizes the edit distance 
introduced by  
Sridharamurthy et al. \cite{SridharamurthyM20} 
(\autoref{edit_distance_definition}) and discusses some of its 
technical 
aspects 
which make it not conducive to interpolation-based geodesics 
(\autoref{edit_distance_interpolation}).

\subsection{Definition}
\label{edit_distance_definition}
The edit distance between two merge trees $\mergetree(f_i)$ and 
$\mergetree(f_j)$, noted $\editdistance\big(\mergetree(f_i), 
\mergetree(f_j)\big)$, is defined as follows.
Let $\revision{N_i}$ be a subset of the nodes of $\mergetree(f_i)$ and $\overline{\revision{N_i}} $
its complement. Let $\phi'''$ be a \emph{partial} assignment  
between $\revision{N_i}$ and a subset $\revision{N_j}$ of the nodes of
$\mergetree(f_j)$ (with complement $\overline{\revision{N_j}}$).
Then $\editdistance\big(\mergetree(f_i), 
\mergetree(f_j)\big)$ is given by:
\begin{eqnarray}
\label{eq_edit_distance}
 \editdistance\big(\mergetree(f_i), \mergetree(f_j)\big) = 
  & \hspace{-.35cm}
  \underset{(\phi''', \revision{\overline{N_i}, \overline{N_j}}) \in \Phi'''}\min
  &  \hspace{-.35cm}
\Big( 
    \label{eq_edit_distance_mapping}
    \sum_{\revision{n_i \in N_i}} \gamma\big(\revision{n_i \rightarrow \phi'''(n_i)}\big)\\
    \label{eq_edit_distance_destroying}
    & + &  \sum_{\revision{n_i \in \overline{N_i}}} \gamma(\revision{n_i} \rightarrow \emptyset)\\
    \label{eq_edit_distance_creating}
    & + & \sum_{\revision{n_j \in \overline{N_j}}} \gamma(\emptyset \rightarrow \revision{n_j})
  \Big)
  \label{eq_editDistance}
\end{eqnarray}
%
%
%
where $\Phi'''$ is the space of \emph{constrained} partial assignments (i.e. 
$\phi'''$ maps disjoint subtrees of $\mergetree(f_i)$ to disjoint subtrees of 
$\mergetree(f_j)$) and where $\gamma$ refers to the cost 
for:
\emph{(i)} mapping a node $\revision{n_i} \in \mergetree(f_i)$ to a node 
$\phi''(\revision{n_i}) =
\revision{n_j}
\in \mergetree(f_j)$ (line \ref{eq_edit_distance_mapping}), 
\emph{(ii)} deleting a node 
$\revision{n_i} \in \mergetree(f_i)$ (line \ref{eq_edit_distance_destroying}) and
\emph{(iii)} creating 
a node $\revision{n_j} \in \mergetree(f_j)$ (line \ref{eq_edit_distance_creating}),
$\emptyset$ 
being the empty tree.
%

%

Zhang \cite{zhang96} introduced a polynomial time algorithm for 
computing a 
constrained sequence of edit operations with minimal edit distance 
(\autoref{eq_edit_distance_mapping}), and showed that the resulting distance is 
indeed a 
metric if each cost $\gamma$ for the above three edit operations is itself 
a metric (non-negativity, identity, symmetry, triangle inequality).
Sridharamurthy et al. \cite{SridharamurthyM20} exploited this property to 
introduce their metric,
by defining the following distance-based cost model,
where $\revision{p_i}$ and $\revision{p_j}$ stand for the persistence pairs \emph{containing} the
nodes $\revision{n_i} \in \mergetree(f_i)$ and $\revision{n_j} \in \mergetree(f_j)$:
%
\begin{eqnarray}
  \nonumber
  \gamma(\revision{n_i \rightarrow n_j}) & = & \min\big(\pointMetric_\infty(\revision{p_i, p_j}),
    \gamma(\revision{n_i} \rightarrow \emptyset) + \gamma(\emptyset \rightarrow \revision{n_j})\big)\\
    \nonumber
  \gamma(\revision{n_i} \rightarrow \emptyset) & = & \pointMetric_\infty\big(\revision{p_i},
\projection(\revision{p_i})\big)\\
\nonumber
  \gamma(\emptyset \rightarrow \revision{n_j}) & = & \pointMetric_\infty\big(
     \projection(\revision{p_j}),
      \revision{p_j}
  \big).
\end{eqnarray}
In our work, 
we introduce an alternative 
edit distance
which further adheres to the $L^2$-Wasserstein distance between 
persistence 
diagrams.

\subsection{Interpolation}
\label{edit_distance_interpolation}
As shown in the main manuscript (Fig. 4), the linear interpolation of 
$\editdistance$'s matchings
does not describe a 
shortest path (i.e. it generates inaccurate midpoints). 
A key technical reason for this is that 
$\editdistance$
involves assignments between \emph{nodes} (of the input merge trees) and not 
\emph{persistence pairs}. This has several consequences.
\cutout{, which prevent a 
simple 
interpolation-based geodesic scheme.}
First, 
given two input trees $\mergetree(f_i)$ and $\mergetree(f_j)$,
$\editdistance$'s matchings may assign 
a saddle node
in $\mergetree(f_i)$ to 
an extremum node
in $\mergetree(f_j)$, resulting in inconsistent interpolations in the data 
(from a valley to a peak).
%
%
%
Second, $\editdistance$'s matchings can possibly 
assign two nodes 
in $\mergetree(f_i)$ belonging to a 
\emph{single}
persistence pair of $f_i$ to nodes 
in 
$\mergetree(f_j)$ belonging to \emph{distinct} persistence pairs in $f_j$. This 
second phenomenon further challenges interpolation-based geodesics.

\section{$\wassersteinTree$ is a metric}
\label{sec_metric}

\begin{figure} 
  \includegraphics[width=\linewidth]{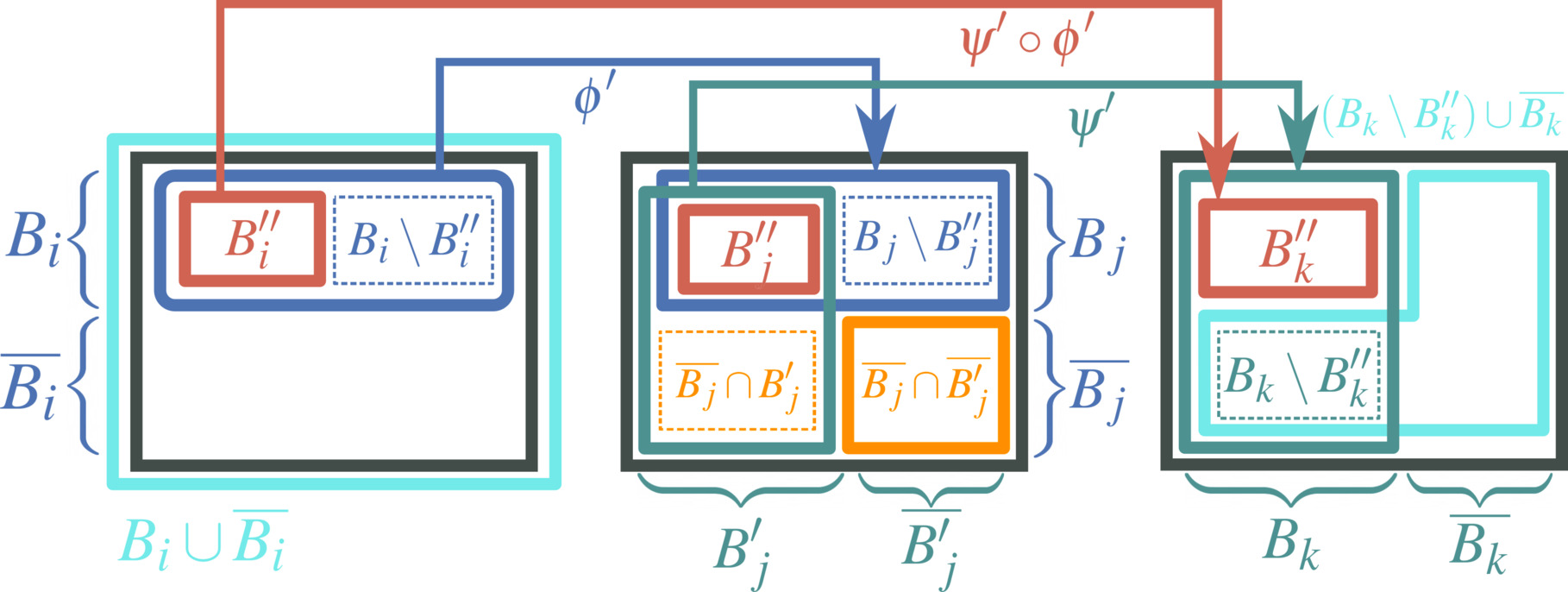}
 \caption{The composition $\psi' \circ \phi'$ of two rooted  
isomorphisms $\phi'$ (blue) and $\psi'$ (green) is itself a rooted 
isomorphism (red). In this schematic view, the involved subtrees
are represented as squares.}
 \label{fig_metricProof}
\end{figure}

As further described in the main manuscript, 
given two merge trees $\mergetree(f_i)$ and $\mergetree(f_j)$ and their branch 
decomposition trees (BDTs) $\branchtree(f_i)$ and $\branchtree(f_j)$,
the dissimilarity measure $\wassersteinTree\big(\branchtree(f_i), 
\branchtree(f_j)\big)$ is 
given by:
\begin{eqnarray}
\nonumber
 \wassersteinTree\big(\branchtree(f_i), \branchtree(f_j)\big) = 
  & \hspace{-.35cm} \underset{(\phi', \revision{\overline{B_i}, \overline{B_j}}) \in
\Phi'}\min &
    \hspace{-.35cm} \Big( 
    \label{eq_our_distance_mapping}
    \sum_{\revision{b_i \in B_i}} \gamma\big(\revision{b_i} \rightarrow \phi'(\revision{b_i})\big)^2\\
    \label{eq_our_distance_destroying}
    & + &  \sum_{\revision{b_i} \in \overline{\revision{B_i}}} \gamma(\revision{b_i} \rightarrow \emptyset)^2\\
    \label{eq_our_distance_creating}
    & + & \sum_{\revision{b_j \in \overline{B_j}}} \gamma(\emptyset \rightarrow \revision{b_j})^2
  \Big)^{1/2}
  \label{eq_editDistance}
\end{eqnarray}
where $\phi'$ is an isomorphism between $\revision{B_i} \subseteq \branchtree(f_i)$ and
$\revision{B_j} \subseteq \branchtree(f_j)$.

~

In this section, we argue that $\wassersteinTree$ is a metric.

$\wassersteinTree\big(\branchtree(f_i), \branchtree(f_j)\big)$ is always 
non-negative (the costs $\gamma$ are squared). 

$\wassersteinTree\big(\branchtree(f_i), \branchtree(f_j)\big)$
is symmetric (destruction and creation costs 
are symmetric, lines \ref{eq_our_distance_destroying} and 
\ref{eq_our_distance_creating}).

$\wassersteinTree\big(\branchtree(f_i), \branchtree(f_j)\big) = 0$ if and only 
if all costs $\gamma = 0$, which only happens if $\branchtree(f_i) =
\branchtree(f_j)$ (the identity is included in $\Phi'$). 


We now argue that $\wassersteinTree$ preserves the triangle inequality, given 
three trees $\branchtree(f_i)$, $\branchtree(f_j)$ and 
$\branchtree(f_k)$.
For this, we follow a classical approach which we detail here for the sake of 
completeness.
First, we argue that a composition of (optimal) partial 
rooted 
isomorphisms (from $\branchtree(f_i)$ to $\branchtree(f_j)$, then from 
$\branchtree(f_j)$ to $\branchtree(f_k)$) is itself a valid partial rooted 
isomorphism (and hence belong to our solution space $\Phi'$) and that its 
associated cost consequently bounds by above 
$\wassersteinTree\big(\branchtree(f_i), \branchtree(f_k)\big)$ 
(\autoref{eq_aboveBound}). Second, we argue that this associated cost is 
itself bounded by above by $\wassersteinTree\big(\branchtree(f_i), 
\branchtree(f_j)\big) + \wassersteinTree\big(\branchtree(f_j), 
\branchtree(f_k)\big)$.

Let $(\phi', \revision{\overline{B_i}, \overline{B_j}})$ be the optimal
solution of the partial assignment problem between $\branchtree(f_i)$ and 
$\branchtree(f_j)$. 
$\phi'$ is a \emph{rooted} isomorphism (i.e. an isomorphism between rooted 
subtrees) between a subtree $\revision{B_i}$ of $\branchtree(f_i)$
and a 
subtree $\revision{B_j}$ of $\branchtree(f_j)$ (blue, \autoref{fig_metricProof}).
Equivalently,
$\phi'$ can also be interpreted as a bijection between 
the \emph{arcs} of $\revision{B_i}$ and those of $\revision{B_j}$.

Let $(\psi', \overline{\revision{B_j}'}, \overline{\revision{B_k}})$ be the
optimal solution of the partial assignment problem between $\branchtree(f_j)$ 
and $\branchtree(f_k)$. $\psi'$ is 
a rooted isomorphism 
between a subtree $\revision{B_j}'$ of $\branchtree(f_j)$
and a subtree $\revision{B_k}$ of $\branchtree(f_k)$ (green, \autoref{fig_metricProof}).

Let $\revision{B_j}''$ be the set of nodes of $\branchtree(f_j)$  involved in \emph{both}
$\phi'$ and $\psi'$ ($\revision{B_j}'' = \revision{B_j} \cap \revision{B_j}'$, in red in \autoref{fig_metricProof}, center).
Let $\revision{B_i}''$ be their pre-image by
$\phi'$ ($\revision{B_i}'' = \phi'^{-1}(\revision{B_j}'')$, in red in \autoref{fig_metricProof}, left)
and $\revision{B_k}''$ their image by $\psi'$
($\revision{B_k}'' =
\psi'(\revision{B_j}'')$, in red in \autoref{fig_metricProof}, right).

Since both $\phi'$ 
and $\psi'$ are rooted isomorphisms, their 
composition $\psi' \circ \phi'$ is also a (rooted) isomorphism between 
the subtrees $\revision{B_i}''$ of $\branchtree(f_i)$ and $\revision{B_k}''$ of $\branchtree(f_k)$
(equivalently, it is a bijection between the arcs of $\revision{B_i}''$ and the arcs of
$\revision{B_k}''$).
Then
$(\psi' \circ \phi', \overline{\revision{B_i}''}, \overline{\revision{B_j}''})$
is itself a rooted partial isomorphism 
and belongs to $\Phi'$.

Then, it follows that:
\begin{eqnarray}
  \label{eq_aboveBound}
 \wassersteinTree\big(\branchtree(f_i), \branchtree(f_k)\big) \leq
  & \hspace{-.35cm}  &
    \hspace{-.35cm} \Big( 
    \sum_{b \in \revision{B_i}''} \gamma\big(b \rightarrow \psi' \circ
      \phi'(b)\big)^2\\
      \nonumber
    & + &  \sum_{b \in \overline{\revision{B_i}''}} \gamma(b \rightarrow \emptyset)^2\\
    \nonumber
    & + & \sum_{b \in \overline{\revision{B_k}''}} \gamma(\emptyset \rightarrow b)^2
  \Big)^{1/2}.
\end{eqnarray}

%

Now, let $U$, $V$, $W$ be scalar functions on the nodes of the set  $\revision{B_{ik} =
B_i \cup \overline{B_i} \cup (B_k \setminus B''_k) \cup \overline{B_k}}$
(cyan subset, \autoref{fig_metricProof})  such 
that:

\begin{eqnarray}
\label{def_U}
U(b) = \begin{cases}\gamma\big(b \rightarrow \psi'\circ \phi'(b)\big) & \text{ 
for } b 
\in \revision{B_i'' = \phi'^{-1}(B_j'')}
\\ \gamma(b \rightarrow \emptyset) & \text{ for } b \in \revision{B_i \setminus
B_i''}\\ \gamma(b \rightarrow \emptyset) & \text{ for } b \in \overline{\revision{B_i}} \\
\gamma(\emptyset \rightarrow b)
& \text{ for } b \in \overline{\revision{B_k}}\\
\gamma(\emptyset \rightarrow b)
& \text{ for } b \in \revision{B_k \setminus B_k''=
\psi'(\overline{B_j} \cap B_j')}. \\\end{cases}
\end{eqnarray}

$U$ describes all the possible individual costs involved in the composition 
$\psi' \circ \phi'$. In particular, we can re-write \autoref{eq_aboveBound} as:
\begin{eqnarray}
\label{eq_costU}
 \wassersteinTree\big(\branchtree(f_i), \branchtree(f_k)\big)  \leq 
 & \|U\|_2  &
 = \big(\sum_{b \in B_{\revision{ik}}} U(b)^2\big)^{1/2}.
\end{eqnarray}

\begin{eqnarray}
\label{def_V}
V(b) = \begin{cases}\gamma\big(b \rightarrow \phi'(b)\big) & \text{ for } b 
\in \revision{B_i'' = \phi'^{-1}(B_j'')}\\
\gamma\big(b \rightarrow \phi'(b)\big) & \text{ for } b \in \revision{B_i \setminus
B_i}''\\
\gamma(b 
\rightarrow \emptyset) & \text{ for } b \in \overline{\revision{B_i}} \\ 0 & \text{ for }
b 
\in \overline{\revision{B_k}}\\ \gamma\big(\psi'^{-1}(b) \rightarrow \emptyset\big) &
\text{ for } 
b 
\in \revision{B_k \setminus B_k'' = \psi'(\overline{B_j} \cap B_j')}\\\end{cases}
\end{eqnarray}

$V$ describes a \emph{subset} of the individual costs involved in the optimal 
rooted partial isomorphism $\phi'$. In particular, only the costs involving 
$\revision{\overline{B_j} \cap \overline{B'_j}}$ (orange square,
\autoref{fig_metricProof}, middle) are excluded. Thus, we have:

\begin{eqnarray}
\label{eq_costV}
 \wassersteinTree\big(\branchtree(f_i), \branchtree(f_j)\big)  \geq
 & \|V\|_2  & = \big(\sum_{b \in B_{\revision{ik}}} V(b)^2\big)^{1/2}.
\end{eqnarray}



\begin{eqnarray}
\label{def_W}
W(b) = \begin{cases}\gamma\big(\phi'(b) \rightarrow \psi'\circ\phi'(b)\big) & 
\text{ for 
} 
b \in \revision{B_i''= \phi'^{-1}(B_j}'')\\ \gamma\big(\phi'(b) \rightarrow \emptyset\big) & \text{ for } b
\in 
\revision{B_i
\setminus B_i''}\\ 0 & \text{ for } b \in \overline{\revision{B_i}} \\ \gamma(b\rightarrow
\emptyset) & \text{ for } b \in \overline{\revision{B_k}}\\
\gamma\big(
\psi'^{-1}(b) \rightarrow b
\big) 
& \text{ for } b \in \revision{B_k \setminus B_k''= \psi'(\overline{B_j} \cap
B_j')}\\\end{cases}
\end{eqnarray}


Similarly to $V$, $W$ describes a \emph{subset} of the individual costs 
involved in the optimal rooted partial isomorphism $\psi'$. In particular, only 
the costs involving \revision{$B_k''$} (red square, \autoref{fig_metricProof},
right) are excluded. Thus:

\begin{eqnarray}
\label{eq_costW}
 \wassersteinTree\big(\branchtree(f_j), \branchtree(f_k)\big)  \geq
 & \|W\|_2  & = \big(\sum_{b \in B_{\revision{ik}}} W(b)^2\big)^{1/2}.
\end{eqnarray}

Now,  
since 
$\gamma$ is defined by the Euclidean distance (Equation 9
of the 
main manuscript), 
we have for each node $b \in B_{\revision{ik}}$:
\begin{eqnarray}
\nonumber
0 \le U(b) \le V(b) + W(b).
\end{eqnarray}
This can be verified 
by comparing the $i^{th}$ line of \autoref{def_U} to the sum of the $i^{th}$ 
lines of \autoref{def_V} and \autoref{def_W}. Then, we have:
\begin{eqnarray}
\label{eq_u_vw}
 \|U\|_2 \leq \|V + W\|_2.
\end{eqnarray}

Now, since the $L^2$ norm between vectors respects itself the triangle 
inequality, we have the following inequality:
\begin{eqnarray}
\label{eq_vw_v_w}
 \|V+ W\|_2 \leq \|V\|_2 + \|W\|_2.
\end{eqnarray}

Then, by combining equations 
\ref{eq_costU},  \ref{eq_u_vw}, \ref{eq_vw_v_w},
\ref{eq_costV}, and \ref{eq_costW},
it 
follows that:
\begin{eqnarray}
\nonumber
 \wassersteinTree\big(\branchtree(f_i), \branchtree(f_k)\big) & 
 \leq & \hspace{-.1cm} \|U\|_2
 \leq \|V + W\|_2
 \leq\|V\|_2 + \|W\|_2\\
 \nonumber
 &  &\leq \wassersteinTree\big(\branchtree(f_i), \branchtree(f_j)\big) 
  + \wassersteinTree\big(\branchtree(f_j), \branchtree(f_k)\big) 
\end{eqnarray}
which concludes the proof.




\section{Comparison to the edit distance algorithm \cite{zhang96}}

In addition to considering squared costs in our edit distance (equations of 
the section 3.3 of the main manuscript), our algorithm for the exploration of 
the search space 
%
indeed simplifies the approach by 
Zhang 
\cite{zhang96} (used by Sridharamurthy et al. \cite{SridharamurthyM20}), as our 
search space is significantly more constrained.

First, since our solution space only considers partial 
isomorphisms between rooted subtrees, this implies that
%
the destruction of a node (a branch) $\revision{b_j} \in \branchtree(f_j)$
necessarily implies the destruction of its 
subtrees, i.e. of its
forest $\forest(f_j, \revision{b_j})$. Thus, the admissible
solutions in \cite{zhang96, SridharamurthyM20} consisting in deleting 
$\revision{b_j}$ and mapping a subtree $\branchtree(f_i,
\revision{b_i})$ to one of the subtrees of \revision{$b_j$} in the forest $\forest(f_j, \revision{b_j})$
are no longer admissible 
given our overall solution space $\Phi'$. The removal of such solutions 
drastically simplifies the evaluation of the distance between subtrees
(being the 
minimum 
of three solutions in  \cite{SridharamurthyM20}, Eq. 12) to the Equation 11 of 
our main manuscript
(containing only one expression to 
evaluate). 

Second, 
our solution space (rooted partial isomorphisms)
also implies that the nodes of 
$\branchtree(f_i)$  can only be assigned to nodes with the same \emph{depth} 
in $\branchtree(f_j)$. This further implies that 
the distance between subtrees (Equation 11 of the main manuscript) only needs 
to be evaluated for subtrees rooted at nodes of identical depth 
(see Fig. 6 of the main manuscript). 

Together, these two simplifications
(\emph{(i)} simpler subtree distance and \emph{(ii)} distance 
evaluation restricted to subtrees of identical depth from the root) 
are the key adaptations of Zhang's algorithm \cite{zhang96} that are required 
for the exploration of our (more constrained) solution space.


%
\section{\revision{Parallel computation of $\wassersteinTree$}}
\revision{
In our work, we express the computation of $\wassersteinTree$ in terms of
\emph{tasks}, to leverage task-based shared memory parallelism.
First,
the Equation 10 of the main manuscript is evaluated.
For this, we initiate a task at
each leaf of $\branchtree(f_i)$. If a task is the last one to compute among all
the direct children of a node $b \in
\branchtree(f_i)$, it is then authorized to continue and estimate
Equation 10
in $b$. Atomic counters in $b$ are implemented
(and atomically incremented by the task of each child) to determine which child
task is the last one to complete, which enables an efficient lightweight
synchronization (Fig. 6 of the main manuscript).
Overall, Equation 10
is completely
estimated  with this strategy in a bottom-up fashion.
Second,
Equation 11 (main manuscript)
is
evaluated similarly, by initiating a task at each leaf $b_i$ of
$\branchtree(f_i)$. In particular, this task will evaluate Equation 11
given $b_i$ against all subtrees of
$\branchtree(f_j)$ of identical depth (again using independent tasks initiated
at the leaves of
$\branchtree(f_j)$,
see Fig. 6).
Similarly to Equation 10,
we employ the
same lightweight synchronization mechanism based on atomic counters to continue
a task over to its parent only when it is the last child task reaching it.
Thus, in both cases (Eqs. 10 and 11),
the number of parallel tasks
is initially
bounded by the number of leaves in $\branchtree(f_i)$ and $\branchtree(f_j)$
(which is typically much
larger than the number of cores) and progressively decreases during the
computation.}

\section{$\branchtree_\alpha(f_i \rightarrow f_j)$ is a geodesic for 
$0 \leq \alpha \leq 1$}


We now argue that $\wassersteinTree$ defines a geodesic space. For this, 
for any two BDTs $\branchtree(f_i)$ and $\branchtree(f_j)$,
we 
describe the existence of a path between 
them
whose length is equal to 
$\wassersteinTree\big(\branchtree(f_i), \branchtree(f_j)\big)$ (and thus minimal).



Let $P = \big(\branchtree_t\big)_{t \in [0, 1]}$ be a path of BDTs 
parameterized by $t$.

We recall that the length $\mathcal{L}(P)$ of $P$ is given by:
\begin{eqnarray}
\nonumber
\mathcal{L}(P) = \sup_{n; 0 = t_0 \leq t_1 \leq \dots \leq t_n = 1} \sum_{k = 
0}^{n-1} \wassersteinTree\big(\branchtree_{t_k}, \branchtree_{t_k + 1}\big).
\end{eqnarray}

Now, let $P_\alpha$ be the path corresponding to the interpolation between 
$\branchtree(f_i)$ and $\branchtree(f_j)$, as defined in 
section 4.1 of the main manuscript. We now argue that $\mathcal{L}(P_\alpha) = 
\wassersteinTree\big(\branchtree(f_i), \branchtree(f_j)\big)$.


Let $(\phi', \revision{\overline{B_i}, \overline{B_j}})$ be the optimal rooted
partial isomorphism between $\branchtree(f_i)$ and $\branchtree(f_j)$. 
Moreover, let $\branchtree_{s}(f_i \rightarrow f_j)$ and $\branchtree_t(f_i 
\rightarrow f_j)$ be two interpolated trees obtained respectively with $\alpha = 
s$ and $\alpha = t$, given $0 \leq s \leq t \leq 1$.
We will note
$\phi'_s$  the application on $\revision{B_i} \cup \overline{\revision{B_i}} \cup \overline{\revision{B_j}}$
defined by
interpolation (section 4.1 of the main manuscript):
\begin{eqnarray}
\label{definition_phi_s}
 \phi'_s(b) =   
  \begin{cases} 
      (1-s)b+ s \phi'(b) & \text{ for } b \in \revision{B_i}\\
    (1-s)b+s\projection(b)  & \text{ for } b \in \overline{\revision{B_i}}\\
    sb+(1-s)\projection(b)  &  \text{ for } b \in \overline{\revision{B_j}}.
  \end{cases}
\end{eqnarray}
$\phi'_t$ is defined similarly for $t$.
Then, as discussed in \autoref{sec_metric} of this appendix, since 
the composition of 
partial rooted isomorphisms is itself a 
partial rooted 
isomorphism,
the composition $\phi'_t \circ {\phi'_{s}}^{-1}$ (which goes from 
$\branchtree_s(f_i \rightarrow f_j)$ to $\branchtree(f_i)$ and then from 
$\branchtree(f_i)$ to $\branchtree_t(f_i \rightarrow f_j)$) does define a valid 
partial rooted isomorphism between 
$\branchtree_s(f_i \rightarrow f_j)$ and $\branchtree_t(f_i \rightarrow f_j)$ 
and we have:

\begin{eqnarray}
\nonumber
 \wassersteinTree\big(
    \branchtree_s(f_i \rightarrow f_j), 
    \branchtree_t(f_i \rightarrow f_j)
  \big)
   &\leq& \\
   \nonumber
     & & \hspace{-1cm}
  \julien{\Big(}
  \sum_{b \in \revision{B_i \cup \overline{B_i} \cup \overline{B_j}}}
    \gamma \big( 
      \phi'_s(b) \rightarrow \phi'_t(b)
    \big)\julien{^{2} \Big)^{1/2}}
\end{eqnarray}
\julien{and we also have by definition of $\phi'_s$ and $\phi'_t$ 
(\autoref{definition_phi_s}):}
\begin{eqnarray}
  \nonumber
  \julien{\Big(}
  \sum_{b \in \revision{B_i \cup \overline{B_i} \cup \overline{B_j}}}
    \gamma \big( 
      \phi'_s(b) \rightarrow \phi'_t(b)
    \big)\julien{^{2} \Big)^{1/2}}    
  \julien{=}
  ~
  (t - s) \wassersteinTree\big(\branchtree(f_i), \branchtree(f_j)\big).
\end{eqnarray}

Now, given the triangle inequality on the path $P_\alpha$,
we have:
\begin{eqnarray}
\nonumber
 \wassersteinTree\big(\branchtree(f_i), \branchtree(f_j)\big) & \leq & \\
\nonumber 
  & & \wassersteinTree\big(\branchtree_0(f_i \rightarrow f_j), 
\branchtree_s(f_i \rightarrow f_j)\big)
\\
\nonumber
  & + & \wassersteinTree\big(\branchtree_s(f_i \rightarrow 
f_j), \branchtree_t(f_i \rightarrow 
f_j)\big)\\
\nonumber
  & + & \wassersteinTree\big(\branchtree_t(f_i \rightarrow 
f_j), \branchtree_1(f_i \rightarrow f_j)\big) \\
  \nonumber
  & & \leq \big(s + (t - s) + (1 - t)\big) 
    \wassersteinTree\big(\branchtree(f_i), \branchtree(f_j)\big)\\
      \nonumber
      & & \quad = \wassersteinTree\big(\branchtree(f_i), \branchtree(f_j)\big).
\end{eqnarray}
If follows that the above inequalities are in fact equalities and we have:
\begin{eqnarray}
\nonumber
 \wassersteinTree\big(
    \branchtree_s(f_i \rightarrow f_j), 
    \branchtree_t(f_i \rightarrow f_j)
  \big) = (t - s) \wassersteinTree\big(\branchtree(f_i), \branchtree(f_j)\big).
\end{eqnarray}

%
%
%

Then, for \emph{any} subdivision $0=t_0\le t_1 \le \dots \le t_n=1$ of 
$P_\alpha$, we have:
\begin{eqnarray}
\nonumber
 \sum_{k = 0}^{n - 1} \wassersteinTree\big( \branchtree_{t_k}(f_i \rightarrow 
f_j),  \branchtree_{t_k + 1}(f_i \rightarrow 
f_j)\big) \\
\nonumber
 = \sum_{k = 0}^{n - 1} (t_{k + 1} - t_k) 
  \wassersteinTree\big(\branchtree(f_i), \branchtree(f_j)\big)\\
  \nonumber
  = (t_n - t_0) \wassersteinTree\big(\branchtree(f_i), \branchtree(f_j)\big)\\
  \nonumber
  = \wassersteinTree\big(\branchtree(f_i), \branchtree(f_j)\big).
\end{eqnarray}


Thus $\mathcal{L}(P_\alpha) =  \wassersteinTree\big(\branchtree(f_i), 
\branchtree(f_j)\big)$.

Hence the space of merge trees equipped with 
$\wassersteinTree$ 
is a geodesic 
space, and 
$\branchtree_\alpha(f_i \rightarrow f_j)$ constructs paths of 
minimal length on it.

\section{$\wassersteinTree$ with normalized costs defines a 
geodesic space}

Let $\normalizedWasserstein\big(\
  \branchtree(f_i), \branchtree(f_j)
\big)$ be a similarity measure between $\branchtree(f_i)$ and 
$\branchtree(f_j)$, defined as:
\begin{eqnarray}
\nonumber
\normalizedWasserstein\big(\
  \branchtree(f_i), \branchtree(f_j)
\big)  = \wassersteinTree\Big(\normalizedLocation\big(\branchtree(f_i)\big), 
\normalizedLocation\big(\branchtree(f_j)\big)\Big)
\end{eqnarray}
where $\normalizedLocation$ is the local normalization described in Section 4.2 
of the main 
manuscript. Since $\normalizedLocation$ is invertible, $\normalizedWasserstein$ 
inherits all the 
properties of $\wassersteinTree$ and is also a distance metric.


%
%
%


The \emph{normalized} interpolation $\branchtree_{s}(f_i \rightarrow f_j)$, $s 
\in [0,1]$ 
between  $\branchtree(f_i)$ and $\branchtree(f_j)$  is defined as the image by 
$\normalizedLocation^{-1}$ of the interpolation between the normalized trees  
$\normalizedLocation(\mathcal{B}(f_i))$ and 
$\normalizedLocation(\mathcal{B}(f_j))$. Then, given $s$ and $t$ such that $0 
\leq s \leq t \leq 1$, it follows 
that:
\begin{eqnarray}
 \nonumber
 \normalizedWasserstein\big(
  \branchtree_s(f_i \rightarrow f_j),
  \branchtree_t(f_i \rightarrow f_j)
 \big) 
 \hspace{-0.25cm}
 &=& 
 \hspace{-0.25cm}
 (t - s) \wassersteinTree\Big(
    \normalizedLocation \big(\branchtree(f_i)\big), 
    \normalizedLocation \big(\branchtree(f_j)\big)\Big)\\
  \nonumber
  \hspace{-0.25cm}
 &=& 
 \hspace{-0.25cm}
  (t - s) \normalizedWasserstein\big(\branchtree(f_i), 
\branchtree(f_j)\big)
\end{eqnarray}
which proves that the space of merge trees equipped with 
$\normalizedWasserstein$ is a geodesic space, and that the above normalized 
interpolation constructs paths of minimal length on it. 


%

%


\section{Minimizing the Fr\'echet energy}
The optimization algorithm described in Section 5.2 of the main manuscript 
constructively decreases the Fr\'echet energy at each iteration.
In particular, once a local minimizer of the Fr\'echet energy is 
obtained for a fixed assignment with the \emph{update} step \emph{(ii)}, the 
subsequent \emph{assignment} step \emph{(i)} does further improve the 
assignments 
hence iteratively decreasing the Fr\'echet energy constructively.



Let $F'$ be a function of an arbitrary BDT $\branchtree$ and of an arbitrary 
(i.e. not necessarily optimal) set of $N$ rooted partial isomorphisms 
$({\phi'}_i, \revision{\overline{B_{\branchtree}}, \overline{B_i}})_{i = 1, \dots, N}$ between
$\branchtree$ and the $N$ BDTs of $\branchset$:
\begin{eqnarray}
\nonumber
F'\big(
\branchtree, ({\phi'}_i, \revision{\overline{B_{\branchtree}},
\overline{B_i}})_{i = 1, \dots, N}
\big) 
:= &
  \underset{\branchtree(f_i) \in \branchset}\sum
    \Big( 
    \sum_{\revision{b_i \in B_i}} \gamma\big(\revision{b_i} \rightarrow
{\phi'_{i}}(\revision{b_i})\big)^2\\
    \nonumber
    & +  \sum_{\revision{b_i \in \overline{B_i}}} \gamma(\revision{b_i} \rightarrow
\emptyset)^2\\
    \nonumber
    & + \sum_{\revision{b_\branchtree} \in \overline{\revision{B_\branchtree}}} \gamma(\emptyset \rightarrow
\revision{b_\branchtree})^2
\Big)^{1/2}.
\end{eqnarray}

Now, let $\branchtree_k$ be the candidate barycenter at the iteration $k$ 
of the 
algorithm and let  $({\phi'_i}^k,\revision{\overline{B_{\branchtree}^k},\overline{B_i^k}} )$
be the optimal 
rooted partial isomorphism 
between $\branchtree_k$ and   $\branchtree(f_i)$, computed by the 
\emph{assignment} step of the iteration. Then, we have:
\begin{eqnarray}
 \nonumber
 F'\big(
    \branchtree_k,
    ({\phi'_i}^k,\overline{\revision{B_{\branchtree}}^k},\overline{\revision{B_i}^k} )_{i =1, \dots, N}
 \big) = \sum_{\branchtree(f_i) \in \branchset} 
\wassersteinTree\big(\branchtree_{k},\branchtree(f_i)\big)^2 .
\end{eqnarray}

Next, the \emph{update}
step of the iteration $k$ consists in moving $\branchtree_k$ to 
$\branchtree_{k+1}$ by placing (in the 2D birth/death space)  each branch $b 
\in 
\branchtree_k$ at the arithmetic mean of the assignments.
Since the 
arithmetic mean generally minimizes sums of Euclidean distances,
we have:
%
%
%
$$ F'(\branchtree_{k+1}, ({\phi'_i}^k,\overline{\revision{B_{\branchtree}}^k},
\overline{\revision{B_i}^k})_{i=1,\dots,N}) \le
F'(\branchtree_k, ({\phi'_i}^k,\overline{\revision{B_{\branchtree}}^k},
\overline{\revision{B_i}^k} )_{i=1,\dots,N}).$$

Now, observe that since the previous 
rooted partial isomorphisms
are not optimal anymore for $\branchtree_{k+1}$, we also have:
$$\sum_{\branchtree(f_i) \in \branchset} 
\wassersteinTree\big(\branchtree_{\julien{k+1}},\branchtree(f_i)\big)^2
\le  F'(\branchtree_{k+1}, ({\phi'_i}^k,\overline{\revision{B_{\branchtree}}^k},
\overline{\revision{B_i}^k} )_{i=1,\dots,N}).$$


Once $\branchtree_{k+1}$ is fixed, 
all the rooted partial isomorphisms are then optimized again with the 
\emph{assignment} step of the iteration $k+1$
to attain:
\begin{eqnarray}
\nonumber
F'\big(
\branchtree_{k+1}, ({\phi'}_i^{k+1}, \overline{\revision{B_{\branchtree}}^{k+1}},
\overline{\revision{B_i}^{k+1}})_{i = 1, \dots, N}
\big) \\
\nonumber=
\sum_{\branchtree(f_i) \in \branchset} 
\wassersteinTree\big(\branchtree_{k+1},\branchtree(f_i)\big)^2.
\end{eqnarray}

The result of these two steps is that:
\begin{eqnarray}
\nonumber
F'\big(
\branchtree_{k+1}, ({\phi'}_i^{k+1}, \overline{\revision{B_{\branchtree}}^{k+1}},
\overline{\revision{B_i}^{k+1}})_{i = 1, \dots, N}
\big)\\
\nonumber\leq 
F'(\branchtree_{k}, ({\phi'_i}^k,\overline{\revision{B_{\branchtree}}^k},
\overline{\revision{B_i}^k})_{i=1,\dots,N})\\
\nonumber
\sum_{\branchtree(f_i) \in \branchset} 
\wassersteinTree\big(\branchtree_{k+1},\branchtree(f_i)\big)^2
\leq 
\sum_{\branchtree(f_i) \in \branchset} 
\wassersteinTree\big(\branchtree_{k},\branchtree(f_i)\big)^2.
\end{eqnarray}

Then, each iteration of our algorithm indeed decreases 
the Fr\'echet energy. 
Since 
there is a finite number of combinations of 
rooted partial isomorphisms
between the barycenter and the $N$ input trees $\branchtree(f_i)$,  
it follows that 
the algorithm 
converges,
in a finite number of steps,
to a local minimum $\branchtree^*$ of the Fr\'echet energy (if multiple, 
equally valued, optimal sets of assignments exist between $\branchtree^*$ and 
$\branchset$, each one needs to be explored with the update step of our 
algorithm).
%
In practice, as described in the 
manuscript, we stop our algorithm when the Fr\'echet energy has decreased by 
less than $1\%$ between consecutive iterations.

\section{Temporal reduction algorithm}
\begin{figure}
\includegraphics[width=\linewidth]{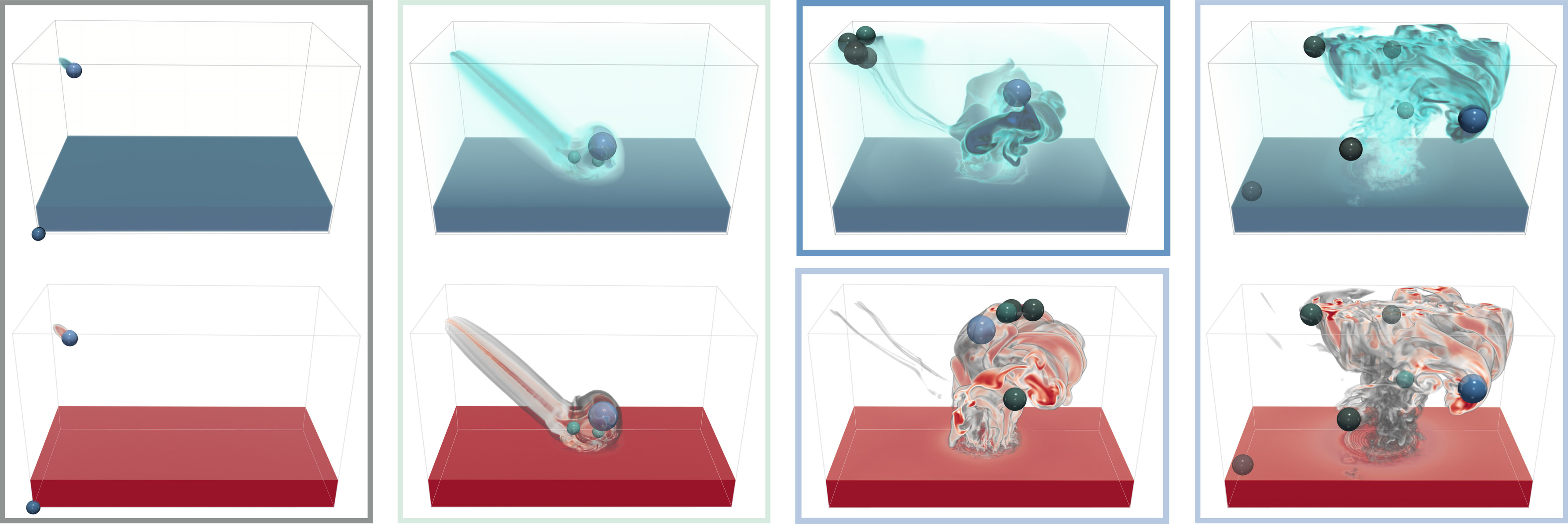}
  \caption{\revision{Comparison between the \emph{key frames} identified by 
 our temporal reduction algorithm, with regard to $\wassersteinTree$ (blue, 
 top) and with regard to $\wasserstein{2}$ (red, bottom). By construction, the 
 reduction algorithm identifies as key frames the first and last time steps, 
 irrespective of the employed metric.}
}
  \label{fig_reductionComparison}
\end{figure}
Let $\mathcal{S} = \{\branchtree(f_1), \branchtree(f_2), \dots, 
\branchtree(f_N)\}$ be the input temporal sequence of 
BDTs
(we assume a regular temporal sampling). Let 
$\mathcal{K} \subseteq 
\mathcal{S}$ be a set of key frames. Let $\mathcal{S}' = \{\branchtree'(f_1), 
\branchtree'(f_2), \dots, 
\branchtree'(f_N)\}$ be a \emph{reduced} temporal sequence, where:
\begin{eqnarray}
 \branchtree'(f_i) = (1 - \alpha_i) \branchtree(f_j) + \alpha_i \branchtree(f_k)
\end{eqnarray}
where $\branchtree(f_j)$ and $\branchtree(f_k)$ are two consecutive trees in 
$\mathcal{K}$, such that $j \leq i \leq k$ and $\alpha_i = (i-j)/(k-j)$. 
$\branchtree'(f_i)$ is then on a geodesic between $\branchtree(f_j)$ and 
$\branchtree(f_k)$. We introduce the following distance
between 
the temporal sequences $\mathcal{S}$ and $\mathcal{S}'$:
\begin{eqnarray}
 \distanceSequence(\mathcal{S}, \mathcal{S}') = \Big(\sum_{i = 0}^{N} 
\wassersteinTree\big(\branchtree(f_i), \branchtree'(f_i)\big)^2\Big)^{1/2}.
\end{eqnarray}
$\distanceSequence$ is indeed a metric since it is a composition of metrics 
(being the $L^2$ norm between vectors of 
BDTs under the metric $\wassersteinTree$). 

Our algorithm for temporal reduction consists in initializing $\mathcal{K}$ 
with the entire input sequence ($\mathcal{K} \leftarrow \mathcal{S}$) and 
then removing greedily, at each iteration, the tree $\branchtree^*$ from 
$\mathcal{K}$
($\mathcal{K} \leftarrow \mathcal{K} - \{\branchtree^*\}$) which minimizes 
$\distanceSequence(\mathcal{S}, \mathcal{S}')$, 
and which, hence, better preserves the input sequence,
until $\mathcal{K}$ reaches a 
target size.

\revision{\autoref{fig_reductionComparison} shows the 
temporal reduction performed by this algorithm on the 
\emph{Asteroid impact} sequence (see Section \ref{sec_asteroidImpact}).
This figure illustrates \emph{key frames}, which correspond to 
time steps for which $\branchtree(f_i) = \branchtree'(f_i)$: these are the 
time steps which have \emph{not} been removed from the sequence through the 
reduction. In particular, this figure compares the usage of two metrics in the 
reduction algorithm: $\wassersteinTree$ (blue, top) and $\wasserstein{2}$ 
(red, bottom, obtained with $\epsilon_1 = 1$). By construction, since our 
reduction algorithm is based on interpolation only, the first and last BDTs in 
the sequence $\mathcal{S}$ are always kept in the reduced sequence 
$\mathcal{S}'$. In other words, the first (leftmost, 
\autoref{fig_reductionComparison}) and last (rightmost, 
\autoref{fig_reductionComparison}) time steps are always identified as key 
frames, irrespective of the employed metric. For this specific 
example, the second key frame (second from left, 
\autoref{fig_reductionComparison}) also happens to be identical for both 
metrics. In contrast to the sequence extremities, the common identification of 
this time step as a key frame by $\wassersteinTree$ and $\wasserstein{2}$ is 
not obtained by construction: the reduction algorithm did select this key frame 
in both configurations. Then, only the third key frame (third from left, 
\autoref{fig_reductionComparison}) is different in this example. In particular, 
when using $\wassersteinTree$, the reduction algorithm identifies one key 
frame per key phase of the simulation (see Section \ref{sec_asteroidImpact}, 
each key phase is represented in \autoref{fig_reductionComparison} with a 
frame of distinct color). In contrast, when using $\wasserstein{2}$, the third 
key frame belongs to the same key phase as the last key frame
(\emph{``Aftermath''}, light blue frame). Then the reduction driven by 
$\wasserstein{2}$ fails at identifying a key frame for the third key phase 
(\emph{``Impact''}, dark blue frame). This is confirmed visually in 
\autoref{fig_reductionComparison}, as the third key frame identified with 
$\wassersteinTree$ (in blue) seems to represent an intermediate step in the 
simulation between the second and fourth key frames. In contrast, the third key 
frame identified by the reduction with $\wasserstein{2}$ (in red) is more 
visually similar to the fourth key frame, and hence possibly more redundant.}
%


\section{\revision{Data specification}}

\revision{This section provides a complete specification of the ensemble
datasets used in the paper. In particular, we document the data provenance,
its representation, its pre-processing when applicable, and we specify the
associated ground-truth classification.}

\revision{All of these ensemble datasets were extracted from public repositories.
We additionally provide
a set of scripts which
automatically download
all of these datasets (at the exception of \emph{Asteroid impact} and
\emph{Cloud processes}, for which the dataset providers need to be contacted
personally), pre-process them with TTK and output them in VTK file format, with
the ground-truth classification attached to the files as meta-data (i.e.
\emph{``Field Data''} in the VTK terminology). For convenience, we also provide
an archive containing the entire curated ensemble datasets (in VTK file
format). All of this new material (scripts and curated data) is located at the
following address:
\url{https://github.com/MatPont/WassersteinMergeTreesData}.}

\revision{Moreover, we also provide in additional material all the ensembles of
merge trees computed from these datasets (in the code archive containing the
implementation of our method).}

\subsection{\revision{Asteroid impact}}
\label{sec_asteroidImpact}
\revision{This ensemble is composed of 7 members, given as 3D regular grids
(sampled at $300\times300\times300$, implicitly triangulated by TTK).
It has been made available in the context of the SciVis contest 2018
\cite{scivis2018}.}
\revision{Each member corresponds to the last time step of
the simulation of
the impact of an asteroid with the sea at the surface of the Earth, for two
configurations of asteroid diameter.
The considered scalar field is the matter density, which is one of the
variables of the simulation which discriminates well the asteroid from the
water and the ambient air.
This ensemble corresponds to a
parameter study (in this case, studying the effect of the asteroid's diameter
on the resulting wave),
which is a typical task in numerical simulation.
In this
application, salient maxima capture well the asteroid and large water
splashes. Thus, each member is represented by the split tree (capturing
maxima).}
\revision{The associated ground-truth classification assigns members computed
with similar asteroid diameters
to the same class. Thus, the corresponding
classification task consists in identifying, for a given 
member, its correct asteroid diameter class. The ground-truth
classification is
as follows:
\begin{itemize}
  \item \textbf{Class 1} (3 members): yA11, yB11, yC11.
  \item \textbf{Class 2} (4 members): yA31, yA32, yB31, yC31
\end{itemize}
}
\revision{Another selection of the original data has been used for the
evaluation of our temporal reduction framework (Fig. 10 of the main
manuscript). For this experiment, we used the asteroid diameter configuration
\emph{``yA31''} and considered the following time steps, organized in 4 phases
(according to the SciVis contest companion documentation \cite{scivis2018}):
\begin{itemize}
 \item \textbf{Phase 1, initial state} (5 time steps): 01141, 03429, 05700,
07920, 09782
  \item \textbf{Phase 2, approach} (5 time steps): 13306, 16317, 18124, 19599,
21255
  \item \textbf{Phase 3, impact} (5 time steps): 28649, 31737, 34654, 37273,
39476
  \item \textbf{Phase 4, aftermath} (5 time steps): 44229, 45793, 47190, 48557,
49978
\end{itemize}
}

\subsection{\revision{Cloud processes}}
\revision{This ensemble is composed of 12 members, given as 2D regular grids
(sampled at $1430\times1557$, implicitly triangulated by TTK). Each member
corresponds to a time step of the simulation of cloud
formations \cite{scivis2017}.
For this application, large
clouds are well captured by the maxima of the pressure variable
(pre-processed with 10 iterations of smoothing).
Thus,
split trees (capturing maxima) are considered for this ensemble.
The associated ground-truth classification assigns each time step to one of the
three key phases of the simulation.
The
corresponding classification task therefore consists in identifying, for each
time step, to which phase it belongs. The ground-truth classification is as
follows:
\begin{itemize}
  \item \textbf{Class 1} (4 members): 0, 5, 10, 15
  \item \textbf{Class 2} (4 members): 500, 505, 510, 515
  \item \textbf{Class 3} (4 members): 1000, 1005, 1010, 1015
\end{itemize}
}

\subsection{\revision{Viscous fingering}}
\revision{This ensemble is composed of 15 members, given as 3-dimensional point
clouds (representing a particle-based flow simulation). Each
point cloud is turned into a Eulerian representation of the variables by
using the \emph{``Gaussian Resampling''} filter of ParaView,
effectively transforming, via interpolation
\cite{gaussianResampling}, each ensemble member into a 3D regular grid
(sampled at $50\times50\times50$, implicitly triangulated by TTK).
The original data
has been made available in the context of the SciVis contest 2016
\cite{scivis2016}.}
\revision{
Each member corresponds to the last time step of
the simulation of a viscous fingering phenomenon, occurring when dissolving
salt in water.
The considered scalar field is the salt concentration, whose
salient maxima capture well the most prominent fingers. Thus, each member is
represented by the split tree (capturing maxima).
 Given the studied physical phenomenon, the
simulation approach is not deterministic, resulting in distinct outputs for
identical initial configurations. In this application, three distinct solver
resolutions have been considered, corresponding to three distinct numbers of
particles (resolution code 20: 194k particles, resolution code 30: 544k
particles, resolution code 44: 1.7M particles). Thus, this ensemble corresponds
to a parameter study (in this case, studying the effect of the input resolution
on the output fingering), which is a typical task in numerical simulation. The
associated ground-truth classification assigns members with the same input
resolution  to the same class. Thus, the corresponding classification task
consists in identifying, for a given ensemble member, its corresponding particle
count. The ground-truth classification is as follows:
\begin{itemize}
  \item \textbf{Class 1, resolution 20} (5 members): 20run1, 20run3, 20run4,
20run5, 20run6.
  \item \textbf{Class 2, resolution 30} (5 members): 30run1, 30run2, 30run3,
30run4, 30run5
  \item \textbf{Class 3, resolution 44} (5 members): 30run1, 30run2, 30run3,
30run4, 30run5
\end{itemize}
}

\subsection{\revision{Dark matter}}
\revision{This ensemble is composed of 40 members, given as 3-dimensional point
clouds (representing a particle-based simulation). Each
point cloud is turned into a Eulerian representation of the variables by
using the \emph{``Gaussian Resampling''} filter of ParaView,
effectively transforming, via interpolation
\cite{gaussianResampling}, each ensemble member into a 3D regular grid
(sampled at $100\times100\times100$, implicitly triangulated by TTK).
The original data
has been made available in the context of the SciVis contest 2015
\cite{scivis2015}.}
\revision{
Each member corresponds to a time step of a simulation of the universe
formation, where regions of high concentration of dark matter form a filament
structure known as the \emph{cosmic web}. The considered scalar field is
therefore dark matter density, whose salient maxima capture well large clusters
of galaxies. Thus, each member is represented by the split tree (capturing
maxima).
The
associated ground-truth classification assigns
each time step to one of the four key phases of the simulation.
The corresponding classification
task therefore consists in identifying, for each time step, to which phase
it belongs. The ground-truth classification is as follows:
\begin{itemize}
  \item \textbf{Class 1} (10 members): 0.0200,
0.0300,
0.0400,
0.0500,
0.0600,
0.0700,
0.0800,
0.0900,
0.1000,
0.1100
  \item \textbf{Class 2} (10 members):
0.2700,
0.2800,
0.2900,
0.3000,
0.3100,
0.3200,
0.3300,
0.3400,
0.3500,
0.3600
  \item \textbf{Class 3} (10 members):
  0.5900,
0.6000,
0.6100,
0.6200,
0.6300,
0.6400,
0.6500,
0.6600,
0.6700,
0.6800
  \item \textbf{Class 4} (10 members):
  0.9100,
0.9200,
0.9300,
0.9400,
0.9500,
0.9600,
0.9700,
0.9800,
0.9900,
1.0000
\end{itemize}
}

\subsection{\revision{Volcanic eruptions}}
\revision{This ensemble is composed of 12 members, given as 2D regular grids
(sampled at $500\times500$, implicitly triangulated by TTK). Each member
corresponds to an observation of a volcanic eruption, obtained by satellite
imaging (as this data exhibits a bit of noise, it has been pre-simplified by
removing all saddle-maxima pairs with a persistence lower than 0.5\% of the
data range).
The original data
has been made available in the context of the SciVis contest 2014
\cite{scivis2014}.
The considered scalar field is the sulfur dioxide concentration, for which
salient maxima correspond to volcanic eruptions.
Thus, each observation is represented by the split tree (capturing maxima).
Each member corresponds to a specific acquisition period, itself corresponding
to the eruption of one particular volcano at the surface of the Earth.
The associated ground-truth classification assigns observations acquired in the
same period of time to the same class.
The corresponding classification task therefore
consists in identifying, for each observation (taken at a specified date), the
erupting volcano it corresponds to.
The ground-truth classification is as
follows:
\begin{itemize}
  \item \textbf{Class 1} (4 members):
  150\_am,
150\_pm,
151\_am,
151\_pm
  \item \textbf{Class 2} (4 members):
  156\_am,
156\_pm,
157\_am,
157\_pm
  \item \textbf{Class 3} (4 members):
  164\_am,
164\_pm,
165\_am,
165\_pm
\end{itemize}
}

\subsection{\revision{Ionization front (3D)}}
\revision{This ensemble is composed of 16 members, given as 3D regular grids
(sampled at $300\times124\times124$, implicitly triangulated by TTK). Each
member
corresponds to a time step of a simulation of ionization front propagation
\cite{scivis2008}.
For this application, large ionization flares are well captured by salient
maxima of the ion concentration. Thus,
split trees (capturing maxima) are considered for this ensemble.
The associated ground-truth classification assigns each time step to one of the
four key phases of the simulation.
The
corresponding classification task therefore consists in identifying, for each
time step, to which phase it belongs. The ground-truth classification is as
follows:
\begin{itemize}
  \item \textbf{Class 1} (4 members):
  0025,
0026,
0027,
0028
  \item \textbf{Class 2} (4 members):
  0075,
0076,
0077,
0078,
  \item \textbf{Class 3} (4 members):
0125,
0126,
0127,
0128
  \item \textbf{Class 4} (4 members):
  0175,
0176,
0177,
0178,
\end{itemize}
}

\subsection{\revision{Ionization front (2D)}}
\revision{This ensemble is a 2D version of the above ensemble, where
the dataset providers have selected a 2D slice in the center of the volume
(sampled at $600\times248$). The associated classification task is therefore
identical.}

\subsection{\revision{Earthquake}}
\revision{This ensemble is composed of 12 members, given as 3D regular grids
(sampled at $375\times188\times50$, implicitly triangulated by TTK). Each member
corresponds to a time step of the simulation of an earthquake at the San
Andreas fault \cite{scivis2006}.
For this application, the shock wave can be tracked with the local maxima of
the wave front velocity magnitude (this scalar field is pre-processed to
pre-simplify all saddle-maxima pairs with a persistence smaller than 0.05\%
of the data range).
Thus,
split trees (capturing maxima) are considered for this ensemble.
The associated ground-truth classification assigns each time step to one of the
three key phases of the simulation.
The
corresponding classification task therefore consists in identifying, for each
time step, to which phase it belongs. The ground-truth classification is as
follows:
\begin{itemize}
  \item \textbf{Class 1} (4 members):
  002700,
002900,
003100,
003300
  \item \textbf{Class 2} (4 members):
  007700,
007900,
008100,
008300
  \item \textbf{Class 3} (4 members):
  011700,
011900,
012100,
012300
\end{itemize}
}

\subsection{\revision{Isabel}}
\revision{This ensemble is composed of 12 members, given as 3D regular grids
(sampled at $250\times250\times50$, implicitly triangulated by TTK). Each member
corresponds to a time step of the simulation of the Isabel hurricane
\cite{scivis2004}. This ensemble has been used in previous work
\cite{favelier2018, vidal_vis19} and the corresponding VTK files are available
at the following address:
\url{https://github.com/julesvidal/wasserstein-pd-barycenter}. In this
application, the eyewall of the hurricane is typically characterized by high
wind velocities, well captured by the the maxima of the flow velocity. Thus,
split trees (capturing maxima) are considered for this ensemble.
The associated ground-truth classification assigns each time step to one of the
three key phases (formation, drift, landfall) of the hurricane simulation. The
corresponding classification task therefore consists in identifying, for each
member, to which key phase it belongs. The ground-truth classification is as
follows:
\begin{itemize}
  \item \textbf{Class 1} (4 members): 2, 3, 4, 5
  \item \textbf{Class 2} (4 members): 30, 31, 32, 33
  \item \textbf{Class 3} (4 members): 45, 46, 47, 48
\end{itemize}
}

\subsection{\revision{Starting vortex}}
\revision{This ensemble is composed of 12 members, given as 2D regular grids
(sampled at $1500\times1000$, implicitly triangulated by TTK). It has been
generated with the
Gerris flow solver \cite{gerris} and was provided in previous work
\cite{favelier2018, vidal_vis19}. It is available at the following address:
\url{https://github.com/julesvidal/wasserstein-pd-barycenter}.}
\revision{The data models flow turbulence behind a wing, for two ranges of wing
inclination angles. The considered
scalar field is the orthogonal component of the curl of the flow velocity.
This ensemble corresponds to a
parameter study (in this case, studying the effect of the wing configuration on
turbulence), which is a typical task in numerical simulation. In this
application, salient extrema are typically considered as reliable estimations of
the center of vortices. Thus, each run is represented by two merge trees (the
join tree -- capturing minima, and the split tree, capturing maxima),
which are
processed independently by our algorithms.}
\revision{The associated ground-truth classification assigns members computed
with similar inclination angles to the same class. The corresponding
classification task therefore consists in identifying, for a given ensemble
member, its correct wing configuration class. The ground-truth classification is
as
follows:
\begin{itemize}
  \item \textbf{Class 1} (6 members): Angle=2, Angle=3, Angle=4, Angle=5,
Angle=6, Angle=8
  \item \textbf{Class 2} (6 members): Angle=38, Angle=39, Angle=40, Angle=41,
Angle=42, Angle=43
\end{itemize}
}

\subsection{\revision{Sea surface height}}
\revision{This ensemble is composed of 48 members, given as 2D regular grids
(sampled at $1440\times720$, implicitly triangulated by TTK). Each member
corresponds to an observation of the sea surface height at the surface of the
Earth, taken in January, April, July and October 2012. The original data can be
found at the following address:
\url{https://ecco.jpl.nasa.gov/products/all/}. This ensemble has been used in
previous work \cite{favelier2018, vidal_vis19} and the corresponding VTK files
are available at the following address:
\url{https://github.com/julesvidal/wasserstein-pd-barycenter}. In this
application, the features of interest are the center of eddies, which can be
reliably estimated with height extrema. Thus, each observation is represented
by two merge trees (the
join tree -- capturing minima, and the split tree, capturing maxima),
which are
processed independently by our algorithms.
The associated ground-truth classification assigns observations acquired in the
same month to the same class. The corresponding classification task therefore
consists in identifying, for each observation (taken at a specified date), the
season in which it has been acquired. The ground-truth classification is as
follows:
\begin{itemize}
  \item \textbf{Class 1} (12 members): 20120111, 20120115, 20120116, 20120117,
20120118, 20120119, 20120120, 20120121, 20120123, 20120128, 20120129
  \item \textbf{Class 2} (12 members): 20120419, 20120420, 20120421, 20120422,
20120423, 20120424, 20120425, 20120426, 20120427, 20120428, 20120429, 20120430
  \item \textbf{Class 3} (12 members): 20120711, 20120712, 20120713, 20120714,
20120715, 20120716, 20120717, 20120718, 20120719, 20120720, 20120721, 20120722
  \item \textbf{Class 4} (12 members): 20121008, 20121009, 20121010, 20121011,
20121012, 20121016, 20121017, 20121018, 20121019, 20121020, 20121022, 20121023
\end{itemize}
}

\subsection{\revision{Vortex street}}
\revision{This ensemble is composed of 45 members, given as 2D regular grids
(sampled at $300\times100$, implicitly triangulated by TTK). It has been
generated with the Gerris flow solver \cite{gerris} and was provided in previous
work \cite{favelier2018, vidal_vis19}. It is available at the following address:
\url{https://github.com/julesvidal/wasserstein-pd-barycenter}.}
\revision{The data models flow turbulence behind an obstacle. The considered
scalar field is the orthogonal component of the curl of the flow velocity, for 5
fluids of different viscosity. This ensemble corresponds to a
parameter study (in this case, studying the effect of viscosity on
turbulence), which is a typical task in numerical simulation. In this
application, salient extrema are typically considered as reliable estimations of
the center of vortices. Thus, each run is represented by two merge trees (the
join tree -- capturing minima, and the split tree, capturing maxima),
which are
processed independently by our algorithms.}
\revision{The associated ground-truth classification assigns members computed
with similar viscosities to the same class. The corresponding
classification task therefore consists in identifying, for a given ensemble
member, its correct viscosity class. The ground-truth classification is as
follows:
\begin{itemize}
  \item \textbf{Class 1} (9 members): Viscosity=100.0, Viscosity=100.1,
Viscosity=100.2, Viscosity=100.3, Viscosity=100.4, Viscosity=100.5,
Viscosity=100.6, Viscosity=100.7, Viscosity=100.9
  \item \textbf{Class 2} (9 members): Viscosity=160.0, Viscosity=160.1,
Viscosity=160.2,
Viscosity=160.3, Viscosity=160.4, Viscosity=160.5, Viscosity=160.6,
Viscosity=160.7, Viscosity=160.8
  \item \textbf{Class 3} (9 members): Viscosity=200.0, Viscosity=200.1,
Viscosity=200.2,
Viscosity=200.3, Viscosity=200.4, Viscosity=200.5, Viscosity=200.6,
Viscosity=200.7, Viscosity=200.8
  \item \textbf{Class 4} (9 members): Viscosity=50.0, Viscosity=50.1,
Viscosity=50.2,
Viscosity=50.3, Viscosity=50.5, Viscosity=50.6,
Viscosity=50.7, Viscosity=50.8, Viscosity=50.9
  \item \textbf{Class 5} (9 members): Viscosity=60.1, Viscosity=60.2,
Viscosity=60.3, Viscosity=60.4, Viscosity=60.5, Viscosity=60.6,
Viscosity=60.7, Viscosity=60.8, Viscosity=60.9
\end{itemize}
}

\section{\revision{Parameter analysis}}

\revision{In this section, we study the practical effect of the parameters of
our approach.
In particular, we extend the empirical stability evaluation of our metric with
regard to all the parameters of our approach and we illustrate their effect on
geodesic computation.}

\subsection{\revision{Interpretation}}
\revision{The first parameter of our approach is $\epsilon_1 \in [0, 1]$. It
dictates the merge of saddles in the input trees, to mitigate saddle swap
instabilities, as previously documented by Sridharamurthy et al.
\cite{SridharamurthyM20}.
Adjacent saddles in the input trees are merged if their
relative
difference in scalar value (relative to the
largest function difference between adjacent saddles)
is 
\emph{smaller} than
$\epsilon_1$.
For $\epsilon_1=0$, no saddle merge  is performed whereas for
$\epsilon_1=1$, all saddles are merged and
$\wassersteinTree$ becomes
equivalent to the $L^2$ Wasserstein distance between persistence
diagrams, noted $\wasserstein{2}$.
}

\revision{The local normalization step  of 
our
framework (Section 4.2 of the main manuscript)
guarantees the topological consistency of the interpolated
merge trees (Fig. 8 of the main manuscript). However, this normalization
shrinks 
the birth/death values of 
all the input branches to the interval $[0, 1]$, irrespective of their
original persistence. To mitigate this effect, the input BDTs are
pre-processed, so that branches with small initial persistence (i.e.
\emph{small branches}) are not given too much importance in the metric. In
particular, small branches are moved up the input BDT if their persistence
relative to their parent is \emph{larger} than $\epsilon_2 \in [0, 1]$.
When $\epsilon_2 = 0$, all branches are moved up to the root of the BDT and
again, $\wassersteinTree$  becomes equivalent to $\wasserstein{2}$.
When $\epsilon_2 = 1$, no branch is moved up the BDT and $\epsilon_2$ has no
effect on the outcome (i.e. the input BDT is left unchanged).
In practice, we
recommend the default value $\epsilon_2 = 0.95$: 
if a branch
$b$ has a nearly identical persistence to that of its parent $b'$, it is moved
higher in the BDT, so that its normalized persistence becomes nearly identical
to that of its parent $b'$ (instead of being artificially larger due to the
local normalization).}

\revision{The parameter $\epsilon_3 \in [0, 1]$ further restricts the 
application of the
above
BDT pre-processing, by only considering (for displacement up the BDT) the
branches with a relative persistence (with respect to the overall data range)
smaller than $\epsilon_3$.
When $\epsilon_3 = 1$, all branches are subject to the above pre-processing
and $\epsilon_2$ fully dictates the BDT pre-processing. When $\epsilon_3 = 0$,
no branch is moved up the BDT and the two parameters $\epsilon_2$ and $\epsilon_3$ have no effect on the outcome.
In practice, we recommend the default value $\epsilon_3 = 0.9$, which
prevents
%
the most persistent branches from moving up the BDT.}

\revision{Overall, when the parameters $(\epsilon_1, \epsilon_2, \epsilon_3)$ 
are set to the values $(0, 1, 1)$, the input trees are \emph{not} pre-processed
by the
above procedures (i.e. they are left unchanged) and their structure has a 
strong impact on $\wassersteinTree$. 
When $\epsilon_1 = 1$ or when $\epsilon_2 = 0$, 
$\wassersteinTree$ becomes equivalent to $\wasserstein{2}$ and 
the structure of the input trees has no impact anymore on the metric. 
In-between values balance the importance of the structure of the trees on the 
metric. We recommend the default values $(0.05, 0.95, 0.9)$, which provides an 
acceptable stability with regard to saddle swaps (mitigated by $\epsilon_1$) and 
which gives a reasonable importance to small branches in the metric (controlled 
by $\epsilon_2$ and $\epsilon_3$, which are dependent parameters).}


\subsection{\revision{Metric stability}}
\label{sec_stability}

\begin{figure}
\includegraphics[width=\linewidth]{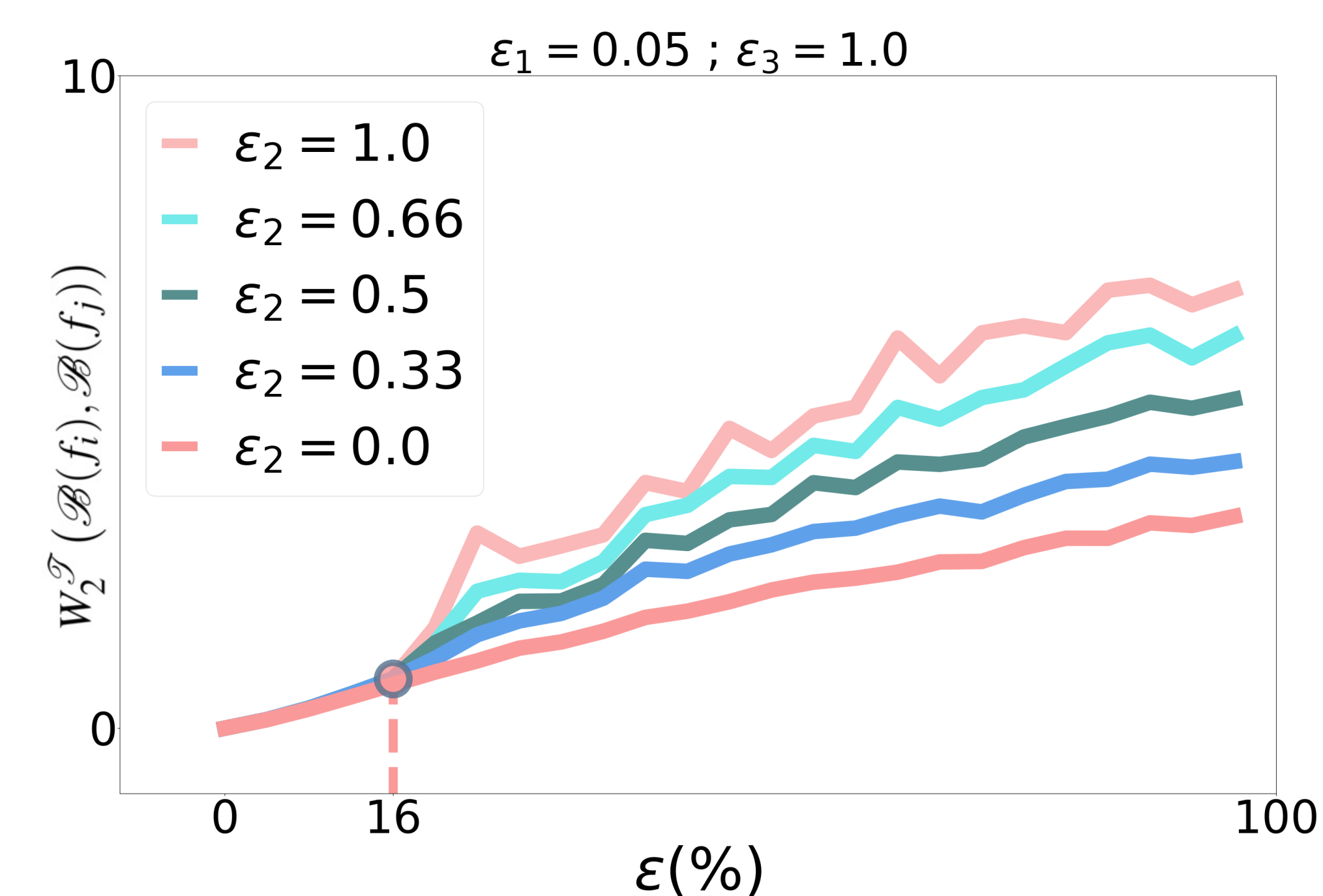}
\caption{\revision{Empirical stability evaluation with regard to $\epsilon_2$.
Given an input scalar field $f_i$, a
noisy version $f_j$ is created by inserting a random noise of increasing
amplitude $\epsilon$ (cf. Figure 14 of the main manuscript). The evolution of
$\wassersteinTree\big(\branchtree(f_i), \branchtree(f_j)\big)$
with $\epsilon$ is reported for varying values of $\epsilon_2$.}
}
\label{fig_stabilityEpsilon2}
\end{figure}

\begin{figure}
\includegraphics[width=\linewidth]{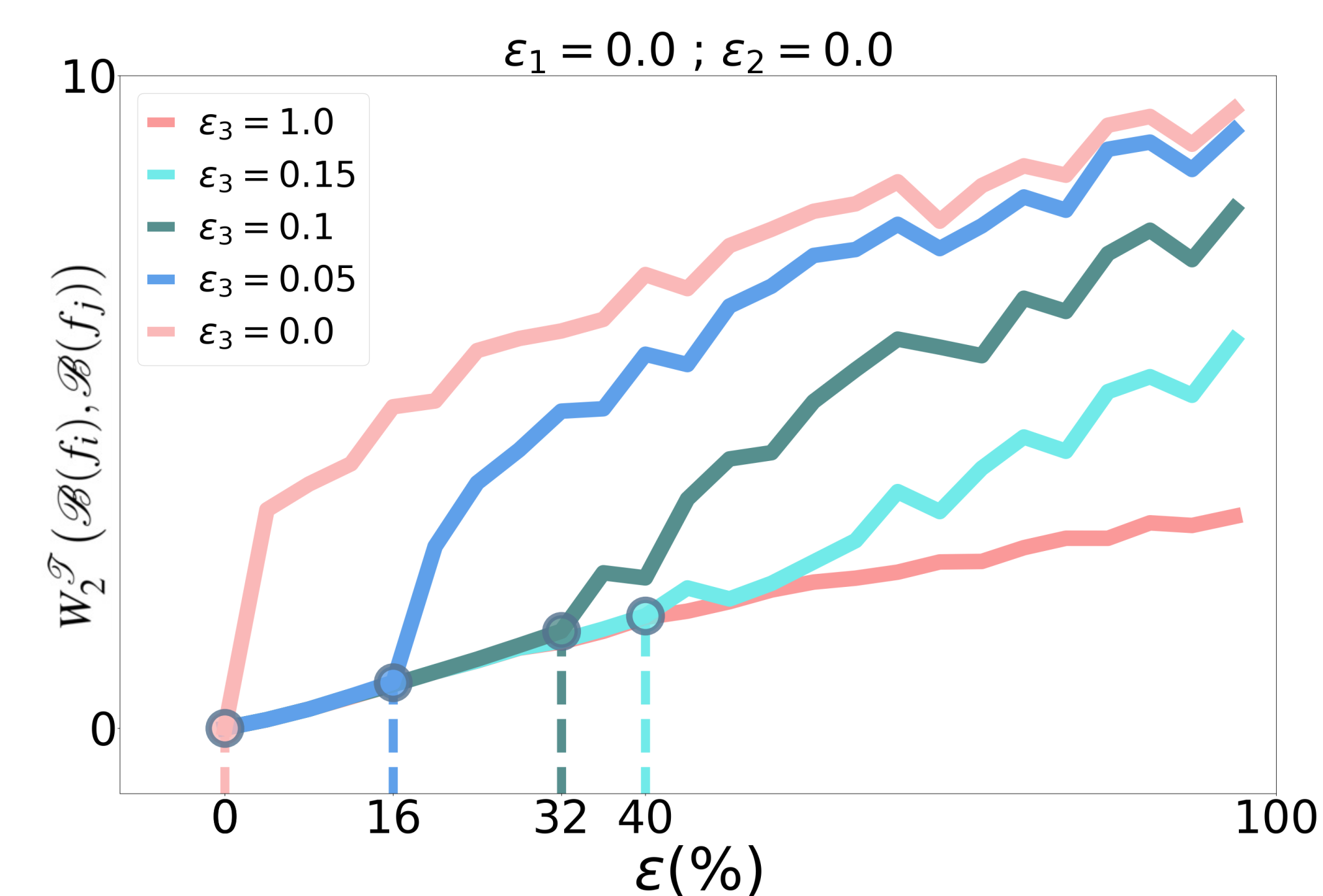}
\caption{\revision{Empirical stability evaluation with regard to $\epsilon_3$.
Given an input scalar field $f_i$, a
noisy version $f_j$ is created by inserting a random noise of increasing
amplitude $\epsilon$ (cf. Figure 14 of the main manuscript). The evolution of
$\wassersteinTree\big(\branchtree(f_i), \branchtree(f_j)\big)$
with $\epsilon$ is reported for varying values of $\epsilon_3$.}
}
\label{fig_stabilityEpsilon3}
\end{figure}

\revision{Figure 14 of the main manuscript provides an empirical stability
evaluation of our new metric
$\wassersteinTree$,
as a function of an input
perturbation, modeled by a random noise of amplitude $\epsilon$. In particular,
this experiment is achieved for several values of
$\epsilon_1$. The conclusion of this experiment is that $\wassersteinTree$ is
not stable when $\epsilon_1=0$ (sudden increase in $\wassersteinTree$ for
small values of $\epsilon$) and that it is stable when $\epsilon_1=1$ (as 
anticipated
\cite{Turner2014}). For in-between values, $\wassersteinTree$ is stable until a
transition point (colored dots in Fig. 14 of the main manuscript), located at 
increasing noise levels ($\epsilon$) for
increasing values of $\epsilon_1$. In particular, for the recommended default
value $\epsilon_1=0.05$, $\wassersteinTree$ is stable up to a perturbation
noise of amplitude $16\%$ (of the overall data range).
}

\revision{In the following, we perform the same study for the other parameters
of our approach, $\epsilon_2$ and $\epsilon_3$. 
\autoref{fig_stabilityEpsilon2} studies the practical stability of 
$\wassersteinTree$, for several values of $\epsilon_2$. For this experiment, 
$\epsilon_3$ has been set to $1$ (then, only $\epsilon_2$ has an impact on the 
BDT pre-processing described in the previous section). Moreover, $\epsilon_1$ 
has been set to its 
recommended 
value, $0.05$. Several curves are 
reported, one per $\epsilon_2$ values. 
For $\epsilon_2 = 0$, all branches are moved up the BDT (irrespective of 
$\epsilon_1$) and $\wassersteinTree$ becomes equivalent to $\wasserstein{2}$ 
and the corresponding curve (red) exactly coincides with the light blue curve 
of the Figure 14 of the main manuscript ($\epsilon_1 = 1$). For $\epsilon_2 = 
1$, the input BDT is not pre-processed at all and the corresponding curve 
(pink) exactly coincides with the cyan curve of Figure 14 of the main 
manuscript (obtained for the default value $\epsilon_1=0.05$). In-between 
values of $\epsilon_2$ result in continuous transitions between these two 
extreme cases (blue, green and cyan curves).}

\revision{\autoref{fig_stabilityEpsilon3} studies the practical 
stability of $\wassersteinTree$, for several values of $\epsilon_3$. For this 
experiment, 
we set
$\epsilon_1 = 0$ and $\epsilon_2 = 0$, to 
better isolate the effect of $\epsilon_3$. When $\epsilon_3 = 1$, all the 
branches of the input BDTs are subject to the BDT pre-processing. Since 
$\epsilon_2=0$, all branches are moved up to the root and $\wassersteinTree$ 
becomes equivalent to $\wasserstein{2}$ and the corresponding curve (red) 
exactly coincides with the light blue curve of the Figure 14 of the main 
manuscript ($\epsilon_1=1$). When $\epsilon_3 = 0$, no branch is moved up in 
the input BDTs and the corresponding curve (pink) exactly coincides with the 
grey curve of the Figure 14 of the main manuscript ($\epsilon_1 = 0$). 
In-between values of $\epsilon_3$ result in transitions between these two
extreme
cases (blue, green and cyan curves), with transition points (similar to the 
Figure 14 of the main manuscript), before which $\wassersteinTree$ is
stable. Note however, that since it is dependent on $\epsilon_2$ (default 
value: $0.95$), $\epsilon_3$ has only a very mild practical impact on the 
metric.}

\subsection{\revision{Geodesic analysis}}

\begin{figure*}
  \centering
\includegraphics[width=.45\linewidth]{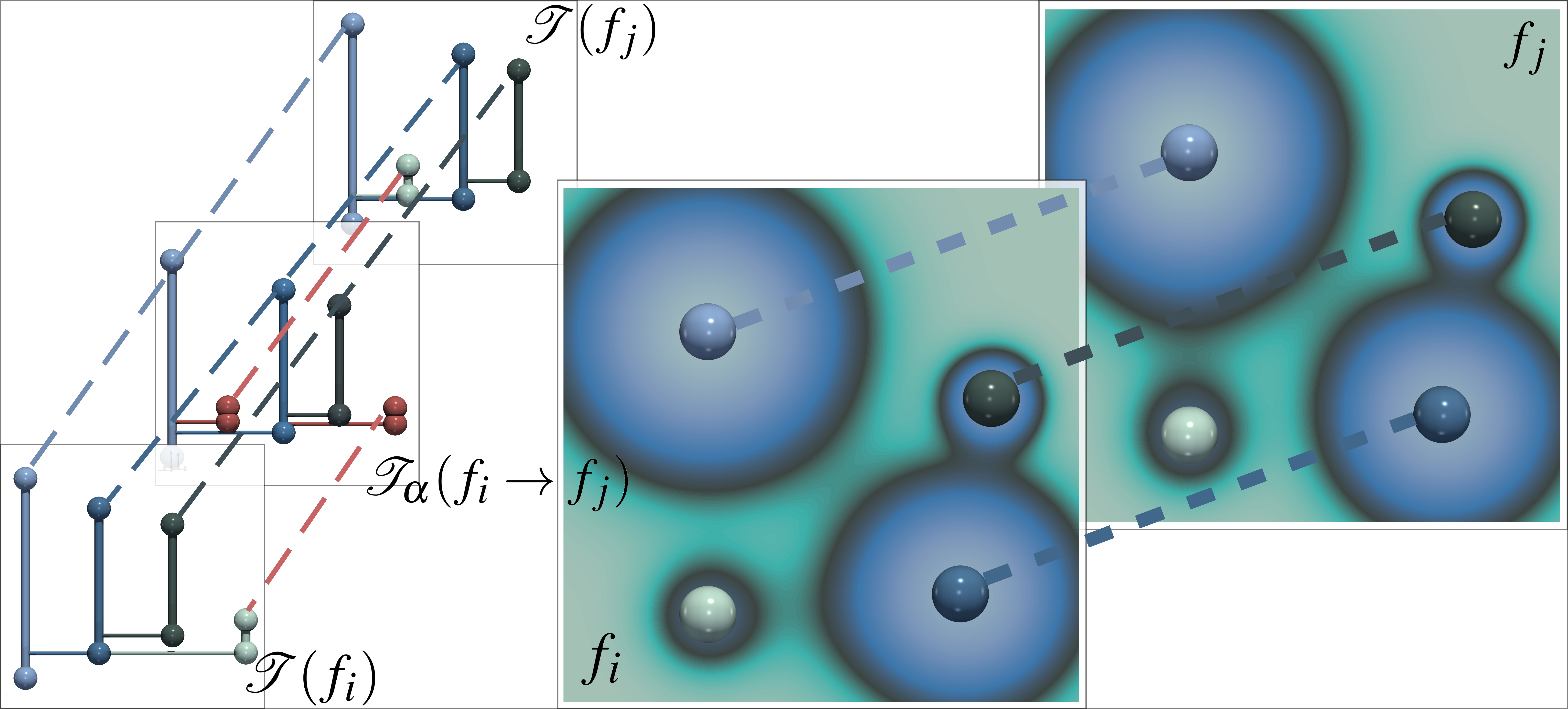}
  \hfill
  \includegraphics[width=.45\linewidth]{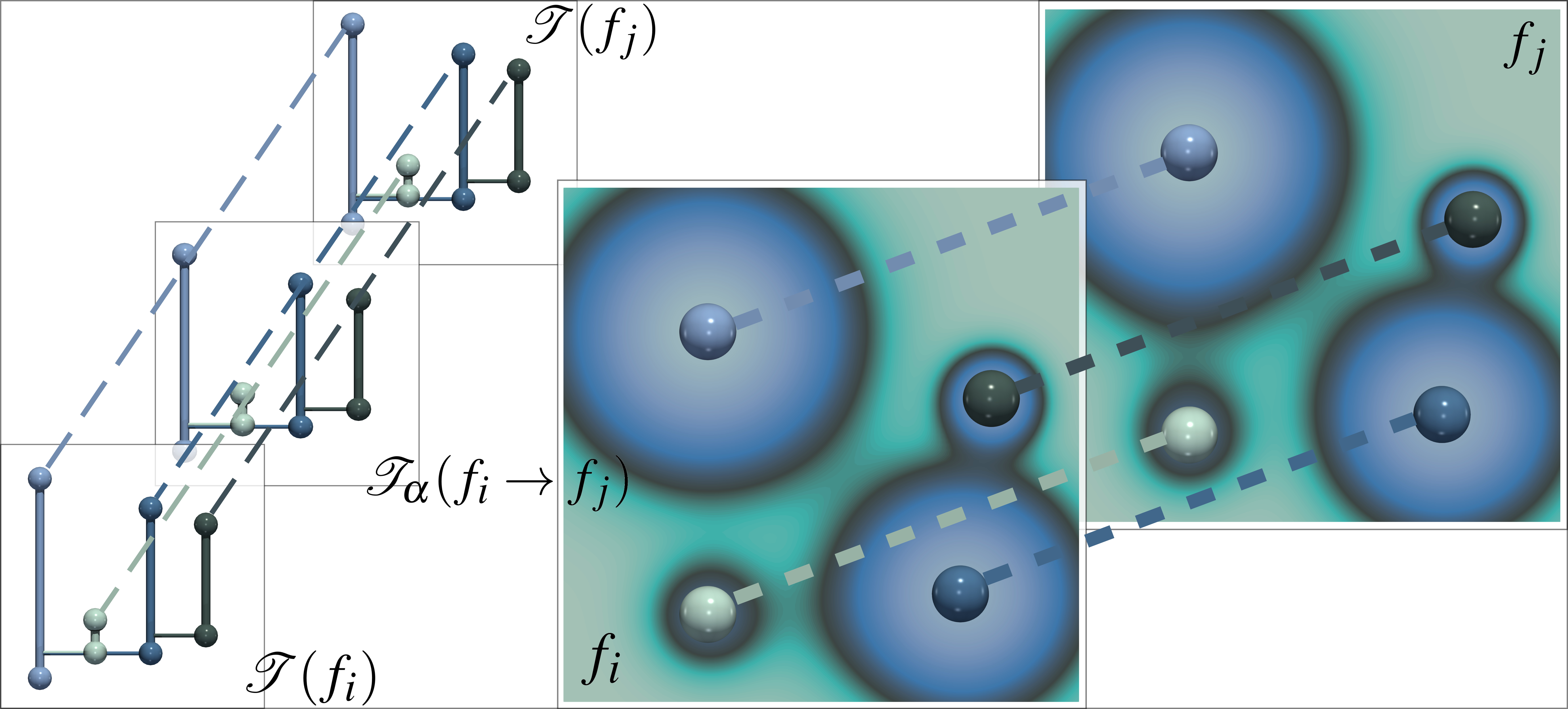}
  \caption{\revision{Impact of the parameter $\epsilon_1$ on geodesic
computation (left: $\epsilon_1 = 0$, right: $\epsilon_1=0.05$). In this
example (left), the white branch in $\mergetree(f_i)$ is not matched to the
white branch in $\mergetree(f_j)$ as they have distinct depths in the
corresponding BDTs (2 versus 1). However, these features are visually similar
in the data (Gaussians with the white maximum in $f_i$ and $f_j$, bottom left
corner of the domain). With
$\epsilon_1 = 0.05$ (right), the saddle of the white branch in
$\mergetree(f_i)$ gets merged with its ancestor saddle (whose $f_i$ value was
less than $\epsilon_1$ away). Consequently, the white branch gets moved up the
BDT (the white branch is attached to the main light blue branch in
$\mergetree(f_i)$, right). Since they now have identical depths in the
corresponding BDTs, the white branches of $\mergetree(f_i)$ and
$\mergetree(f_j)$ can now be matched together (right), which results in an
overall matching (and geodesic) between these two trees which better conveys the
resemblance between the two scalar fields $f_i$ and $f_j$. Equivalently, one
can interpret this procedure of saddle merge in the input
trees as a modification of the input scalar field, turning $f_i$ into $f_j$.
In particular, this field modification disconnects the Gaussian with the white
maximum from the Gaussian with the dark blue maximum ($f_i$) and reconnects it
to the Gaussian with the light blue maximum ($f_j$).
}
  }
  \label{fig_interpolation1}
\end{figure*}

\begin{figure*}
  \centering
\includegraphics[width=.45\linewidth]{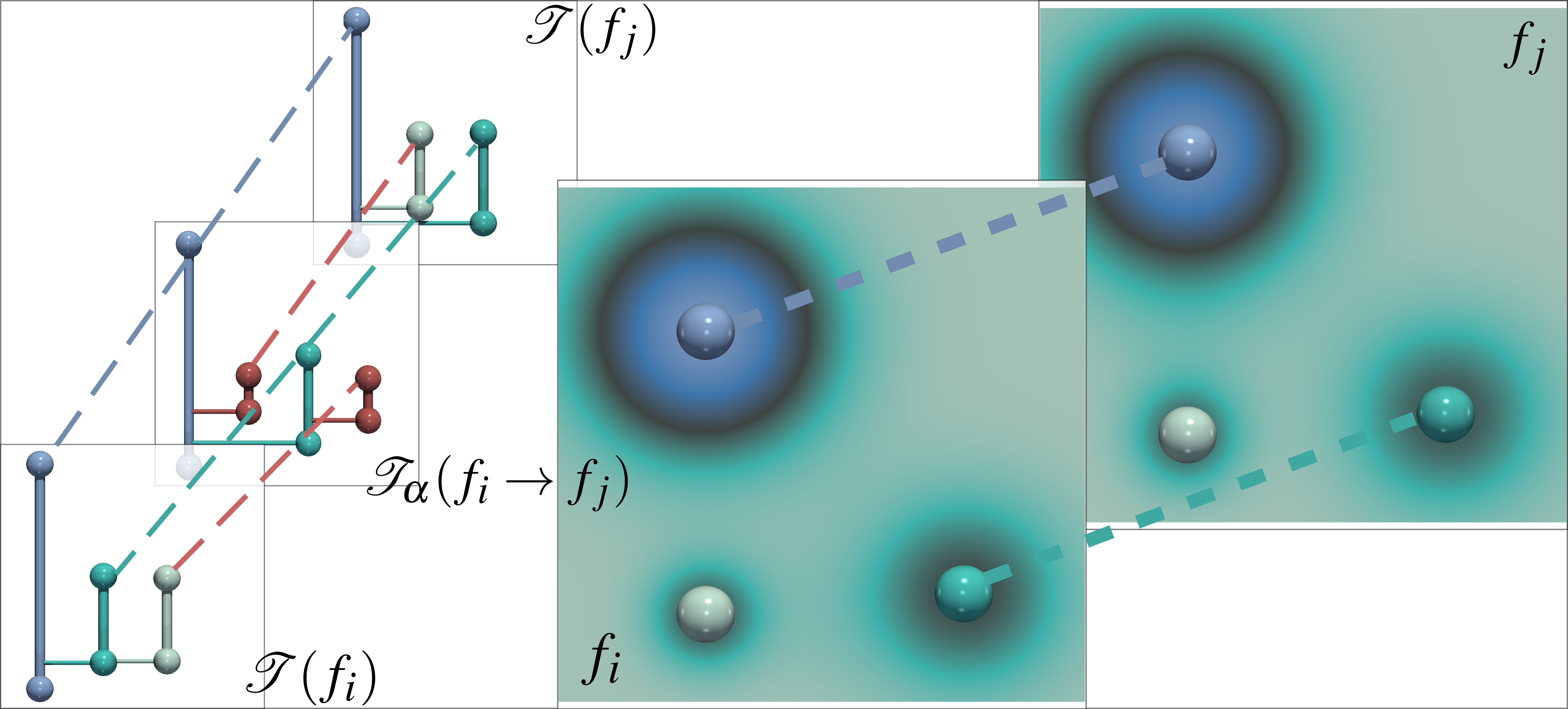}
  \hfill
  \includegraphics[width=.45\linewidth]{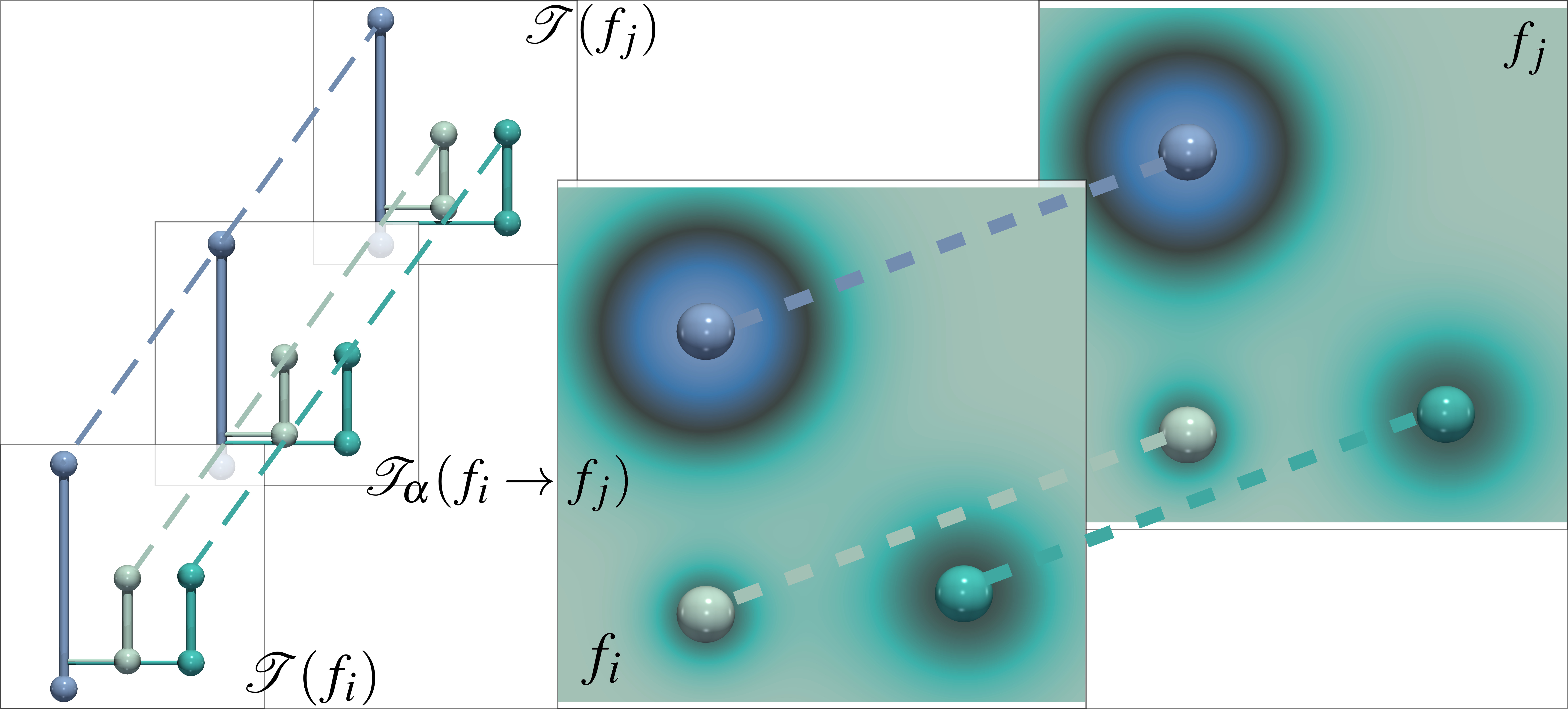}
  \caption{\revision{Impact of the parameter $\epsilon_2$ on geodesic
computation (left: $\epsilon_2 = 1$, right: $\epsilon_2 = 0.95$).
In this example (left), the white branch in $\mergetree(f_i)$ is not
matched to the
white branch in $\mergetree(f_j)$ as they have distinct depths in the
corresponding BDTs (2 versus 1). Moreover, given the function difference
between the white branch's saddle and its ancestor, that branch cannot be moved
up the BDT under the effect of the $\epsilon_1$ procedure (above).
The white branch in $\mergetree(f_i)$ has a
persistence nearly identical to its parent (cyan). Thus, after local
normalization (necessary to guarantee the topological consistency of the
interpolated trees), its normalized persistence would become artificially high,
which can have an undesirable effect on the metric.
The BDT pre-processing
addresses this issue and
moves up the BDT branches with a relative persistence to their parent larger
than $\epsilon_2$ (recommended default value: $0.95$). In this example (right),
the white branch in $\mergetree(f_i)$ moves up the BDT and becomes adjacent to
the main light blue branch in $\mergetree(f_i)$.
Since they now have identical depths in the corresponding BDTs, the white
branches of
$\mergetree(f_i)$ and $\mergetree(f_j)$ can now be matched together (right),
which better conveys the resemblance between the two scalar fields $f_i$ and
$f_j$.
Equivalently, one
can interpret this procedure of BDT pre-processing
as a modification of the input scalar field, turning $f_i$ into $f_j$.
In particular, this field modification disconnects the Gaussian with the white
maximum from the Gaussian with the cyan maximum ($f_i$) and reconnects it
to the Gaussian with the light blue maximum ($f_j$).
}
  }
  \label{fig_interpolation2}
\end{figure*}

\begin{figure*}
  \centering
\includegraphics[width=.45\linewidth]{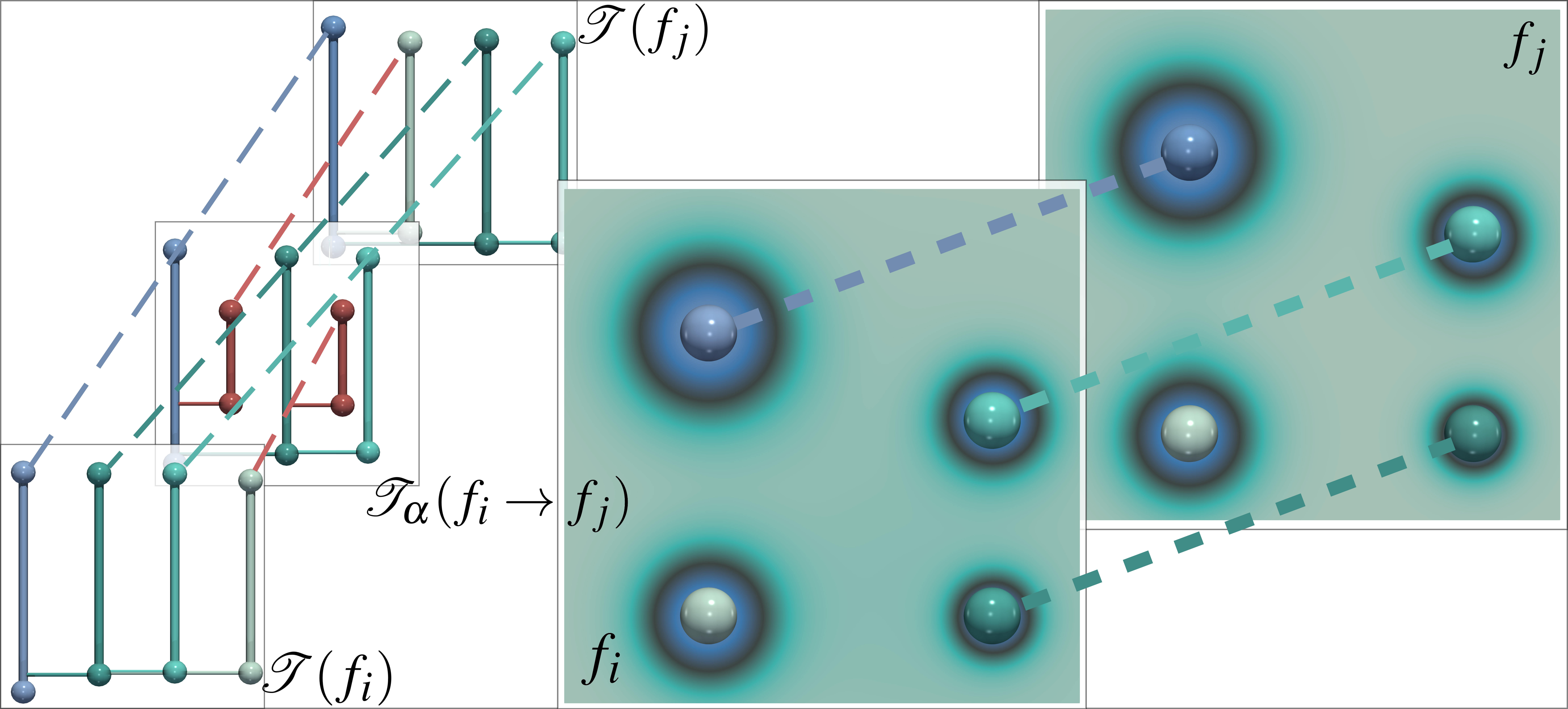}
  \hfill
  \includegraphics[width=.45\linewidth]{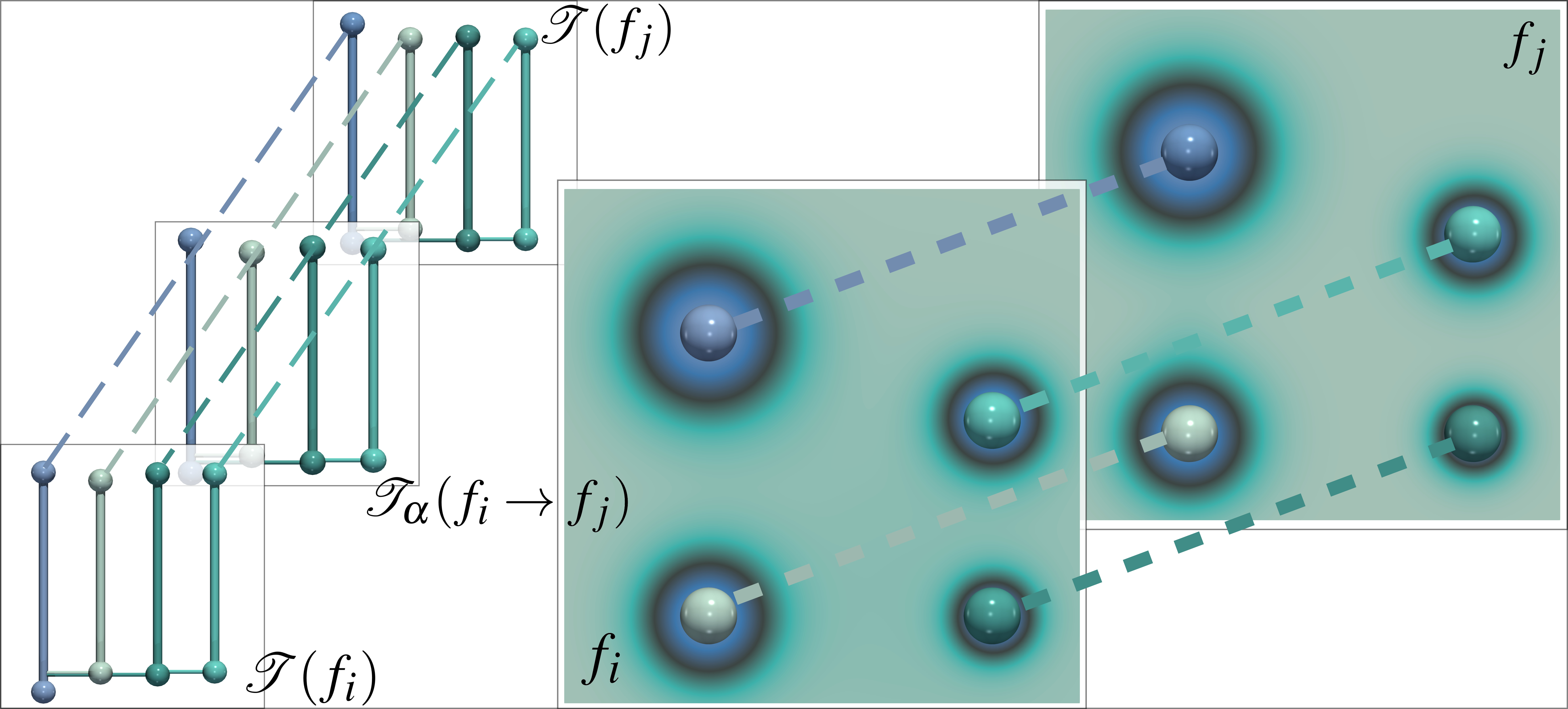}
  \caption{\revision{Effect of the parameter $\epsilon_3$ on geodesic
computation (left: $\epsilon_3 = 0$, right: $\epsilon_3 = 0.9$). In this
example (left), the white branch in $\mergetree(f_i)$ is not matched to the
white branch in $\mergetree(f_j)$ as they have distinct depths in the
corresponding BDTs (3 versus 1). Applying the above BDT pre-processing
($\epsilon_2$) to all branches would move the cyan branch in $\mergetree(f_i)$
up the BDT, which would prevent it to match to the cyan branch in
$\mergetree(f_j)$. The parameter $\epsilon_3$ restricts the application of the
above BDT pre-processing and prevents the movement of
the most persistent
branches (relative persistence larger than $\epsilon_3$, default: $0.9$).
In this example (right), the white branch in
$\mergetree(f_i)$ moves up the BDT and becomes adjacent to the main light blue
branch in $\mergetree(f_i)$. Since they now have identical depths in the
corresponding BDTs, the white branches of $\mergetree(f_i)$ and
$\mergetree(f_j)$
can now be matched together (right), which results in an
overall matching (and geodesic) between these two trees which better conveys the
resemblance between the two scalar fields $f_i$ and $f_j$. Equivalently, one
can interpret this procedure on the BDTs
as a modification of the input scalar field, turning $f_i$ into $f_j$.
In particular, this field modification disconnects the Gaussian with the white
maximum from the Gaussian with the dark green maximum ($f_i$) and reconnects it
to the Gaussian with the light blue maximum ($f_j$).
  }
  }
  \label{fig_interpolation3}
\end{figure*}

\revision{Figures \ref{fig_interpolation1}, \ref{fig_interpolation2} and 
\ref{fig_interpolation3} respectively illustrate the effect of the parameters 
$\epsilon_1$, $\epsilon_2$ and $\epsilon_3$ on the geodesics between merge 
trees. In particular, each figure shows, on the left, the geodesic obtained 
with a disabling value of the parameter (no effect on the 
computation). 
In contrast, the right side of each figure shows the geodesic obtained with the 
recommended default value of the parameter, to clearly visualize its impact.}

\revision{Overall, as discussed in the detailed captions, these three 
parameters have the effect of moving branches up the input BDTs, hence reducing 
the structural impact of the trees on the metric, but also improving its 
stability (as discussed in Section \ref{sec_stability}). In the data, moving a 
branch up the BDT corresponds to only slight modifications,
which
consist in reconnecting maxima to distinct saddles. For each parameter, the 
resulting pre-processing
addresses
cases where nearby saddles have very
close function values, which impacts the stability of the metric. Similarly to 
Sridharamurthy et al. \cite{SridharamurthyM20}, we mitigate this effect with 
$\epsilon_1$, but we also introduce $\epsilon_2$ and $\epsilon_3$ to 
specifically limit the importance in the metric of branches with persistence 
close to that of their parent.}

\end{document}